\documentclass[pdflatex,sn-mathphys-num]{sn-jnl}

\usepackage{graphicx}
\usepackage{multirow}
\usepackage{amsmath,amssymb,amsfonts}
\usepackage{amsthm}
\usepackage{mathrsfs}
\usepackage[title]{appendix}
\usepackage{xcolor}
\usepackage{textcomp}
\usepackage{manyfoot}
\usepackage{booktabs}
\usepackage{algorithm}
\usepackage{algorithmicx}
\usepackage{algpseudocode}
\usepackage{listings}
\usepackage{siunitx}
\usepackage[
  group-separator = {\,},
  group-minimum-digits = 4
]{siunitx}
\usepackage{subcaption}
\usepackage{todonotes}
\usepackage[export]{adjustbox}

\theoremstyle{thmstyleone}

\theoremstyle{thmstyletwo}

\theoremstyle{thmstylethree}

\begin{document}

\title[On the Impact of Intra-node Communication on Supercomputers Interconnection Network]{On the Impact of Intra-node Communication in the Performance of Supercomputer and Data Center Interconnection Networks}

\author*[1]{\fnm{Joaquin} \sur{Tarraga-Moreno}}\email{antonioj.tarraga@uclm.es}

\author[1]{\fnm{Jesus} \sur{Escudero-Sahuquillo}}
\author[1]{\fnm{Pedro Javier} \sur{Garcia}}
\author[1]{\fnm{Francisco J.} \sur{Quiles}}

\affil[1]{\orgdiv{Department of Computing Systems}, \orgname{Universidad de Castilla-La Mancha}, \country{Spain}}

\abstract{
In the last decade, specific-purpose computing and storage devices, such as GPUs, TPUs, or high-speed storage, have been incorporated into server nodes of Supercomputers and Data centers. The development of high-bandwidth memory (HBM) enabled a much more compact form factor for these devices, thus allowing the interconnection of several of them within a server node, typically using an intra-node interconnection network (e.g., PCIe, NVLink, or Infinity Fabric). These networks allow scaling up the number of specific computing and storage devices per node. Furthermore, the inter-node networks communicate thousands of these devices placed in different server nodes in a Supercomputer or Data Center. Unfortunately, the intra- and inter-node networks may become the system's bottleneck due to the increasing communication demand among accelerators of applications such as generative AI. Although current intra-node network designs alleviate this bottleneck by increasing the bandwidth of the intra-node network, we show in this paper that such a high bandwidth for intra-node communication may hinder the inter-node communication performance when traffic from outside the node arrives at the intra-node devices, resulting in interference with intra-node traffic. To analyze the impact of this interference, we have studied the communication operations of realistic traffic patterns exploiting intra-node communication. We have developed a generic intra- and inter-node simulation model based on OMNeT++ and modeled the mentioned communication operations. We have also performed extensive simulation experiments that confirm that increasing the intra-node network bandwidth and the number of computing devices per node (i.e., accelerators) is counterproductive to the inter-node communication performance.
}


\maketitle

\section{Motivation}
\label{sec:motivation}
Supercomputers and Data Centers are composed of thousands of server nodes, which are increasing their computing power, memory, and storage capacity to cope with the growing requirements of applications and services such as scientific computing, generative AI, or social networks. While computing power requirements are currently being met using more powerful processors and specific-purpose accelerators (GPUs, TPUs, etc.), innovative technologies, such as 3D-stacked high-bandwidth memory (HBM), Non-Volatile Memory (NVMe), or Storage Class Memory (SCM), are increasing the memory and storage capacity of server nodes and the bandwidth access to these devices. These technologies allow placing unprecedented amounts of data close to where they are processed, reducing inter-node communication latency since data processing is done at multiple accelerators plugged into the same server node.

Accelerators and other devices are interconnected within server nodes through an intra-node interconnection network, which must be designed so that it does not become a bottleneck when these devices communicate. Typically, this is achieved by over-dimensioning the aggregated bandwidth of the intra-node network links. For instance, PCIe is the \emph{de facto} intra-node network in contemporary servers, offering 1~GB/s per lane for PCIe 3.0, 2~GB/s for PCIe 4.0, and 4~GB/s for PCIe 5.0. As communication operations may involve devices beyond a single node, other intra-node technologies exist, such as NVLink (NVIDIA), which is specially designed to interconnect multiple accelerators at either the same or different server nodes (or end nodes). NVLink provides speeds of 4~GB/s, 7.5~GB/s, and 15.1~GB/s per lane for Gen5, Gen6, and Gen7, respectively. Infinity Fabric (AMD) is also used to interconnect specific AMD devices for both CPU (i.e., Zen) and graphics (e.g., Vega). Therefore, the increasing link speed in the intra-node network and the variety of devices within a server node challenge the design of intra-node network technologies.

This design needs to identify potential bottlenecks, assuming that millions of communication operations are generated in the intra- and inter-node networks involving thousands of devices at different server nodes. The most straightforward communication operation (i.e., peer-to-peer or P2P) involves two end-node devices (e.g., accelerators) and can be divided into three phases. First, the sender device generates a communication transaction with another device. This transaction is split into packets at the device's intra-node network interface, which are sent throughout the intra-node network to another device within the same end node or to the network interface (NIC) connecting that end node to other end nodes. Secondly, if the transaction is addressed to a remote device, the intra-node packets are aggregated at the mentioned NIC into inter-node network packets, which are sent through the inter-node network to the destination end node. Third, at the destination end-node NIC, packets are split again into intra-node network packets and sent through the intra-node network to the destination device. Note that if a transaction is exchanged between devices within the same end node, the communication remains in the intra-node network of a single end node.

By contrast, collective operations (e.g., \emph{AllReduce} or \emph{Broadcast}) used in parallel applications, such as the deep neural networks (DNNs) training, may involve multiple computing devices at different end nodes when the model to train does not fit into a single end node~\cite{EfficientLargeScale}. When this happens, the model is stored in a distributed manner in the local memory of multiple end-node devices (e.g., the local HBM memory of accelerators). The training process of DNNs requires updating the weights stored in the memory of the involved devices, so that collective communication operations encapsulate the weight information in network transactions exchanged among these devices, and may generate a high rate of traffic the intra- and inter-node network needs to handle efficiently. Although high-performance interconnection networks have been widely investigated in the last decades, the impact of communication phases first and third, mentioned above, in the intra- and inter-node network performance is still a wide-open field of study~\cite{DeSensi24}. Indeed, it is critical to accurately analyze the network traffic patterns produced from communication operations at multiple end-node devices and their impact on the intra- and inter-node network performance. This analysis will help identify potential bottlenecks in the interconnection network.

This paper addresses the following research question: \emph{To what extent does the inter-node network traffic impact the intra-node traffic and vice-versa?} To answer this question, we have characterized a generic intra-node network architecture and the communication operations (intra- and inter-node) generated at multiple end-node devices. We have modeled the network architecture and generic but realistic traffic patterns into an OMNeT++-based simulation tool. Note that there are several proposals for system-level simulators (e.g., \textit{gem5}~\cite{gem5}), which accurately model the processor, memory, accelerators, and I/O. Other solutions model the specific accelerator-to-accelerator communication (e.g., TraceR-CODES~\cite{Bhowmik21,Jain16}). However, these models lack scalability or packet-level granularity for the combined intra-node and inter-node communication model. We have validated our simulation model using a real cluster infrastructure, where we ran specific micro-benchmarks (e.g., InfiniBand Perftest) and obtained latency and throughput metrics, which we have compared to those obtained using our simulation model. Moreover, we have performed simulation experiments configuring intra- and inter-node networks with different speeds that communicate $32$, $128$, and $512$ end nodes (with $256$, \num{1024}, and \num{4096} accelerators, respectively) using realistic communication patterns, such as those used in the training of Large Language Models (LLMs). The scale-up and scale-out simulation analyses show that the intra-node network can be a bottleneck due to the interference between intra-node and inter-node communication and the overhead when aggregating intra-node packets into inter-node packets and splitting inter-node packets into intra-node packets. We also show that these issues can be alleviated, including specific features at network devices.

The rest of the paper is organized as follows. Section~\ref{sec:background} overviews the background of inter- and intra-node interconnects and communication patterns. Section~\ref{sec:description} describes our proposed simulation model to analyze the impact of intra- and inter-node communication interference. Section~\ref{sec:evaluation} describes the simulation experiments and discusses the obtained results. Finally, Section~\ref{sec:conclusions} draws some conclusions.

\section{Background}
\label{sec:background}

\subsection{End-node devices}
\label{sec:background:components}

The Central Processing Unit (CPU) at server nodes in supercomputers or data centers manages all processes or jobs and interprets and executes instructions to coordinate software and hardware operations. For instance, in generative AI applications based on Large Language Models (LLMs), the CPU prepares and preprocesses vast datasets before sharing them with GPUs or TPUs for training. 

The Graphics Processing Unit (GPU) is a specialized processor optimized for computations requiring massive parallelism. Unlike CPUs, which excel in sequential task execution, GPUs and other accelerators feature thousands of smaller cores designed for simultaneous operations. This architecture makes GPUs critical for training and inference in LLMs. Leading manufacturers such as NVIDIA, AMD, and Intel have developed GPUs tailored for intensive AI workloads. For instance, the NVIDIA Hopper GPU~\cite{H100} combines high memory bandwidth (HBM), increased computational power, and seamless communication via the specific NVLink intra-node network. The NVIDIA Hopper GPU is integrated into the Grace Hopper Superchip~\cite{GraceHopper}, which combines the NVIDIA Hopper GPU with the NVIDIA Grace CPU. It delivers up to \num{989.4} TFLOPS in FP32 and features up to $144$GB of HBM3e memory, making it a famous solution for AI training and inference. Alternatively, the Intel Gaudi GPU~\cite{Gaudi3} is a powerful GPU designed for large-scale AI training and inference. It supports massive datasets and seamless integration into various systems and delivers up to $229$ TFLOPS with $128$GB of HBM memory. 

Solid State Drives (SSDs) and other high-speed storage technologies (e.g., NVMe or SCM) are also key components in data centers and supercomputers. They provide the necessary storage capacity and data retrieval speeds to support the large datasets typical of generative AI and HPC workloads. The performance of memory disks directly impacts data transfer rates, latency, and overall system efficiency. Emerging technologies, such as NVMe over Fabrics (NVMe-oF), enhance storage performance, bridging the gap between compute and data requirements in large-scale systems.

\subsection{Inter-node interconnection networks}
\label{sec:background:internode}

They are a critical subsystem in supercomputers and data centers to provide high-performance communication operations among thousands of end nodes at a reduced cost and reliability, since large-scale interconnection networks increase fault probability and energy efficiency to minimize the network energy fraction.

In the past, inter-node networks for Supercomputers and Data centers have differed in multiple ways. Nonetheless, in recent years, the architecture of these networks in these two different systems has converged~\cite{Hoefler22Convergence} thanks to standard communication requirements of the applications run in these systems. However, the trend nowadays is designing high-performance inter-node networks interchangeable between Supercomputers and Data Centers. Indeed, inter-node interconnection networks for these systems share critical network design aspects, such as the topology, routing algorithm, flow control, quality of service (QoS), power management, and congestion control. Although the research and innovations in these networks have been pervasive during the last decades to overcome these design aspects, the post-Exascale era and Generative AI models pose new challenges to the interconnection network design~\cite{Exascale}.

Numerous manufacturers provide fast and efficient network technologies, such as NVIDIA and its InfiniBand-based network devices, HPE and Cray, which promote the Slingshot technology, or all the partners composing the recently created Ultra Ethernet Consortium (UEC), whose primary goal is to optimize the Ethernet network technology for high-performance AI and HPC. For instance, the RDMA over Converged Ethernet (RoCE) is an advanced protocol implemented in NVIDIA devices that enables Remote Direct Memory Access (RDMA) over Ethernet networks. RoCE is designed to reduce CPU overhead and latency, and to improve throughput by bypassing traditional TCP/IP stacks. It is widely used in inter-node communication within Data Centers, especially for storage and high-performance computing (HPC). An interesting implementation of RoCE can be found in Intel's Gaudi3 system, which leverages RoCE to optimize the performance of AI training workloads. By integrating RDMA capabilities directly into the system architecture, Gaudi3 achieves efficient inter-node communication while minimizing latency and maximizing data throughput.

The different inter-node network technologies mentioned above share common design goals. For instance, regarding link speed, they feature new signaling methods, form factors, and single- and multi-mode fiber at speeds of $200$, $400$, and $800$ Gbps, which are growing exponentially, so that it is expected they will reach Terabit speed (TbE technology) within this decade.

\subsection{Intra-node interconnection networks}
\label{sec:background:intranode}

They are also a fundamental subsystem at the end nodes of Supercomputers and Data centers. Intra-node networks must offer ultra-low latency to facilitate communication between CPUs, accelerators, and memory; high bandwidth to support data-intensive workloads; and energy efficiency to ensure sustainable performance. Furthermore, as end nodes become increasingly heterogeneous due to the combination of CPUs, GPUs, and specialized accelerators, the intra-node networks cope with new communication patterns. Historically, these patterns have been simple, so the intra-node networks were based on bus-based or ring-based interconnects. However, modern intra-node networks have evolved significantly to accommodate the demand for higher performance and scalability. Nowadays, intra-node networks support emerging workloads, such as generative AI models that demand unprecedented levels of parallelism and data movement efficiency~\cite{Borkar19Heterogeneous}. Consequently, intra-node interconnection networks incorporate innovations such as in-network hardware accelerators for specific communication patterns, advanced routing algorithms, novel flow control techniques, power-aware communication strategies, etc.

PCIe is the mainstream interconnect between CPUs, GPUs, network cards, and storage devices. The PCIe bandwidth and low latency enable efficient data transfers crucial for modern high-performance workloads. PCIe speeds have evolved significantly, with PCIe 5.0 and PCIe 6.0 offering per-lane speeds of up to 3.9 and 7.5 GB/s, respectively. These advancements improve the throughput and reduce communication bottlenecks. PCIe is used, for instance, to scale up Intel's HL-325L Habana Gaudi 3 accelerators within a single node. Furthermore, PCIe is a foundational interconnect for inter-node network interface cards (NICs), directly impacting the overall system performance by reducing data movement delays and enhancing throughput. By contrast, NVLink (NVIDIA) provides significantly higher bandwidth than PCIe, with the latest versions supporting up to 900 GB/s of bidirectional bandwidth. This capability enables seamless data sharing between GPUs, CPUs, and memory subsystems, making it particularly effective for AI/ML workloads and other compute-intensive applications. NVLink is also employed in certain inter-node contexts, such as NVIDIA DGX systems, where it enhances the scalability and performance of multi-node configurations by creating tightly coupled GPU clusters.

Other intra-node network solutions, such as Infinity Fabric (AMD) or Quick Path Interconnect (Intel), and specifications such as the UALink, play an important role in the design of intra-node networks. Furthermore, emerging standards and protocols, such as the Compute Express Link (CXL), aim to unify memory and compute interconnects, enhancing intra-node communication efficiency.

\subsection{Communication Patterns}
\label{sec:background:communication}

The communication operations generated by emerging applications may threaten the performance of intra- and inter-node interconnection networks. With the rapid growth of generative AI applications, dominated by Large Language Models (LLMs), the size of these models is expanding so they no longer fit on a single accelerator or even in the accelerators of a single end node. Therefore, parallelization is needed to accelerate LLM training and inference. The parallel programming model used to leverage the computing power and storage capacity of these systems is currently based on two main types of parallelism: data parallelism and model parallelism. 

\emph{Data Parallelism} (DP)~\cite{DataParallelism} distributes the training process across multiple computational devices by dividing the dataset into smaller chunks (or batches). Each device (or group of devices) processes its assigned data portion independently, using an identical model copy. DP enables efficient scaling and faster processing. During training, communication between model copies occurs after a batch is processed when gradients are computed and updated. At this stage, each model copy exchanges gradient updates with all other copies using an \emph{AllReduce} communication operation. This communication step ensures consistent parameter updates across all model replicas. 

By contrast, \emph{Model Parallelism} (MP) distributes large-scale models across multiple computational devices, when these models do not fit in a single device. For instance, LLMs, such as GPT-3, GPT-4, and PaLM, have billions of parameters, making them too large to fit into the memory of a single accelerator. Several approaches exist to MP, such as Tensor Parallelism~\cite{TensorParallelism} and Pipeline Parallelism~\cite{PipelineParallelism}. Specifically, \emph{Tensor Parallelism} (TP) divides individual operations (e.g., matrix multiplications or tensor computations) across multiple devices. Each device performs a portion of the computation, working on tensor subsets. In the training phase, communication is required during the execution of these operations, as the intermediate results need to be shared between devices. Forward and backward passes involve frequent communication to synchronize gradients, typically using an \emph{AllReduce} or \emph{AllGather} operation for each gradient update. Due to the high communication frequency and the need for low latency and high throughput, tensor parallelism is most effective when implemented within a single computing node. By contrast, \emph{Pipeline Parallelism} (PP) splits the model into stages, such as groups of layers, assigning each stage to a separate device. Data flow sequentially through these stages, enabling pipelined execution where different input computations overlap. Communication occurs when intermediate activations or gradients are passed between stages (i.e., accelerators) during forward and backward propagation. Unlike TP, PP involves point-to-point communication between devices at stage boundaries. While PP benefits from low-latency communication, its lower communication intensity makes it feasible to implement across multiple end nodes.

\section{Modeling intra- and inter-node communication}
\label{sec:description}

This section describes the intra- and inter-node network simulation model developed to study the impact of intra-node communication on the performance of the intra- and inter-node interconnection networks, the characterization of the communication among PCIe devices, the generic intra- and inter-node simulation model, and the network traffic model reproducing realistic communication operations in Large Language Models (LLMs) parallel training.

\subsection{Baseline end node architecture}
\label{sec:description:node-arch}

\begin{figure}[!htb]
     \centering
     \includegraphics[width=.65\columnwidth]{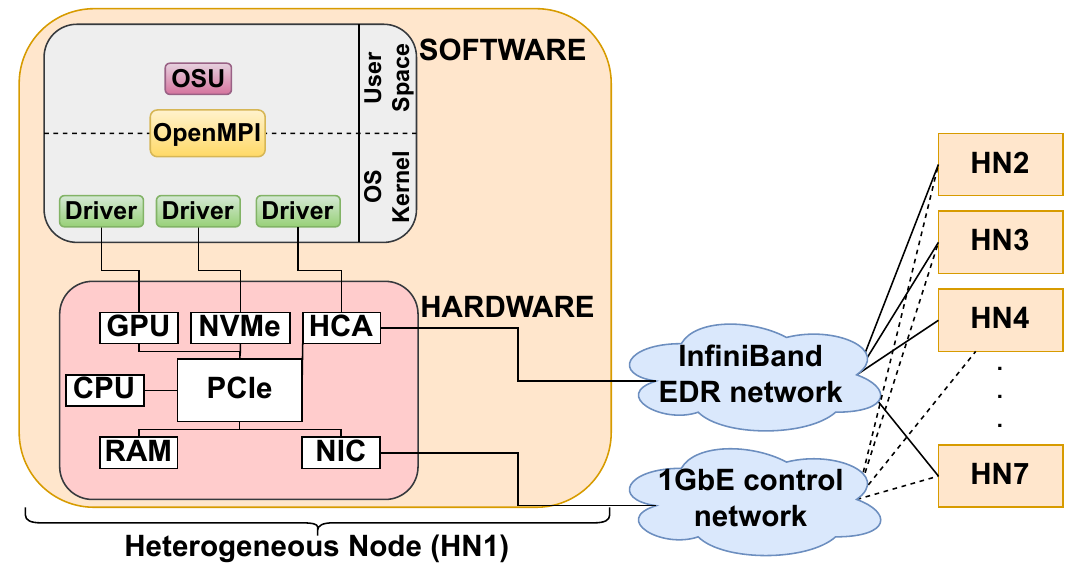}
     \caption{Node configuration in the CELLIA cluster.}
     \label{fig:desc:BigPictureTFG}
 \end{figure}

In the simulation tool, we have modeled an end node architecture based on a real system using PCIe 3.0 to interconnect a GPU, an NVMe disk, and an InfiniBand NIC, which will be used for validation purposes. Figure~\ref{fig:desc:BigPictureTFG} illustrates this infrastructure known as \emph{Cluster for the Evaluation of Low-Latency Architectures} (CELLIA) \footnote{Further details on: \url{https://www.i3a.uclm.es/raap/?page_id=1650}.}.

The CELLIA cluster comprises several computing nodes; each end node has two CPUs, Intel Xeon Silver 4116, connected to the PCIe Gen3, and $192$GB of RAM. These nodes also have an EDR InfiniBand network interface (a.k.a. HCA) MCX556A-ECAT ConnectX-5 dual-port. The InfiniBand EDR technology provides a data rate of 100~\si{Gbps} per HCA port (i.e., $12.5$ GB/s), with packets of $4$~\si{\kibi\byte} of MTU. These packets contain a $64$B header and a \num{4032}B payload. CELLIA has a management 1GbE network, so each node has a network interface (NIC) for management purposes. In addition, each node has installed an NVIDIA Tesla T4 GPU over PCIe Gen3 x16 and an NVMe disk Ultrastar SN200 Series NVMe SSD over PCIe Gen3 x8. Note that this heterogeneous node architecture has several devices attached to the PCIe (e.g., the CPU, GPU, NVMe, and HCA), so the intra-node network could be a bottleneck when inter-node communication involves other devices placed in different nodes.

\subsection{PCIe communication characterization}
\label{sec:description:analysis}

\begin{figure}[!hbt]
    \centering
        \centering
        \includegraphics[width=0.5\columnwidth]{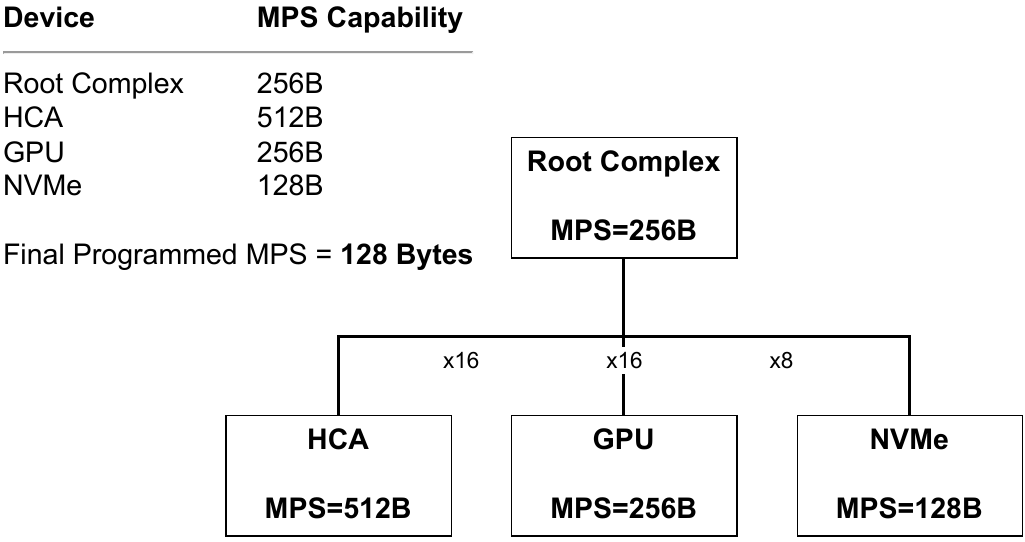}
        \caption{PCIe cluster maximum payload size (MPS).}
        \label{fig:desc:PCIeMaxPayload}
\end{figure}

To accurately model the PCIe subsystem communication, we first need to understand the specific characteristics of the PCI-Express system. We assume PCI-Express version 3.0, which can transfer data at 8 Gbps per lane. Since our HCA\footnote{We use the term \emph{Host Channel Adapter} (HCA) from the InfiniBand jargon to distinguish it from the 1GbE NIC in our assumed end node model.} transmits over 16 PCI lanes, the total data transfer rate from the intra-node network to the HCA can reach up to 128 Gbps (i.e., $16$GB/s). Considering the PCIe version and its $128$b/$130$b encoding, we can transfer accurate data close to $126$ Gbps. At the same time, the HCA can inject traffic into the inter-node network at a maximum rate of $100$Gbps since it is InfiniBand EDR technology. Thus, the intra-node communication may become a bottleneck when the HCA is saturated. Figure~\ref{fig:desc:PCIeMaxPayload} shows the number of PCIe lanes and maximum payload size (MPS) for the different end-node devices. We assume that our PCIe communication model adjusts the MPS to $128$B.

We assume that all end-node devices are connected through a PCIe Root Complex (RC) without using PCIe switches (see Figure~\ref{fig:desc:BigPictureTFG}). This PCIe configuration can impact communication performance between two different endpoints (E1 and E2), as the communication model follows the following steps: $E1\rightarrow RC\rightarrow CPU\rightarrow RC\rightarrow E2$. As end nodes in our real cluster follow this approach, and we want to validate our simulation model compared to these nodes, we have modeled this PCIe communication. Note that the CPU and RC involvement in the communication impact performance, especially when communicating between two end-node devices that do not require CPU involvement. The most efficient path typically uses a PCIe Switch, through which these devices communicate directly. By contrast, the CPU can communicate with the HCA in our tests by traversing the RC only once.

Our intra-node baseline model is focused on PCIe Gen3 parameters, including its data rate, \emph{AckFactor}, and encoding. These parameters determine the time required to process each PCIe message in our system. The PCIe Gen3 communication time is determined by the processing time taken by the end-node devices to compute each packet composition and decomposition, and the time taken to process and transmit each PCIe packet (Transaction Layer Packets or TLPs and Data Link Layer Packets or DLLPs). The resulting intra-node latency for each message in a PCIe communication is expressed in the following set of equations:

\begin{equation}
    \small
    \centering
    \begin{aligned}
        BytesPerNs &= \frac{Width}{\frac{1}{Datarate \times Encoding}} \\ \\
        TLPTime &= \frac{TLPOverhead + MaxPayloadSize}{BytesPerNs} \\ \\
        DLLPTime &= \frac{DLLPOverhead + DLLPSize}{BytesPerNs} \\ \\
        NumberTLPs &= \frac{MessageSize}{MaxPayloadSize} \\ \\
        NumberACKs &= \frac{NumberTLPs}{AckFactor} \\ \\
        LatencyTime &= NumberTLPs \times TLPTime \ + \\
        &\quad NumberACKs \times DLPPTime \ \\
    \end{aligned}
    \label{eq:validationFormula}
\end{equation}

Where \emph{BytesPerNs} is the number of bytes that a PCIe link can transmit per nanosecond (\si{\nano\second}), determined by \emph{Width} that indicates the number of lanes of the link, \emph{DataRate} that indicates the transmission rate determined by the PCIe version, and \emph{Encoding} that express the number of bits of data that are encoded in each transmission, also determined by the PCIe version. Next, the \emph{TLPTime} represents the time required to transmit a TLP over a PCIe link, given by the \emph{TLPOverhead} that is the overhead introduced in each TLP, the \emph{MaxPayloadSize} that is the maximum data contained in each TLP, and the \emph{BytesPerNs} value. The \emph{DLLPTime} is the time required to transmit a DLLP over a PCIe link, which is given by \emph{DLLPOverhead} and \emph{DLLPSize} that represent the overhead and data in each DLLP, and \emph{BytesPerNs}. We also need to compute the \emph{NumberTLPs}, which is the number of TLPs in which it is split each message to be sent over the intra-node network, determined by the message size (\emph{MessageSize}) and the MPS (\emph{MaxPayloadSize}). The \emph{NumberACKs} is the number of ACKs to send over the intra-node network, determined by dividing \emph{NumberTLPs} by \emph{AckFactor}, which is the maximum number of TLPs that can be received before sending an ACK message. Finally, the \emph{LatencyTime} of our communication model sums the time it takes to send all TLPs and ACKs.

This communication model allows us to reproduce the PCIe transactions with a high level of detail. As described in section~\ref{sec:evaluation:cluster}, we have validated this model compared with the communication operations generated in a real cluster.

\subsection{Generic intra-node network model}
\label{sec:description:simulation}

\begin{figure}[!htb]
    \centering
    \begin{minipage}{\textwidth}
        \centering
        \includegraphics[width=0.7\textwidth]{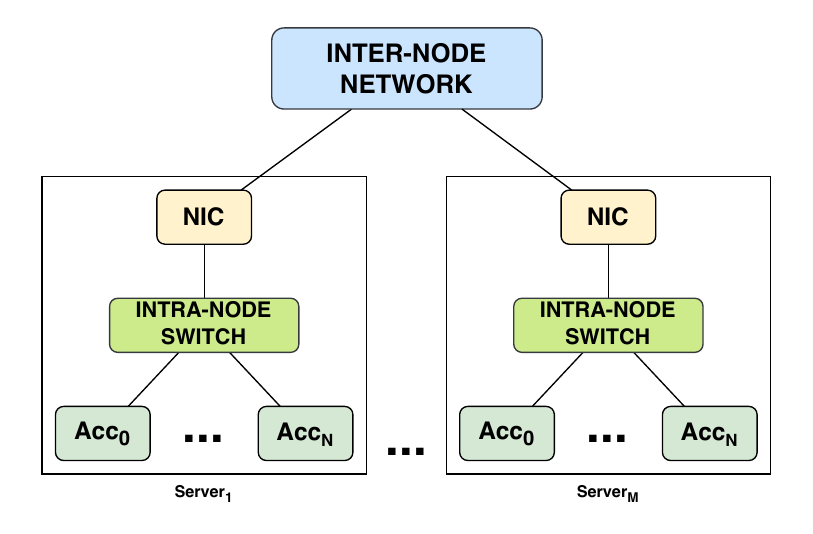}
        \caption{Generic intra- and inter-node network architecture.}
        \label{fig:desc:HostArchitecture}
    \end{minipage}
\end{figure}

Apart from integrating the PCIe model in our SAURON simulator~\cite{SAURON}, we have extended this tool to model a generic intra-node network and realistic communication patterns. SAURON is a highly scalable simulator modeling inter-node communication patterns, topologies, and routing configurations. Figure~\ref{fig:desc:HostArchitecture} shows the intra- and inter-node architecture modeled in the SAURON simulator.

We assume a configurable number of accelerators and other functional units (e.g., CPUs or storage devices) interconnected ``all-to-all'' using a switch in the intra-node network. This switch can be configured to provide the desired bandwidth so the intra-node network can process the incoming traffic addressed to the end-node devices and the outgoing traffic they generate. Indeed, we can also configure the bandwidth between this switch and the end-node NIC, i.e., the bandwidth between the intra- and inter-node network. Note that this generic model permits configuring the intra-node network according to current and emerging technologies  (e.g., PCIe 5.0, 6.0, 7.0, NVLink, UALink, etc.). Indeed, we can configure the number of lanes, the bandwidth for the different end-node devices, their maximum payload size (MPS), etc.

We assume messages are generated at accelerators (GPUs, TPUs, etc.) and functional units (CPUs, NVMe, etc.), which need to split them into packets (according to the intra-node network MPS) that are injected into the intra-node network. Intra-node packets can be addressed to other devices within the same end node or placed at different end nodes. In the latter case, end-node NICs convert the intra-node packets to inter-node packets and inject them into the inter-node network. This conversion involves populating inter-node packets with the headers and payloads of intra-node network packets, introducing additional latency and overhead.

As later described, in our simulation model, we can configure the message size at accelerators and functional units, the destination distribution, and the generation rate. This flexibility permits reproducing realistic communication patterns, such as those observed in the training phase of LLMs, where several types of parallelism are used.

\subsection{Generic LLM communication modeling}
\label{sec:description:llm-modeling}

In our simulator, we have included a generic network traffic model based on the communication operations generated by Large Language Models (LLMs), which are commonly used in Supercomputers and Data centers, as described in Section~\ref{sec:background:communication}. Specifically, training in AI-based applications follows a cyclic pattern, where communication operations repeat across data batches provided to the LLM when splitting a dataset into chunks. During training, accelerators (e.g., GPUs) perform computation tasks until reaching different blocks (e.g., Multi-Head Attention or Feed-forward Neural Network), triggering new communication operations. We assume that accelerators have identical computational capabilities, processing the same amount of data at the same speed and initiating their next communication operation simultaneously.

Our model also distinguishes between Data Parallelism (DP) and Model Parallelism (MP), as described in Section~\ref{sec:background:communication}. We assume the initial LLM model is already stored in the accelerators' memory, eliminating the need to exchange messages with external memory resources (e.g., NVMe disks). Instead, all communication occurs among accelerators to share information during the Forward and Backward phases in MP or to update gradients in DP. We assume five traffic-pattern configurations that may be generated in the network during the LLM training process: $C1$, $C2$, $C3$, $C4$, and $C5$. From $C1$ to $C4$, we model a massive LLM that requires MP to distribute the model across multiple accelerators. The key difference between these traffic patterns is the predominant form of MP: Tensor Parallelism (TP) or Pipeline Parallelism (PP). 

$C1$ assumes that TP is used extensively, i.e., most block communications follow a TP pattern (e.g., \emph{AllReduce}, \emph{AllGather} operations). This scenario generates significant traffic due to the need to update gradients during forward and backward propagation between devices (see Section~\ref{sec:background:communication}). As a result, this traffic pattern generates $20\%$ of the communication to the inter-node network and $80\%$ to the intra-node network. Next, $C2$ and $C3$ also employ MP; however, the transferred data size between devices is smaller as these configurations rely more on PP than TP. This reduces inter-node traffic, as PP primarily uses Point-to-Point operations. Specifically, these traffic patterns generate $15\%$ and $10\%$ of the traffic to the inter-node network, while $85\%$ and $90\%$ remain within the intra-node network, respectively. $C4$ uses MP without TP, relying exclusively on PP. In this case, inter-node communication is minimized (e.g., $5\%$), as fewer updates are required between stages. Consequently, intra-node network communication is $95\%$.

By contrast, $C5$ does not implement MP. We assume the LLM is not large enough to require multiple accelerators. Since it fits within a single accelerator, only DP is used to accelerate training, and no inter-node communication is needed. As a result, $100\%$ communication occurs intra-node.

\subsection{Intra-node overhead modeling}
\label{sec:description:overhead}

As described in Section~\ref{sec:description:simulation}, the conversion between intra- and inter-node packets introduces extra latency and overhead, which leads to communication bottlenecks~\cite{HeaderOverhead}. Specifically, an accelerator generates a message addressed to another accelerator placed on a remote node; that message is split into several intra-node packets that are later aggregated into one or several inter-node packets, which usually are larger in size. Thus, the intra-node packet headers will waste payload space in the inter-node packets. Moreover, when an inter-node packet arrives at a given end node, it is split again into multiple intra-node packets; the header space of the latter ones introduces an overhead in the communication, regardless of whether or not those headers were transported in the inter-node packets' payload. In the following, we analyze the modeling of this overhead by performing a set of simulations. Moreover, we formalize this overhead through linear regression to adapt the modeling to any possible network configuration.

We have performed several simulations using the same scenarios that will be later described in Section~\ref{sec:evaluation}. We assume an intra-node network with $8$ accelerators (Accs) per end node, where each Acc NIC generates up to $512$ Gbps of traffic. We assume a $32$-node fat-tree inter-node network topology, where each end node has a NIC connected to the intra-node and the inter-node network (see Figure~\ref{fig:desc:HostArchitecture}). We assume end-node NICs can inject traffic into the inter-node network at $400$ Gbps. We analyze two different intra-node packet sizes: a commonly used intra-node packet of 148B (128B corresponding to the payload and 20B corresponding to the header), and an intra-node packet of 4KB, which is the same as the one used in the inter-node network (i.e., 64B for the header and 4032B for the payload). Note that this second packet configuration allows for measuring the network performance when there is no overhead due to intra- and inter-node packet conversion.

\begin{figure}[!htb]
    \centering
    \begin{subfigure}{0.4\textwidth}
        \centering
        \includegraphics[width=1\columnwidth]{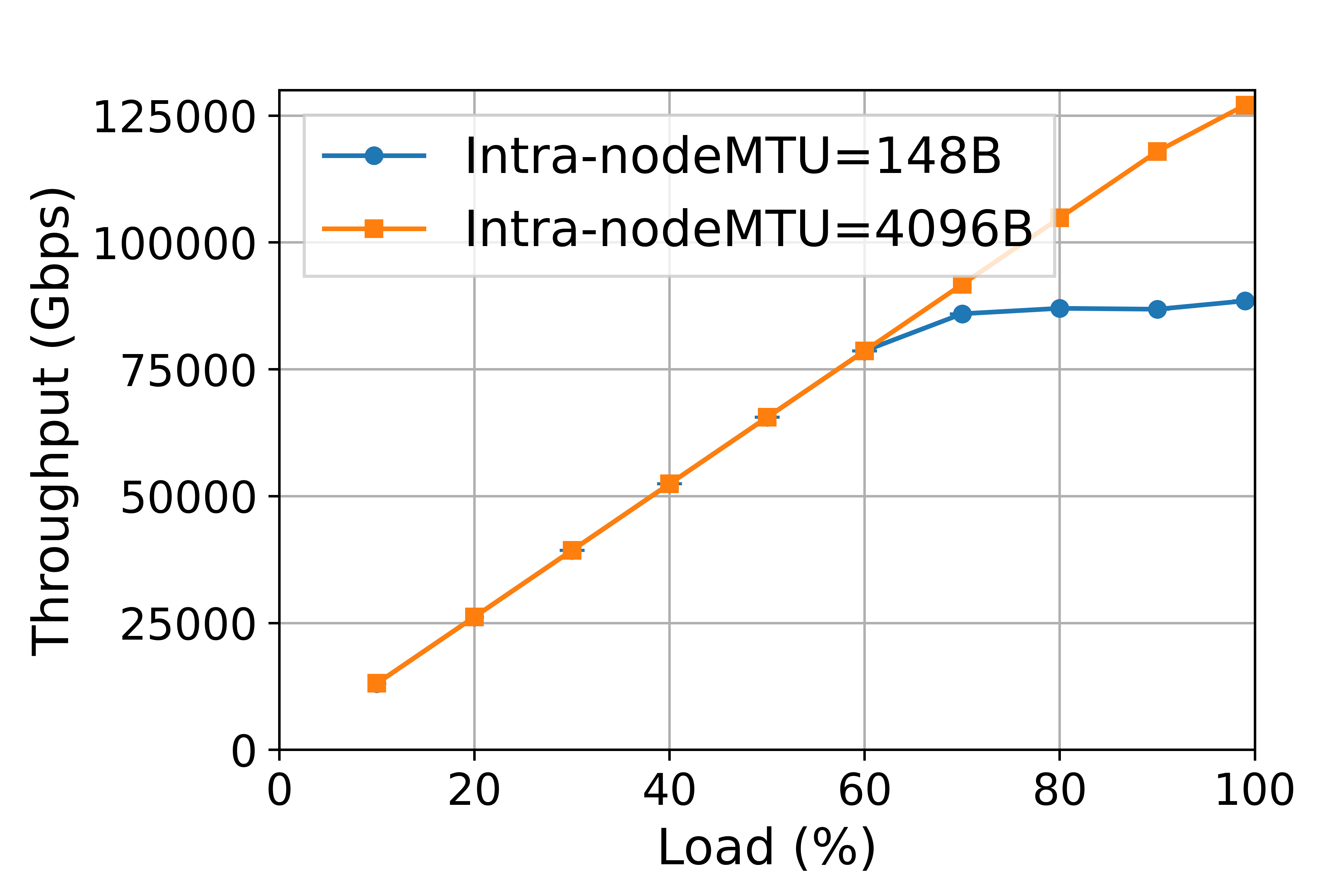}
        \caption{C4 Traffic Configuration.}
        \label{fig:desc:overhead:C4}
    \end{subfigure}
    \begin{subfigure}{0.4\textwidth}
        \centering
        \includegraphics[width=1\columnwidth]{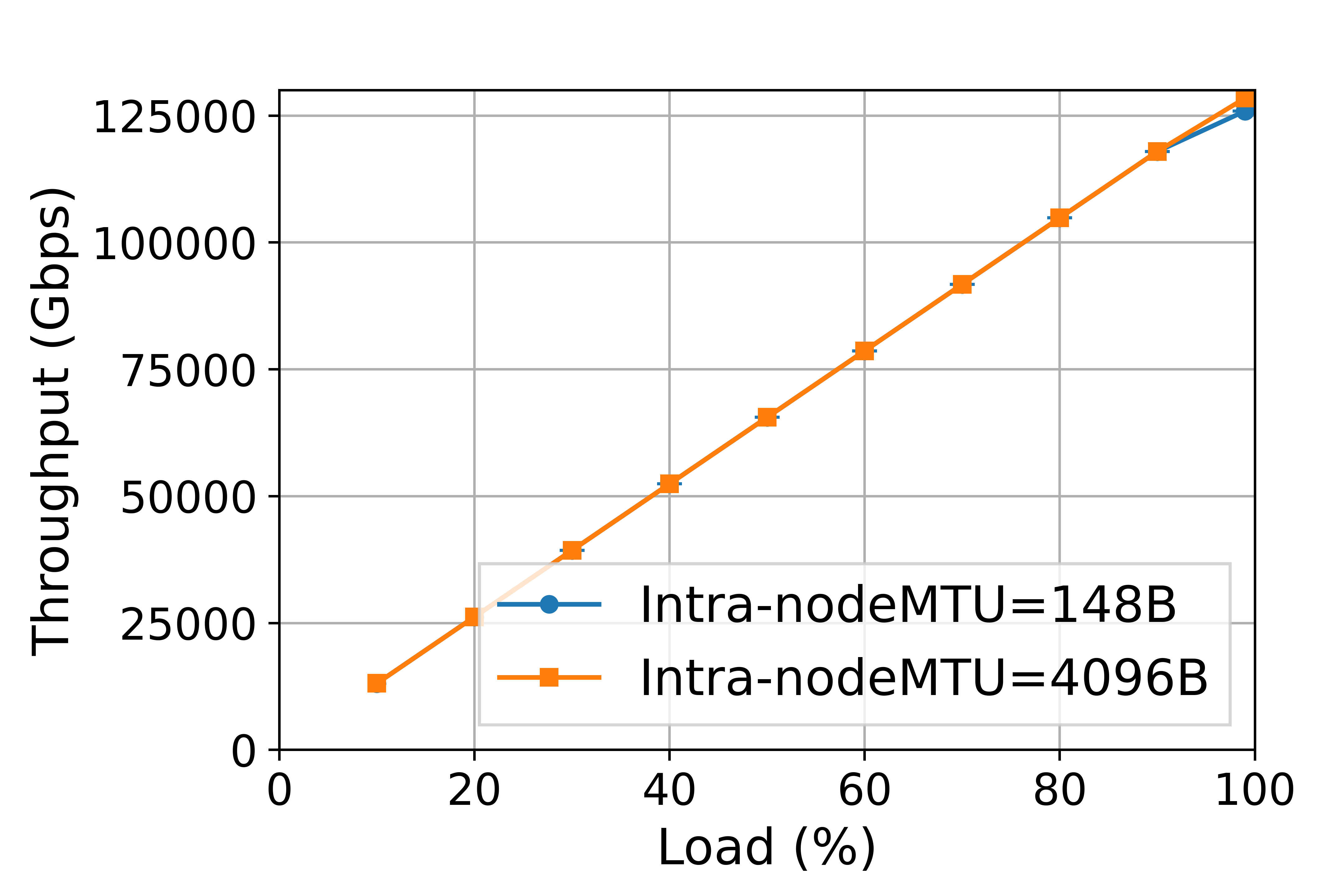}
        \caption{C5 Traffic Configuration.}
        \label{fig:desc:overhead:C5}
    \end{subfigure}
    \caption{Overall Network Throughput versus Traffic Generation Load for different intra-node packets MTUs and traffic patterns in a $32$-node fat-tree inter-node network that communicates $256$ accelerators ($8$ accelerators per node).}
    \label{fig:desc:overhead}
\end{figure}

We use traffic configurations C4 and C5 described in Section~\ref{sec:description:llm-modeling}. In C4, where each Acc generates 5\% of traffic addressed to the inter-node network, each Acc will send $25.6$ Gbps to the inter-node network. So, $8$ Accs per node will generate $204.8$ Gbps in total to the inter-node network, which is less than the $512$ Gbps of inter-node links and the $400$ Gbps of end-node NICs to the inter-node network. By contrast, C5 only generates traffic to the intra-node network (i.e., no inter-node traffic) to show a scenario with no conversion overhead impact. We have simulated these traffic patterns varying the traffic generation rate from a value close to 0\% of the traffic load (i.e., a very small amount of packets are generated at the Acc's NIC) to 100\% of the maximum rate that those NICs can generate (i.e., $512$ Gbps). Note that for C4, we keep 5\% of intra-node packets addressed to the inter-node network regardless of the traffic load rate.
Figure~\ref{fig:desc:overhead} shows the throughput results when these traffic scenarios are generated in the simulated configurations. 

The maximum traffic load all the network Accs can generate is $512~Gbps~\times~8~Acc/node~\times~32~nodes = $\num{131072}$~Gbps$. For the C4 traffic pattern (see Figure~\ref{fig:desc:overhead:C4}), we can see that the network saturates with 70\% of the traffic generation rate when $148$B intra-node packets are used. By contrast, when \num{4096}B intra-node packets are used, there is no saturation, as the network delivers \num{126000}$~Gbps$ in total, i.e., 96\% of the maximum network throughput. In the former case, the saturation occurs due to the aggregation of several $148$B packets in the payload of every inter-node packet. Since a \num{4096}B inter-node packet uses $64$B for the header and \num{4032}B for the payload, this means that each inter-node packet can store up to \num{4032}$/128\simeq32$ inter-node packets (each one consuming $20$B for the header). The header information of the intra-node packets aggregated into a single inter-node packet is replicated in those intra-node packets. For the C5 traffic pattern (see Figure~\ref{fig:desc:overhead:C5}), there is no performance degradation regardless of the intra-node packet size, since there is only intra-node communication and no overhead introduced due to packet conversion.

\begin{figure}[!htb]
    \centering
    \begin{subfigure}{0.8\textwidth}
        \centering
        \frame{\includegraphics[width=1\textwidth]{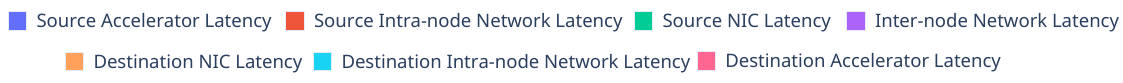}}
    \end{subfigure}
    \\
    \begin{subfigure}{0.49\textwidth}
        \centering
        \adjincludegraphics[width=1\columnwidth,trim={{.03\width} {.05\width} {.13\width} {.13\width}},clip]{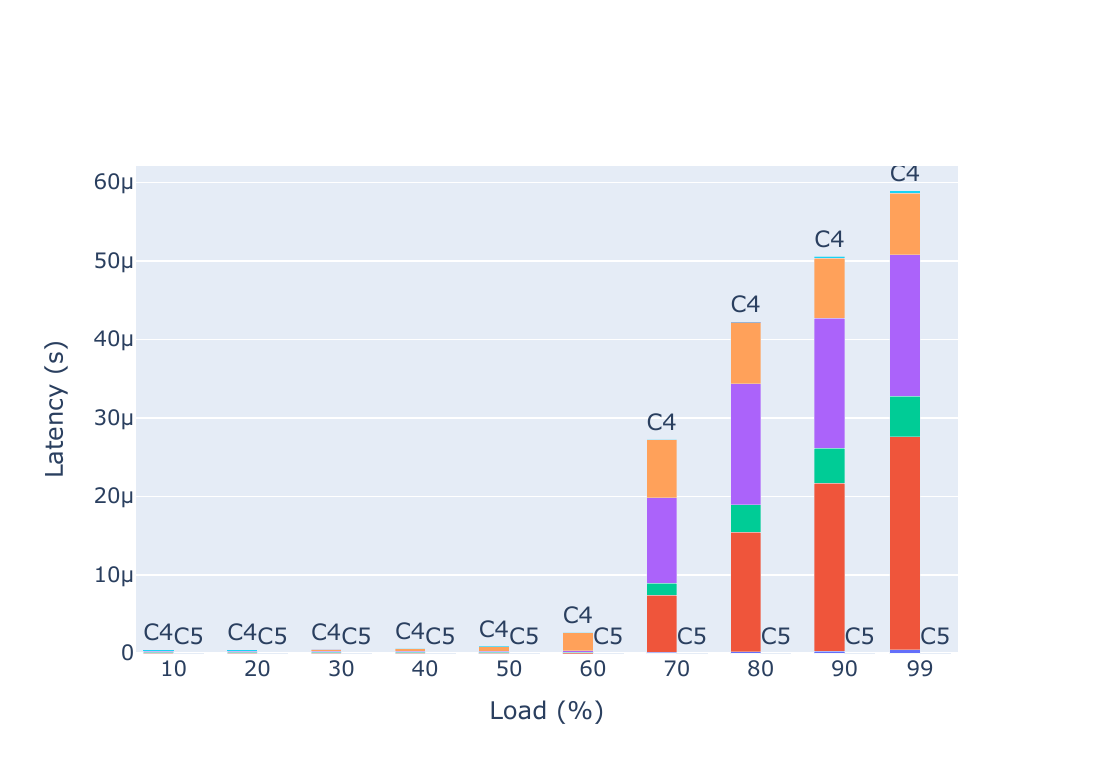}
        \caption{Intra-node MTU of 148B.}
        \label{fig:desc:overhead:latencies:148}
    \end{subfigure}
    \begin{subfigure}{0.49\textwidth}
        \centering
        \adjincludegraphics[width=1\columnwidth,trim={{.03\width} {.05\width} {.13\width} {.13\width}},clip]{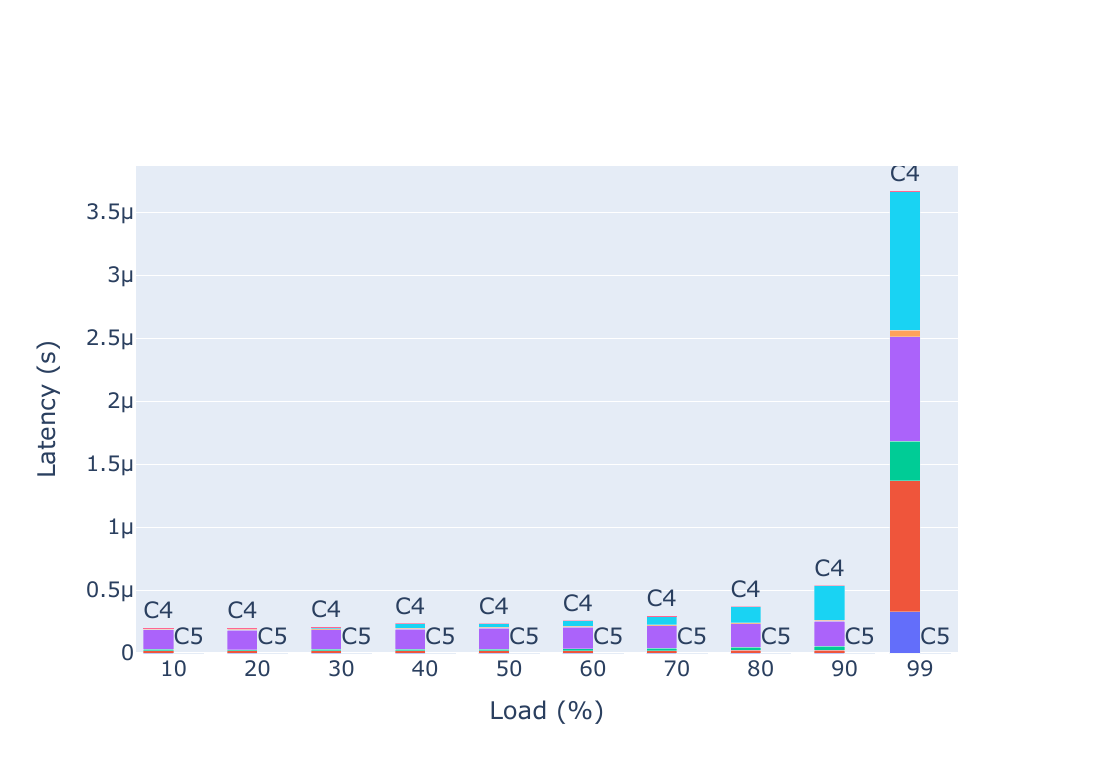}
        \caption{Intra-node MTU of 4096B.}
        \label{fig:desc:overhead:latencies:4032}
    \end{subfigure}
    \caption{Packet Latency divided into components versus traffic load for the traffic patterns C4 and C5 with different intra-node MTUs.}
    \label{fig:desc:overhead:latencies}
\end{figure}

Apart from the throughput metric, we have analyzed the network latency of inter-node packets and the components that compose the latency measurements. Figure~\ref{fig:desc:overhead:latencies} shows the latency results for the same scenarios described above, with packet latency divided into components (later explained in Section~\ref{sec:evaluation:config}). 

For C5, we can see that, for both MTU configurations, the latencies are negligible compared to those of C4, since there is no contention when only intra-node communication is generated. Regarding C4, we can see in Figure~\ref{fig:desc:overhead:latencies:148} that latency significantly augments when the traffic load is 70\%. After saturation, several latency components consume most of the time (e.g., source intra-node, source NIC, inter-node network, and destination NIC). However, the cause of saturation happens in the destination NIC, where inter-node packets are split into intra-node packets. This process takes time and accumulates incoming traffic at destination NICs, which generates backpressure to the inter-node network that reaches source end-node NICs and their intra-node network. Note that for $148$B packets, the aggregated latency is much higher than that of \num{4096}B packets (see Figure~\ref{fig:desc:overhead:latencies:4032}), due to the mentioned overhead $148$B packets.

On the other hand, Figure~\ref{fig:desc:overhead:latencies:4032} shows that when there is no packetization overhead, the maximum latency in the communication is by far lower than in the previous case ($3.7\mu s$ versus $58\mu s$). In this scenario, note that the latency components are not affected by the Destination NIC, since there is no packetization overhead, but at the destination intra-node network, which will be the bottleneck under saturation whenever there is no overhead in the inter-to-intra-node packets conversion.

Due to the important effects of packetization overhead, we developed a mathematical analysis that establishes a relationship between the amount of inter-node traffic (i.e., the packets that accelerators send to remote end nodes) and the overhead caused by different header and payload sizes. This analysis is based on the following steps:

\begin{enumerate}
    \item We selected four different mathematical functions to compare and determine if any of them fit our model (linear, quadratic, cubic, and power-law). 

    \item We collected all the necessary data sets for a 148B intra-node MTU, which includes the inter-node traffic and the throughput obtained by our simulator. We run experiments modeling a $32$-node interconnection network and eight accelerators per node, using $512$ Gbps of link speed each. So the maximum throughput per node is \num{4096}~Gbps.

    \item We fit the four models to determine which one performs better and identify the possible parameters of each model. Our mathematical tool used Python packages, such as \textit{numpy} and \textit{scipy}, to fit the explained models. We can see the used code (i.e., \texttt{modelAdjustment.py}) and data from the performed simulations in the following repository\footnote{\url{https://github.com/ajtarraga/modelAdjustment/tree/main}}.

    \item With this data, we evaluate the quality of each model from the four selected to determine which one represents the better option.

    \item We select the best one and represent it. 
\end{enumerate}

\begin{figure}[ht]
    \centering
    \includegraphics[width=0.5\columnwidth]{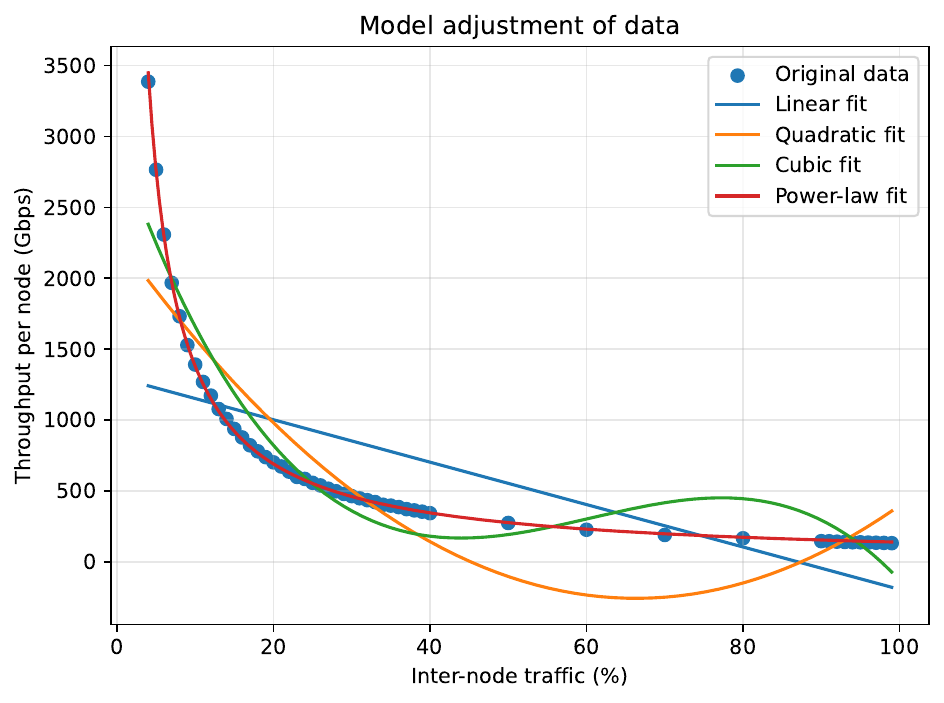}
    \caption{Comparison of mathematical models used to fit throughput degradation.}
    \label{fig:desc:overheadMathematical}
\end{figure}

Figure~\ref{fig:desc:overheadMathematical} illustrates this representation. After performing the mathematical analysis steps, we see that the power-law model best fits the experimental results. Based on this model, we have defined Equations~\ref{eq:trafficOverhead} and~\ref{eq:throughput}, which allow computing the throughput behavior (considering the overhead) under different network configurations.

Equation~\ref{eq:trafficOverhead} sets a relationship between the header and the payload size and between the intra- and inter-node networks. If the intra-node header impacts more than the inter-node header size, the \textit{TrafficOverhead} will be higher than 1, which implies that there could be an impact (see Figure~\ref{fig:desc:overhead}). However, if the impact of the intra-node header is lower than that of the inter-node ones, the value will be less than 1. If they are equal, the value will be 1, which means it does not impact.

\begin{equation}
    \text{TrafficOverhead} = \frac{\frac{\text{IntraNodeHeaderSize}}{\text{MaxIntraNodePayload}} + 1}
    {\frac{\text{InterNodeHeaderSize}}{\text{MaxInterNodePayload}} + 1}
    \label{eq:trafficOverhead}
\end{equation}

Equation~\ref{eq:throughput} represents the function obtained when applying mathematical analysis. We have selected the power-law function as it is the one that best fits our simulation results. This formula is used to compute the expected network throughput, assuming a percentage of inter-node traffic generated at accelerators and a certain percentage of traffic overhead (computed using Equation ~\ref{eq:trafficOverhead}).
\begin{equation}
    \text{Throughput (Gbps)} = 
    \frac{\text{Model\_adjustment}}{\text{TrafficOverhead} \times \text{TrafficInterNode (\%)}} \times \text{NumNodes}
    \label{eq:throughput}
\end{equation}

Note that he \emph{Model\_adjustment} field has been calculated by the \texttt{modelAdjustment.py} script that performs the mathematical analysis applying the power-law function. The analysis of this specific value is out of scope in this paper and is left for future work.

\section{Evaluation}
\label{sec:evaluation}

This section describes the experiments conducted to validate our simulation model compared to a real cluster environment, and to evaluate the results of scale-up and scale-out simulations using realistic network configurations and communication patterns.

\subsection{Simulation model validation}
\label{sec:evaluation:cluster}

We configured our simulation model to mimic the PCIe intra-node network behavior based on the baseline end node architecture described in Section~\ref{sec:description:node-arch}), which is available in a real computing cluster. We have used the Infiniband performance tests~\cite{perftest} (IB Perftests), since they generate direct communication at the user space, directly accessing the InfiniBand hardware thanks to the \emph{verbs} primitives. More precisely, we measure bandwidth and latency using \texttt{ib\_read}, \texttt{ib\_write}, and \texttt{ib\_send} tests with messages of sizes ranging from 128 Bytes to 4 MiB generated at the host-side user space. These messages are packetized in intra-node packets of 128B (i.e., the PCIe Maximum Payload Size or MPS in the CELLIA cluster).Table~\ref{tab:exp:cluster} shows the results for these tests run in two end nodes of the CELLIA cluster (see Section~\ref{sec:description:node-arch}).

\begin{table}[!htb]
    \caption{Results when communicating two nodes in the real cluster}
    \label{tab:exp:cluster}
    \begin{tabular*}{\textwidth}{@{\extracolsep\fill}lcccccc}
    \toprule
    & \multicolumn{3}{@{}c@{}}{Bandwidth results (\si{\gibi\byte}/\si{\second})} & \multicolumn{3}{@{}c@{}}{Latency results (in \SI{}{\micro\second})} \\\cmidrule{2-4}\cmidrule{5-7}
    Msg. Size & ib\_read & ib\_write & ib\_send & ib\_read & ib\_write & ib\_send \\
    \midrule
    128\ \si{\byte} & 0.37 & 0.44 & 0.41 & 2.03 & 1.12 & 1.20 \\
    256\ \si{\byte} & 0.79 & 0.87 & 0.77 & 2.07 & 1.56 & 1.59 \\
    512\ \si{\byte} & 1.51 & 1.75 & 1.64 & 2.02 & 1.58 & 1.64 \\
    \midrule
    1\ \si{\kibi\byte} & 2.74 & 3.30 & 3.10 & 2.15 & 1.70 & 1.77 \\
    2\ \si{\kibi\byte} & 6.63 & 7.35 & 6.22 & 2.43 & 1.95 & 2.02 \\
    4\ \si{\kibi\byte} & 9.90 & 11.02 & 11.00 & 2.88 & 2.46 & 2.56 \\
    8\ \si{\kibi\byte} & 11.38 & 11.58 & 11.55 & 3.40 & 2.84 & 2.94 \\
    16\ \si{\kibi\byte} & 11.78 & 11.53 & 11.63 & 4.28 & 3.88 & 3.86 \\
    32\ \si{\kibi\byte} & 11.80 & 11.60 & 11.67 & 5.68 & 5.41 & 5.32 \\
    64\ \si{\kibi\byte} & 11.81 & 11.62 & 11.60 & 8.38 & 8.06 & 7.97 \\
    128\ \si{\kibi\byte} & 12.09 & 11.90 & 11.90 & 13.66 & 13.39 & 13.25 \\
    256\ \si{\kibi\byte} & 12.09 & 11.92 & 11.93 & 24.25 & 24.27 & 24.10 \\
    512\ \si{\kibi\byte} & 12.09 & 11.93 & 11.92 & 45.40 & 45.73 & 45.41 \\
    \midrule
    1\ \si{\mebi\byte} & 12.09 & 11.93 & 11.93 & 87.73 & 88.95 & 88.46 \\
    2\ \si{\mebi\byte} & 12.06 & 11.93 & 11.94 & 173.31 & 174.65 & 173.74 \\
    4\ \si{\mebi\byte} & 12.03 & 11.86 & 11.94 & 343.93 & 345.97 & 344.31 \\
    \botrule
    \end{tabular*}
\end{table}

As we can see, the maximum link bandwidth for all tests is around $12.1$ GiB/s out of $12.12$ GB/s\footnote{Our InfiniBand cluster uses links at EDR speed and 64b/66b encoding, so the maximum performance for 12.5GB/s is 12.12GB/s.}. This maximum bandwidth is first achieved with $128$KiB messages. Note that the CELLIA cluster where we run the experiments achieves speeds of over $12.5$~\si{\gibi\byte}/\si{\second} on the Infiniband inter-node network when transmitting raw data (i.e., without considering packet headers) and $16$~\si{\gibi\byte}/\si{\second} on the PCIe intra-node network (see Sections~\ref{sec:description:node-arch} and~\ref{sec:description:analysis}). Similar bandwidth results can be seen in the \texttt{ib\_write} and \texttt{ib\_send} tests. Regarding the latency tests in the right part of the table, we can see that latency increases linearly on messages between $128$B and $128$KB. This increment is produced by the packet header overhead (i.e., $60$B) introduced when messages are split into TLPs (and later into MTUs). Note that latency skyrockets exponentially on messages larger than $128$KB because the inter-node InfiniBand link is working at full speed (i.e., 12.1 GB/s) and messages accumulate waiting to be forwarded first in the intra-node buffering space and later in the host-side memory. These results demonstrate that the inter-node network may be a bottleneck when end-node devices demand a bandwidth that is enough for the intra-node network (i.e., 16 GB/s) but not enough for the inter-node network.

\begin{figure}[!htb]
    \centering
    \begin{subfigure}{0.45\columnwidth}
        \centering
        \includegraphics[width=1\columnwidth]{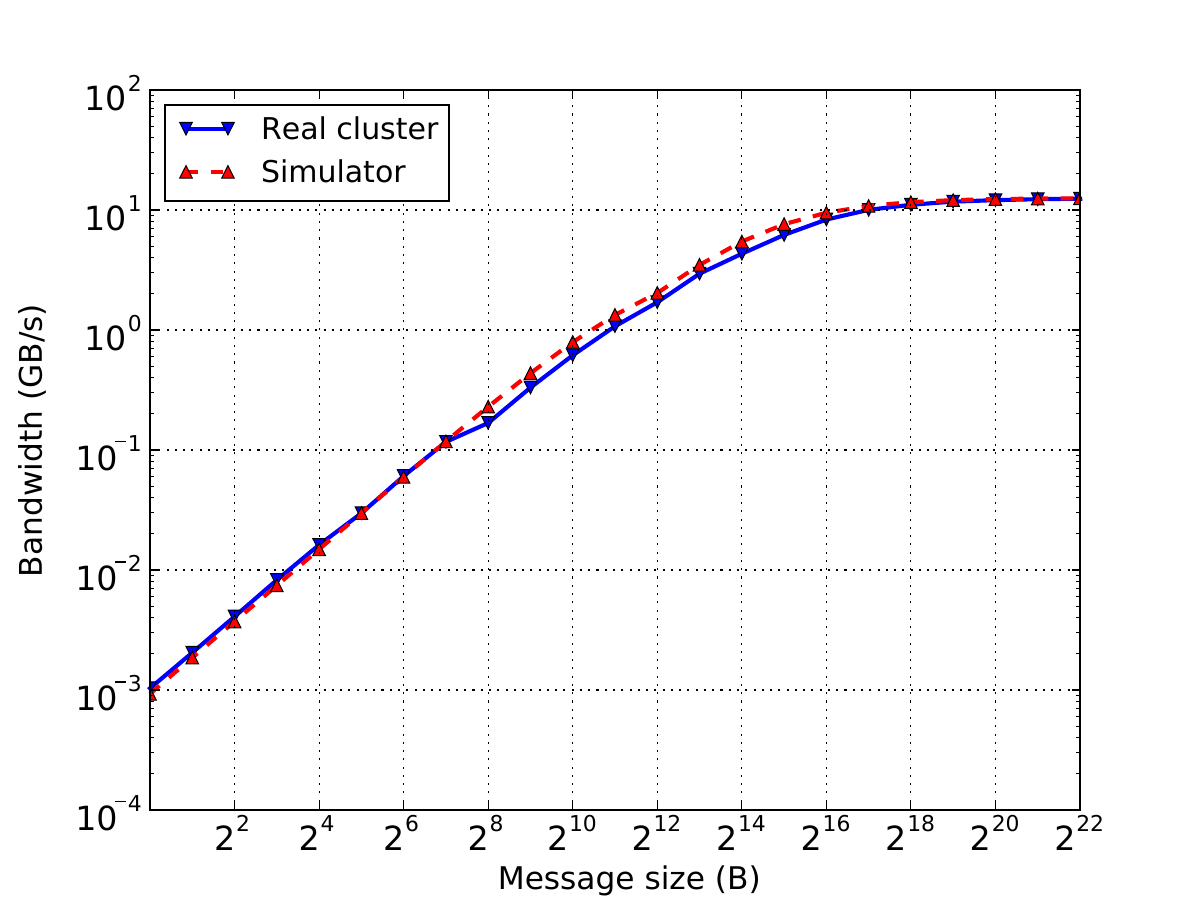}
        \caption{\texttt{ib\_write} bandwidth.}
        \label{fig:evaluation:validationBw}
    \end{subfigure}
    \begin{subfigure}{0.45\columnwidth}
        \centering
        \includegraphics[width=1\columnwidth]{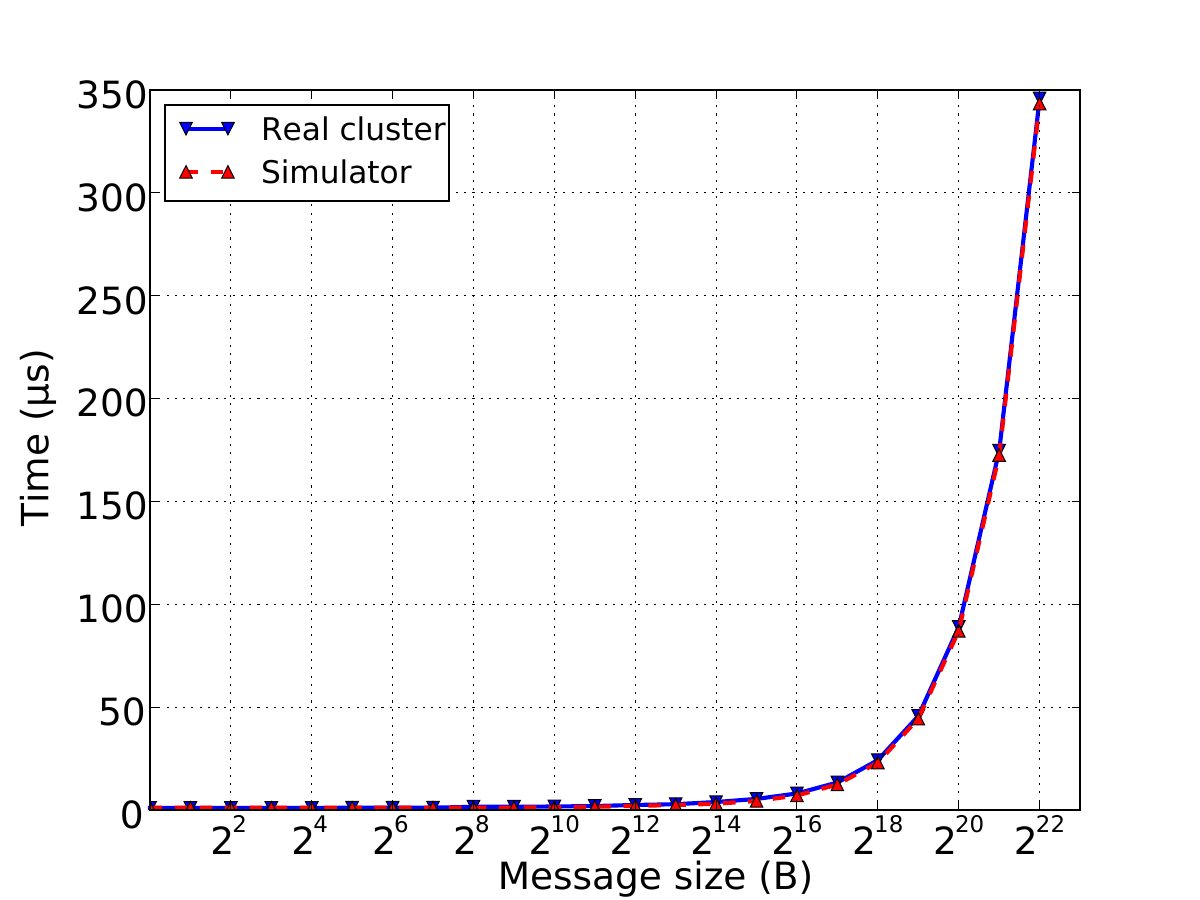}
        \caption{\texttt{ib\_write} latency.}
        \label{fig:evaluation:validationLat}
    \end{subfigure}

    \caption{Experiment results in the CELLIA cluster and the simulator when communicating two end-node devices.}
    \label{fig:evaluation:validation}
\end{figure}

To validate the intra-node network model, we modeled the \texttt{ib\_write} Infiniband test in our simulator. We have configured the intra-node network in the simulator with the same parameters as the PCIe baseline architecture, described in Section~\ref{sec:description:node-arch}. Figure~\ref{fig:evaluation:validation} shows the bandwidth and latency results obtained running the \texttt{ib\_write} test both in the simulator and the real cluster. Figure~\ref{fig:evaluation:validationBw} compares the bandwidth results, which are virtually identical between the two series. Figure~\ref{fig:evaluation:validationLat} shows the latency results that follow the same trend. These results show that, in this simple scenario, the PCIe intra-node model essentially behaves as expected. It splits the end-node CPU transactions into TLPs (i.e., intra-node packets), transferring them through the PCIe network to the NIC device. The NIC composes network packets, each with MTU$~=4$KB, which include several TLPs. These packets are sent through the InfiniBand network and, when received at the destination NIC, they are transformed again into TLPs and forwarded to the receiving buffer at the host-side user space application, where the network performance metrics are measured.

Note that the intra-node network model is fully configurable to behave not only as PCIe but also to be as generic as possible. In the following sections, we extend the evaluation to scale-up and scale-out simulations using generic but realistic intra- and inter-node configurations.

\subsection{Experiments configurations}
\label{sec:evaluation:config}

In the simulator, we have configured the end-node architecture and the intra- and inter-node networks similarly to modern Supercomputers and Data Centers composed of heterogeneous end nodes. At each end node, we assume up to $8$ accelerators (e.g., GPUs) interconnected by a high-speed intra-node network. Each accelerator has a single NIC to enable data generation to the intra-node network. Also, a NIC connects the intra-node network to the inter-node network. We assume three configurations for the intra-node network using different speeds at accelerator NICs:

\begin{itemize}
    \item Configuration \#1 - $128$ Gbps/accelerator. Each accelerator NIC has a maximum generation rate of $128$ Gbps, so the aggregated intra-node network bandwidth among all accelerators ranges from $128$ ($1$ accelerator) to \num{1024} Gbps ($8$ accelerators).
    \item Configuration \#2 - $256$ Gbps/accelerator. Each accelerator NIC can generate traffic at a maximum rate of $256$ Gbps, so the aggregated intra-node network bandwidth varies from $256$ ($1$ accelerator) to \num{2048} Gbps ($8$ accelerators).
    \item Configuration \#3 - $512$ Gbps/accelerator. Each accelerator NIC can generate traffic at a maximum rate of $512$ Gbps, so the aggregated intra-node network bandwidth varies from $512$ ($1$ accelerator) to \num{4096} Gbps ($8$ accelerators).
\end{itemize}

These link speeds and aggregated bandwidth are consistent with several intra-node network technologies, such as PCIe or Infinity Fabric~\cite{schieffer2024}, which utilize a bandwidth per link similar to that of the inter-node network link. Note that the aggregated bandwidth of the intra-node network can be much larger than that of a single inter-node network NIC, which is assumed to be $400$ Gbps (i.e., 50 GB/s) according to current commercial speeds, such as InfiniBand or Gigabit Ethernet. This clearly shows that end-node NICs can be the bottleneck if all the end-node devices want to communicate at full speed with devices at different end nodes.

To measure the impact of a varying number of accelerators in generating bottlenecks, we assume $1$, $2$, $4$, or $8$ accelerators per end node connected to a single intra-node network switch. That switch is also connected to the end-node NIC via an intra-node link. At this switch, we assume buffering only at input ports of size 128~\si{\kibi\byte} (i.e., input queued IQ architecture), virtual cut-through (VCT) switching, and credit-based flow control. Additionally, switch arbitration is made utilizing the iSlip algorithm~\cite{iSlip}, so that each switch output port independently selects which crossing request of packets at input-port buffers is chosen, allowing that packet to cross through that output port. This decision is based on a round-robin (RR) policy among all the input ports or choosing the oldest crossing request (aging). Note that the number of ports of this intra-node switch depends on the number of accelerators per end node. The link bandwidth of each switch port is identical to that of the accelerator NIC.

The inter-node network architecture is similar to that of the intra-node network. Likewise, we have configured high-speed IQ switches of size 128~\si{\kibi\byte} using VCT, credit-based flow control, and RR- or Age-based iSlip scheduling. We have selected a Real-Life Fat-Tree (RLFT) network topology~\cite{RLFT}, widely used in current Supercomputers and data centers. We assume the $D$-mod-$K$ deterministic routing algorithm~\cite{Zahavi12dmodk}, which efficiently balances traffic flow paths among available shortest-path routes in the network, and achieves similar performance to adaptive routing~\cite{Routing}. Finally, regarding the link flying time of packets, we assume that the latency for the first flit of any packet traversing the inter-node network is set to $25$ ns/meter.

\begin{figure}[!htb]
    \centering
    \includegraphics[width=.9\columnwidth]{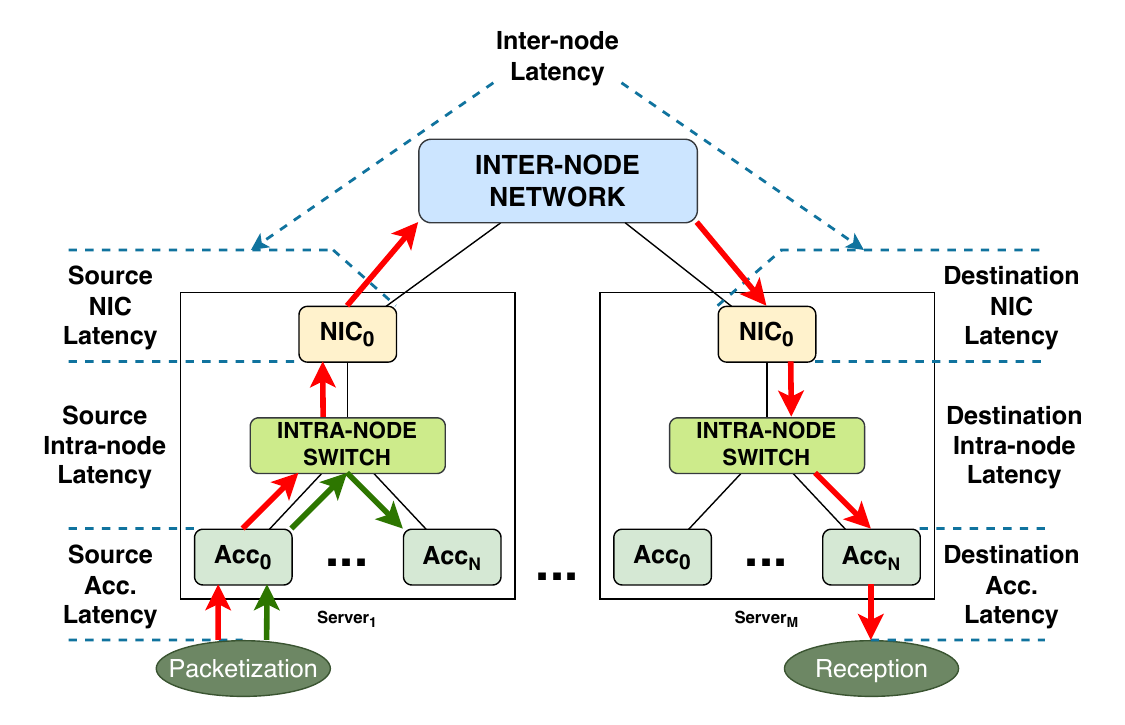}
    \caption{Latency metrics experienced by a traffic flow traversing the intra-node (red flow) and inter-node (green flow) networks.}
    \label{fig:evaluation:latencymetrics}
\end{figure}

Regarding the performance metrics, we have modeled the \emph{intra-node throughput} that measures the average data rate per time unit (in Gbps) of the traffic that is only exchanged among accelerators within the intra-node network. We also measure the \emph{inter-node throughput}, the data rate transmitted through the inter-node network. These two metrics will allow us to analyze what happens in each scenario. Regarding latency, we decompose the latency for a complete message (i.e., before packetization) into several components and measure different types of latency. Figure~\ref{fig:evaluation:latencymetrics} represents the seven latency metrics we measured.

This Figure shows a red flow from $Acc_0$ of $Server_1$ to $Acc_N$ of $ Server_M$, which illustrates inter-node communication. Also, we show a green flow from $Acc_0$ to $Acc_N$ within $Server_1$, illustrating intra-node communication. First, we can observe the \textit{Source Accelerator Latency}, which indicates the latency from the creation of the packet to its transmission. The \textit{Intra-node Network Latency} models the time since a packet leaves the Source Acc. until it reaches the end-node NIC. At the Source end node, we finally have the \textit{Source NIC Latency}, which indicates the time required to receive enough intra-node packets to be aggregated into an inter-node packet MTU. Then, it indicates the time between the arrival of the first intra-node packet and the sending of the inter-node MTU. At the destination end node, there are the same latencies, but in the opposite direction. Finally, we can observe the \textit{Inter-node Network Latency}, which represents the time needed to travel through the inter-node network.

Regarding the traffic patterns, we want to evaluate the impact of intra- and inter-node communication on the overall network performance. We assume the traffic model described in Section~\ref{sec:description:llm-modeling} based on LLMs training, which combines several configurations for the data parallelism (DP), Tensor Parallelism (TP), and Pipeline Parallelism (PP) into five communication patterns: $C1$, $C2$, $C3$, $C4$, $C5$. They differ in the traffic rate generated at the accelerators addressed to the intra- and inter-node network. Note that this configuration does not imply that these communication patterns belong to any specific LLM model. Modeling specific LLMs' traffic patterns, the different types of parallelism (e.g., DP, PP, or TP), or determining the most efficient and optimal configuration for those LLMs is beyond the scope of this paper. For such purposes, existing tools such as Calculon~\cite{calculon} can provide valuable insights, but this analysis and modeling effort is left for future work.

Accelerators NICs generate $4$\si{\kibi\byte} messages to the intra-node network at rates ranging from $0$ to $100$\% of their maximum bandwidth  (i.e., $128$, $256$, and $512$ Gbps for intra-node network configurations \#1, \#2, and \#3, described in Section~\ref{sec:evaluation:config}). We simulated $10$ different traffic load values per traffic pattern, each comprising $2.5$ms of simulated time where messages are generated at a specific load. For intra-node traffic, the destination of each message generated at an accelerator NIC is chosen randomly among the accelerators within an end node. For inter-node traffic, destinations are selected randomly among all the possible end-node devices distinct from the one where a specific message is generated. Note that messages are packetized into $148$\si{\byte}-packets in the intra-node network. Then, for those messages addressed to remote servers, the intra-node packets from each message are gathered at the source server NIC to build $4$\si{\kibi\byte}, which are sent through the inter-node network to the destination end-node NIC, which splits them again into $148$\si{\byte} intra-node packets to reach the destination accelerator. The packetization produces an intense overhead that limits the intra- and inter-node network performance, as we analyze in the next section.

\subsection{Intra-node overhead analysis}
\label{sec:evaluation:overhead}

As described in Section~\ref{sec:description:overhead}, there is an overhead produced when messages generated at a given accelerator are split into packets in the intra-node network, then recomposed again into larger inter-node packets, then split again into intra-node packets at the receiving end node, and finally received in the remote accelerator. In our experiments, we assume intra-node packets with a header size of 20~\si{\byte}, and a payload of 128~\si{\byte}, and inter-node packets whose header is 64~\si{\byte} and the payload is \num{4032}~\si{\byte}. Substituting these values into Equation~\ref{eq:trafficOverhead}, we obtain the following traffic overhead factor:

\[
\text{TrafficOverhead} = \frac{\frac{20}{128} + 1}{\frac{64}{4032} + 1} = \frac{1.15625}{1.01587} \approx 1.14
\]

By applying this overhead to Equation~\ref{eq:throughput}, where we have adjusted the \emph{Model\_adjustment} field to \num{15703.9} as provided by the mathematical analysis made as described in Section~\ref{sec:description:overhead}), we can express the formula as follows:

\[
\text{Throughput (Gbps)} = \frac{\num{15703.9}}{1.14 \times \text{TrafficInterNode (\%)}} \times \text{NumNodes}
\]

This formula can be generalized for any number of nodes and used to estimate the network performance bound, assuming the overhead introduced by the conversion of packets across intra- and inter-node networks. This relationship remains unchanged regardless of the intra-node network speed, as the header and payload sizes are fixed.

\subsection{Scale-Up simulation results}
\label{sec:evaluation:scaleUp}

Next, we evaluate the experiment results as the number of accelerators per end node increases while keeping the number of nodes in the system. We have configured a $32$-node system interconnected using a $2$-stage RLFT topology comprising twelve $8$-port switches. We increase the number of accelerators per server node ($1$, $2$, $4$, and $8$), so the total number in the system ranges from $32$ to $256$. We configure the rest of the inter- and intra-node network parameters as described in Section~\ref{sec:evaluation:config}. We have also configured different speeds for the accelerator NICs (e.g., $128$, $256$, and $512$ Gbps). Table~\ref{tab:exp:scaleup:config} summarizes the intra-node configurations.

\begin{table}[!htb]
    \centering
    \caption{Scale-Up network configurations.}
    \begin{tabular}{@{} c | c | c @{}}
        \toprule
        \textbf{\#}         & \textbf{Intra-node link speed}     & \textbf{Total \# of Accelerators} \\
        \midrule
        1                   & $128$ Gbps/accelerator                   & $32$, $64$, $128$, $256$      \\
        2                   & $256$ Gbps/accelerator                   & $32$, $64$, $128$, $256$      \\
        3                   & $512$ Gbps/accelerator                   & $32$, $64$, $128$, $256$      \\
        \bottomrule
    \end{tabular}
    \label{tab:exp:scaleup:config}
\end{table}

\subsubsection{Analysis of switch arbiter policies}
\label{sec:evaluation:scaleUp:arbiter}

\begin{figure}[!htb]
    \centering
    \begin{subfigure}{0.6\textwidth}
        \centering
        \frame{\includegraphics[width=1\textwidth]{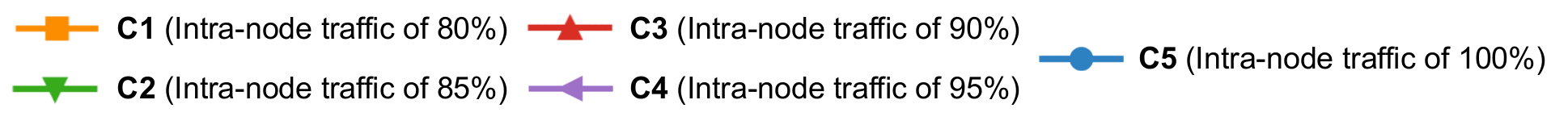}}
    \end{subfigure}
    \\
    \begin{subfigure}{0.32\textwidth}
        \centering
        \includegraphics[width=1\columnwidth]{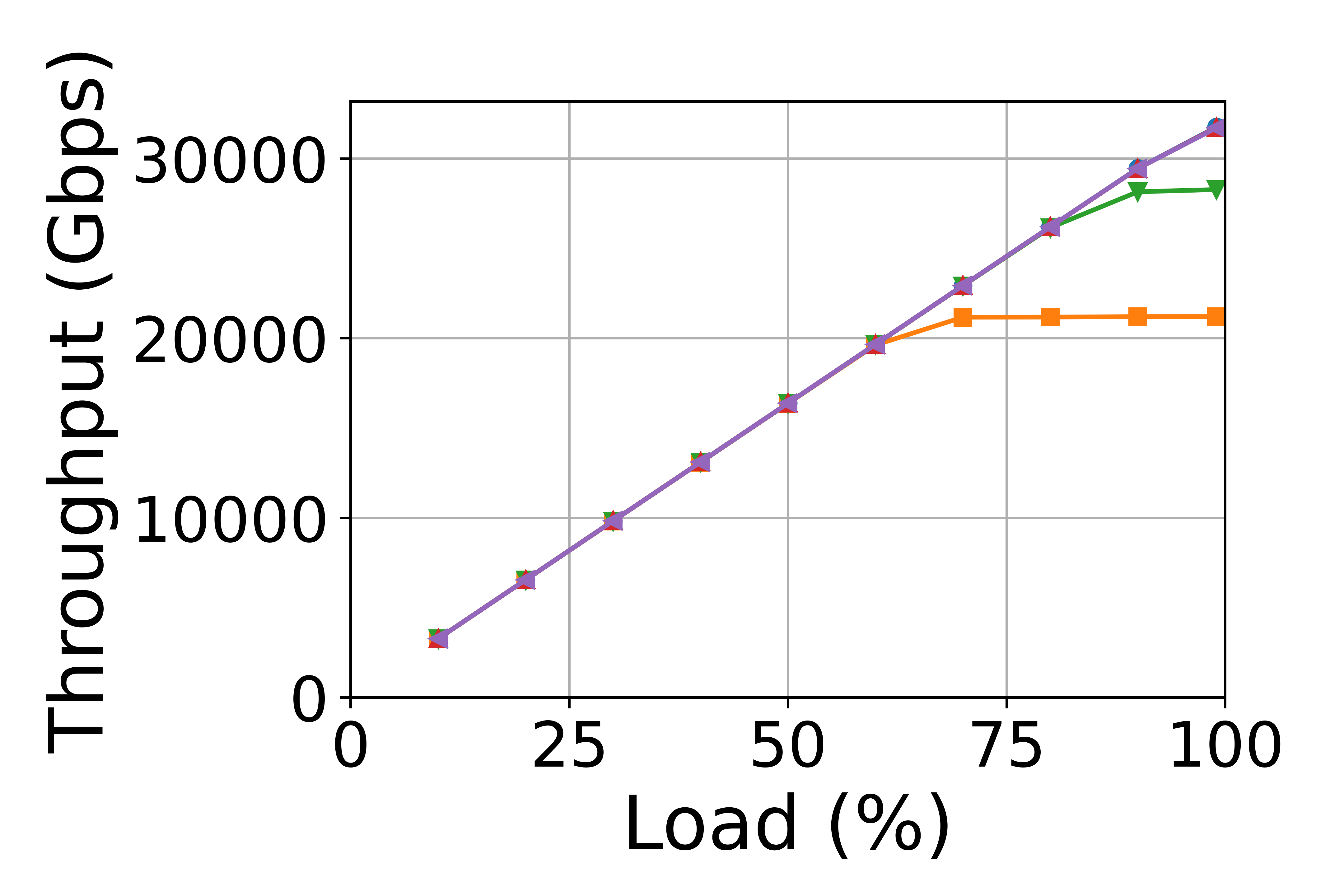}
        \caption{Age Based - conf \#1.}
        \label{fig:exp:arb:pcie3:agebased}
    \end{subfigure}
    \begin{subfigure}{0.32\textwidth}
        \centering
        \includegraphics[width=1\columnwidth]{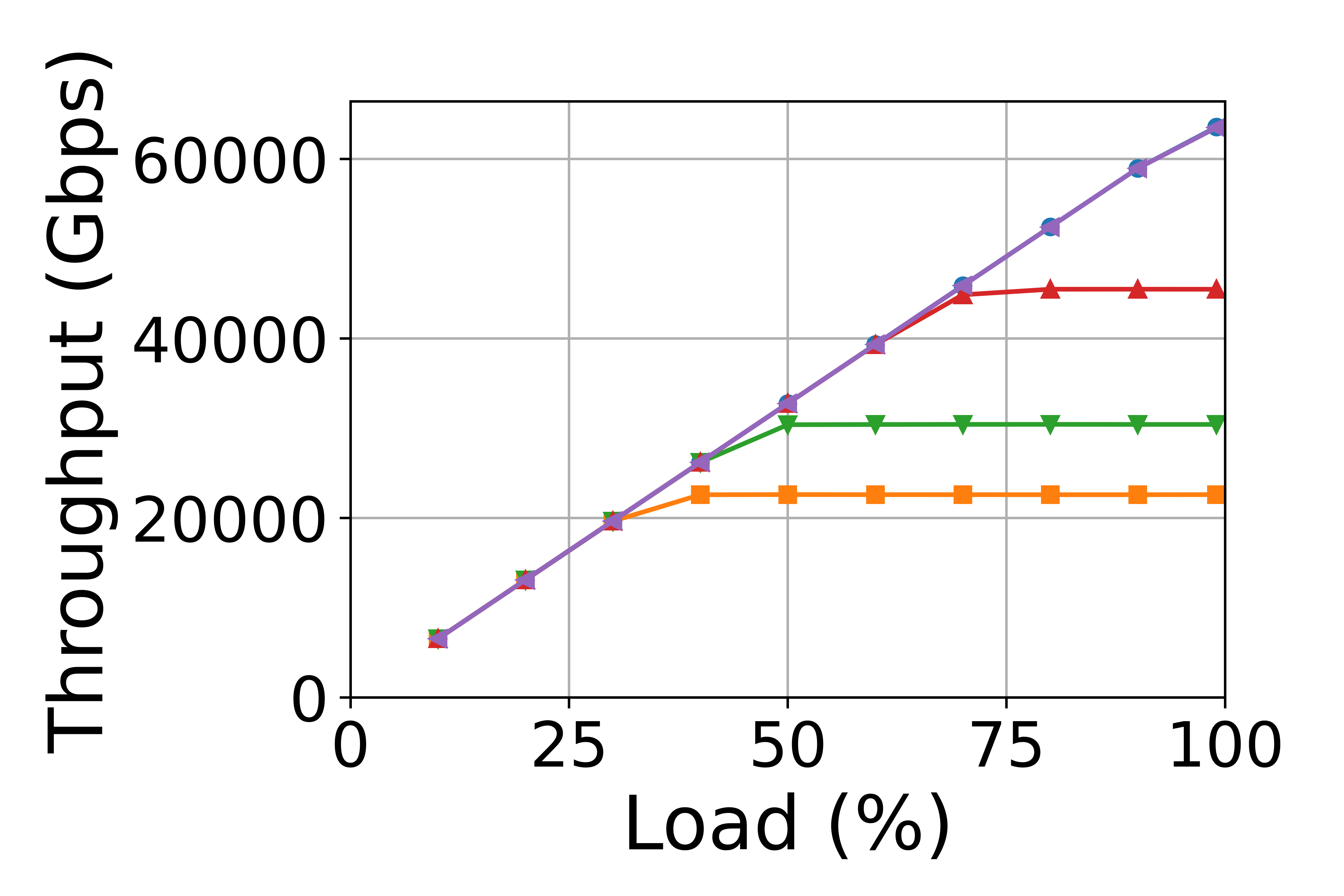}
        \caption{Age Based - conf \#2.}
        \label{fig:exp:arb:pcie4:agebased}
    \end{subfigure}
    \begin{subfigure}{0.32\textwidth}
        \centering
        \includegraphics[width=1\columnwidth]{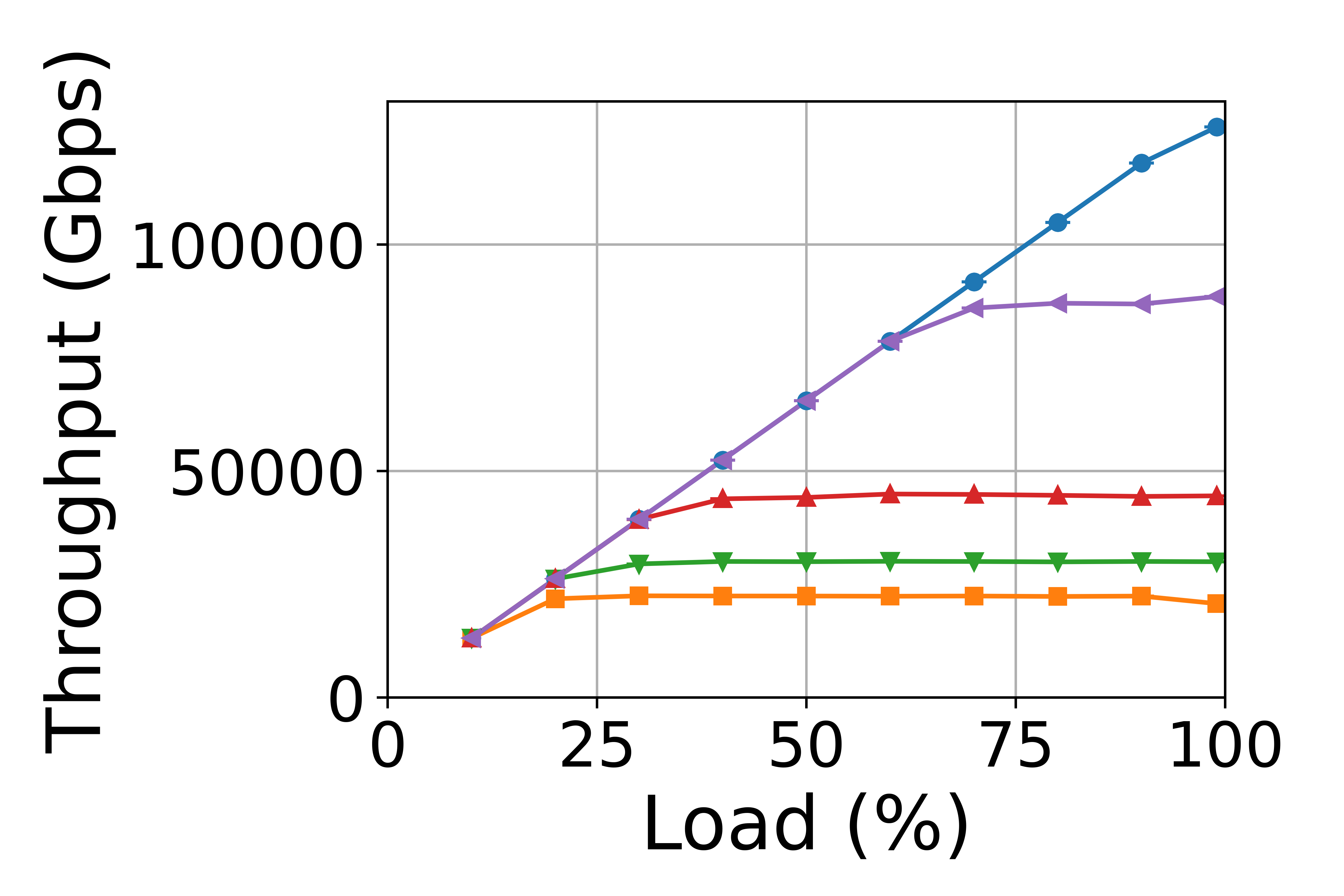}
        \caption{Age Based - conf \#3.}
        \label{fig:exp:arb:pcie5:agebased}
    \end{subfigure}
    \begin{subfigure}{0.32\textwidth}
        \centering
        \includegraphics[width=1\columnwidth]{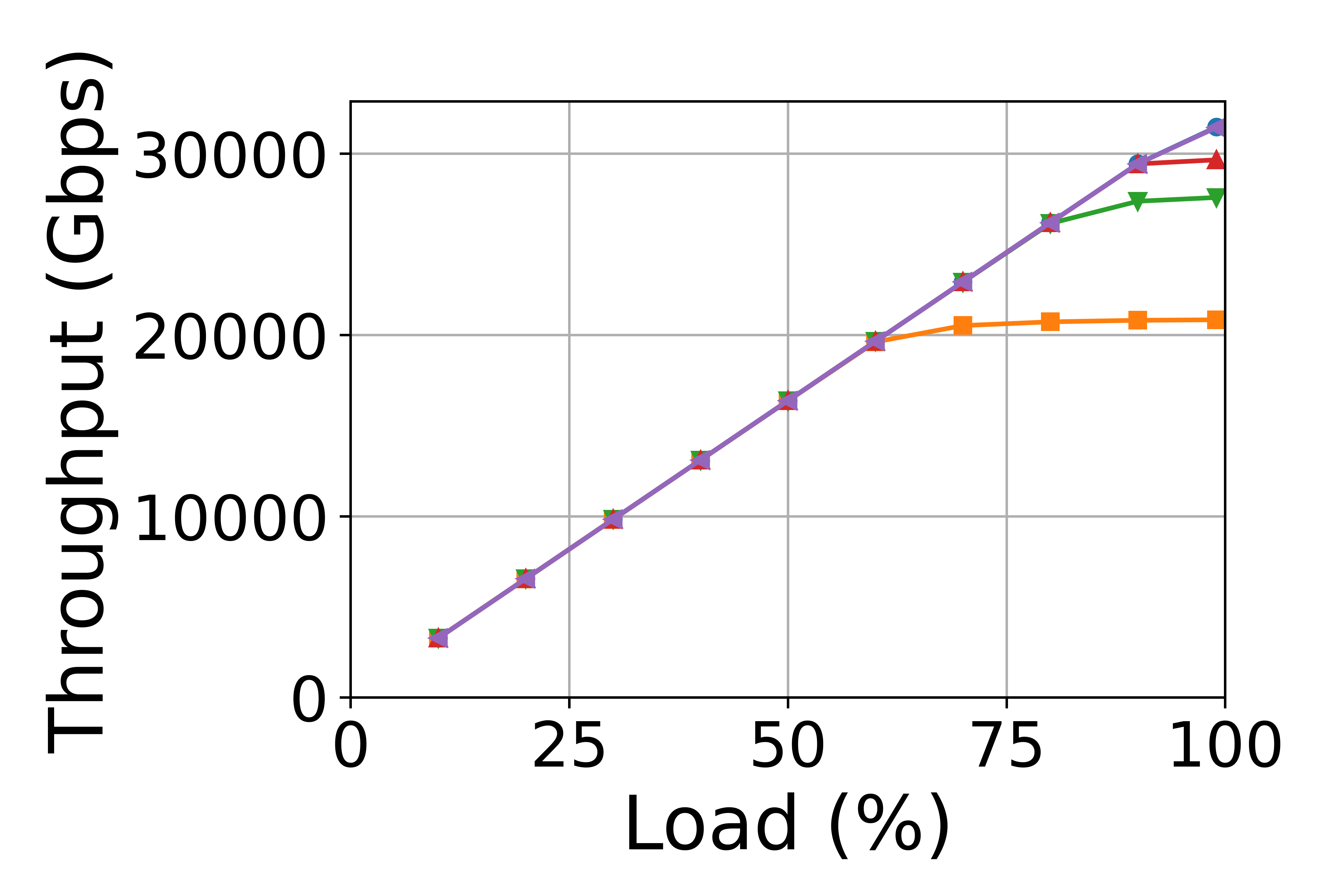}
        \caption{Round Robin - conf \#1.}
        \label{fig:exp:arb:pcie3:rr}
    \end{subfigure}
    \begin{subfigure}{0.32\textwidth}
        \centering
        \includegraphics[width=1\columnwidth]{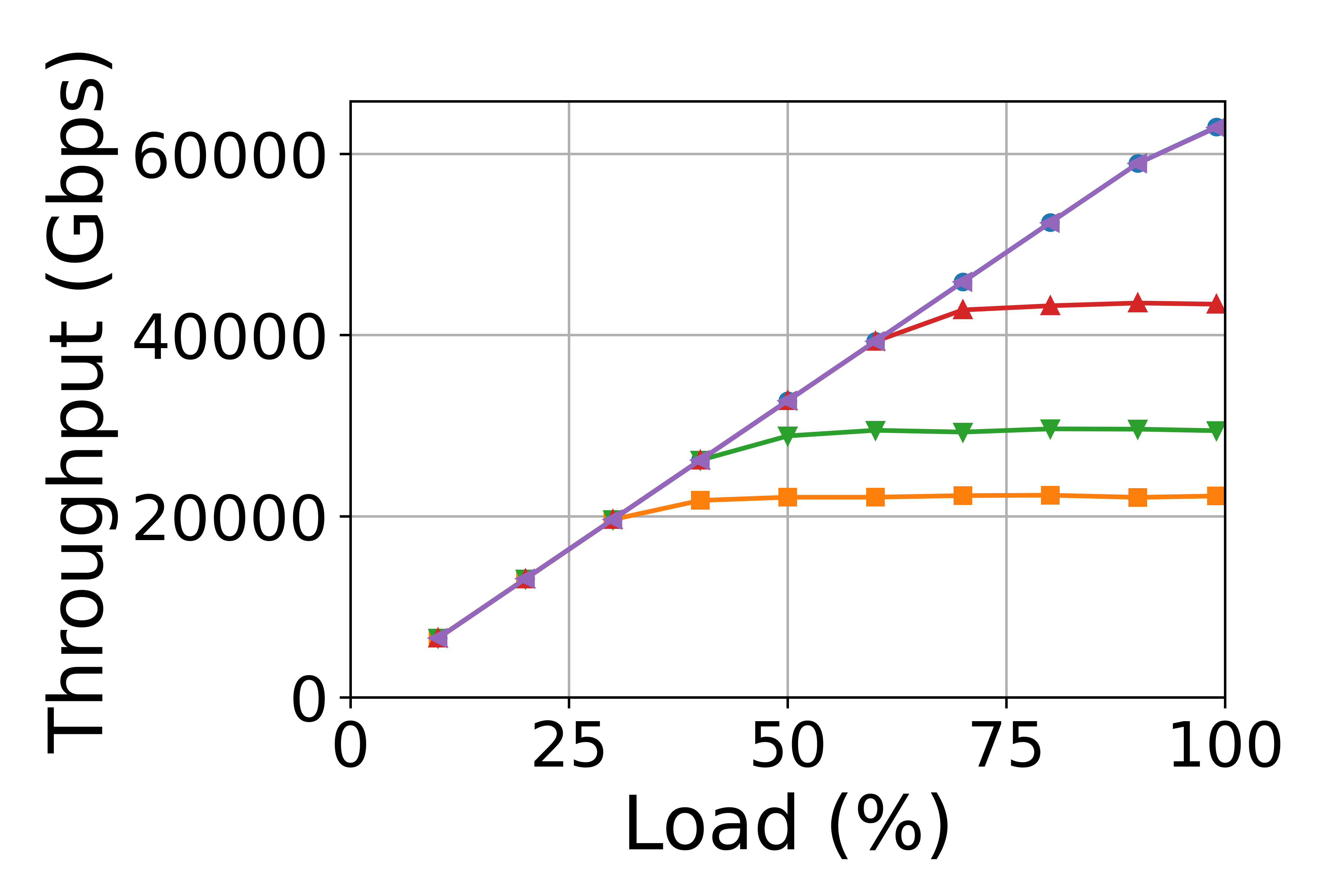}
        \caption{Round Robin - conf \#2.}
        \label{fig:exp:arb:pcie4:rr}
    \end{subfigure}
    \begin{subfigure}{0.32\textwidth}
        \centering
        \includegraphics[width=1\columnwidth]{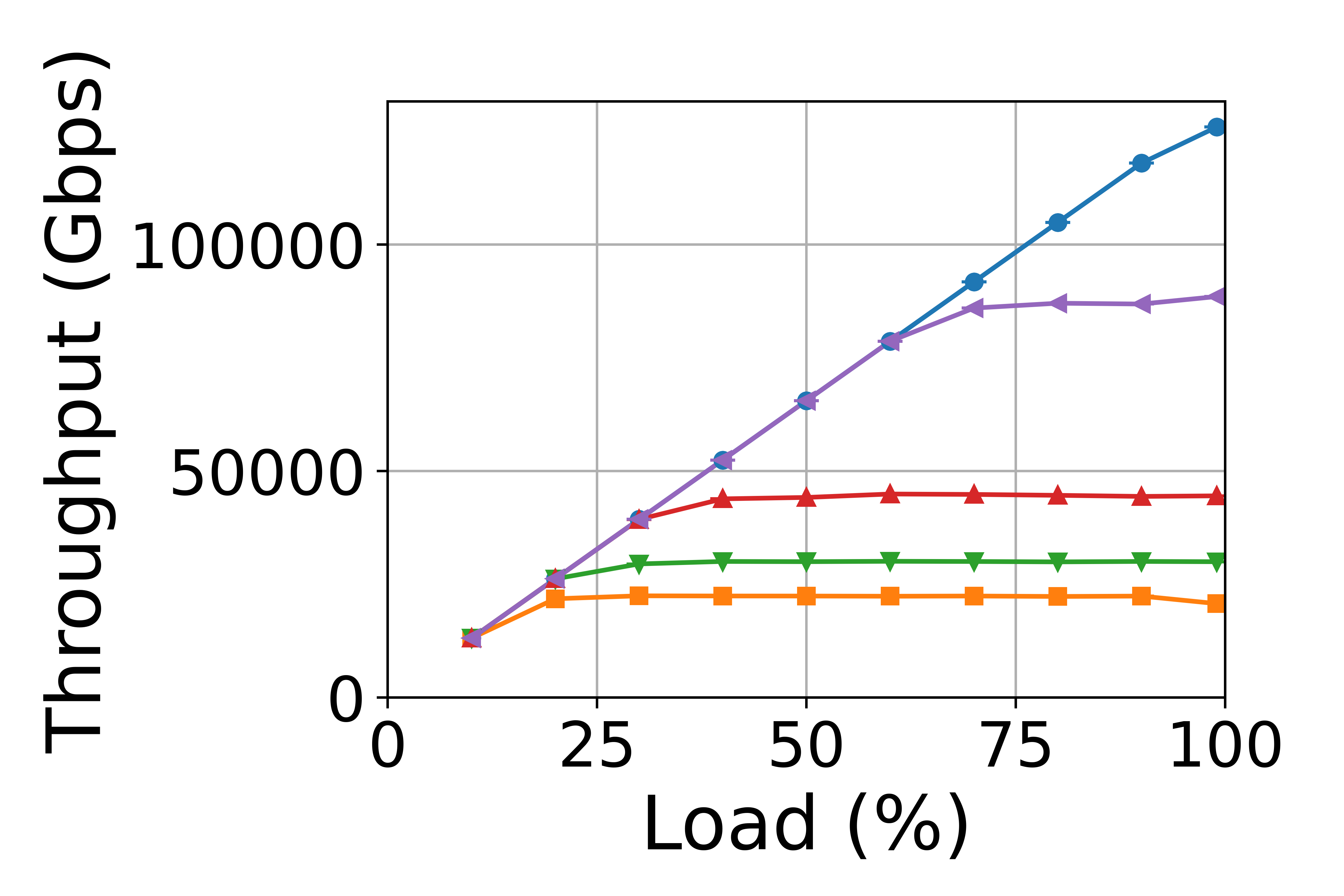}
        \caption{Round Robin - conf \#3.}
        \label{fig:exp:arb:pcie5:rr}
    \end{subfigure}
    \caption{Intra-node network performance with different arbiters as a function of traffic load (\%) in a 32-node RLFT topology.}
    \label{fig:exp:arb}
\end{figure}

First, we analyze whether the switch arbitration policy impacts overall system performance. We compare two arbitration schemes: Round-Robin (RR) and Age-Based for all the intra- and inter-node switches. We assume the intra-node network configurations \#1, \#2, and \#3 from Table~\ref{tab:exp:scaleup:config} only using $8$ accelerators per node, so there are $256$ accelerators in the system. Figure~\ref{fig:exp:arb} shows the average network throughput in the intra-node network versus the traffic load (between 0\% and 100\%) generated at accelerators' NICs, when the five traffic patterns C1, C2, C3, C4, and C5 are generated. As we can see, the performance differences between the two arbitration schemes are negligible across all configurations. This indicates that the arbitration policy does not significantly affect the system throughput in the assumed scenarios.

It is also worth mentioning that, as expected, when the intra-node link speed augments, the throughput saturation point is reached with a smaller aggregated traffic load for those traffic patterns generating a higher percentage of inter-node traffic. For instance, the traffic pattern C1, which generates 80\% of intra- and 20\% of inter-node traffic, saturates the network with 60\% of aggregated generation load in network configuration \#1 (see Figures~\ref{fig:exp:arb:pcie3:agebased} and~\ref{fig:exp:arb:pcie3:rr}), with 30\% of traffic load for network configuration \#2, and with 10\% of traffic load for network configuration \#3. 

\begin{figure}[!htb]
    \centering
    \begin{subfigure}{0.8\textwidth}
        \centering
        \frame{\includegraphics[width=1\textwidth]{Figures/JSC/leyendaBarras.pdf}}
    \end{subfigure}
    \\
    \hfill
    \begin{subfigure}{0.49\textwidth}
        \centering
        \adjincludegraphics[width=1\columnwidth,trim={{.03\width} {.05\width} {.13\width} {.13\width}},clip]{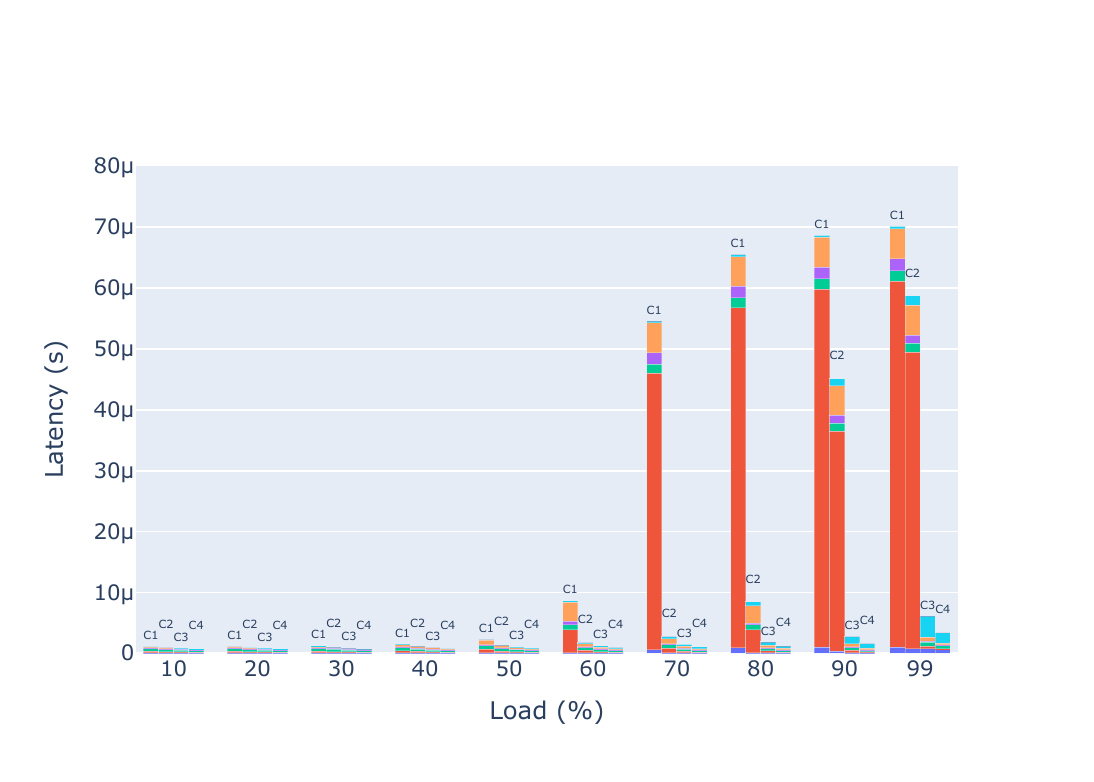}
        \caption{Age Based - conf \#1.}
        \label{fig:exp:arb:latencias:pcie3:agebased}
    \end{subfigure}
    \begin{subfigure}{0.49\textwidth}
        \centering
        \adjincludegraphics[width=1\columnwidth,trim={{.03\width} {.05\width} {.13\width} {.13\width}},clip]{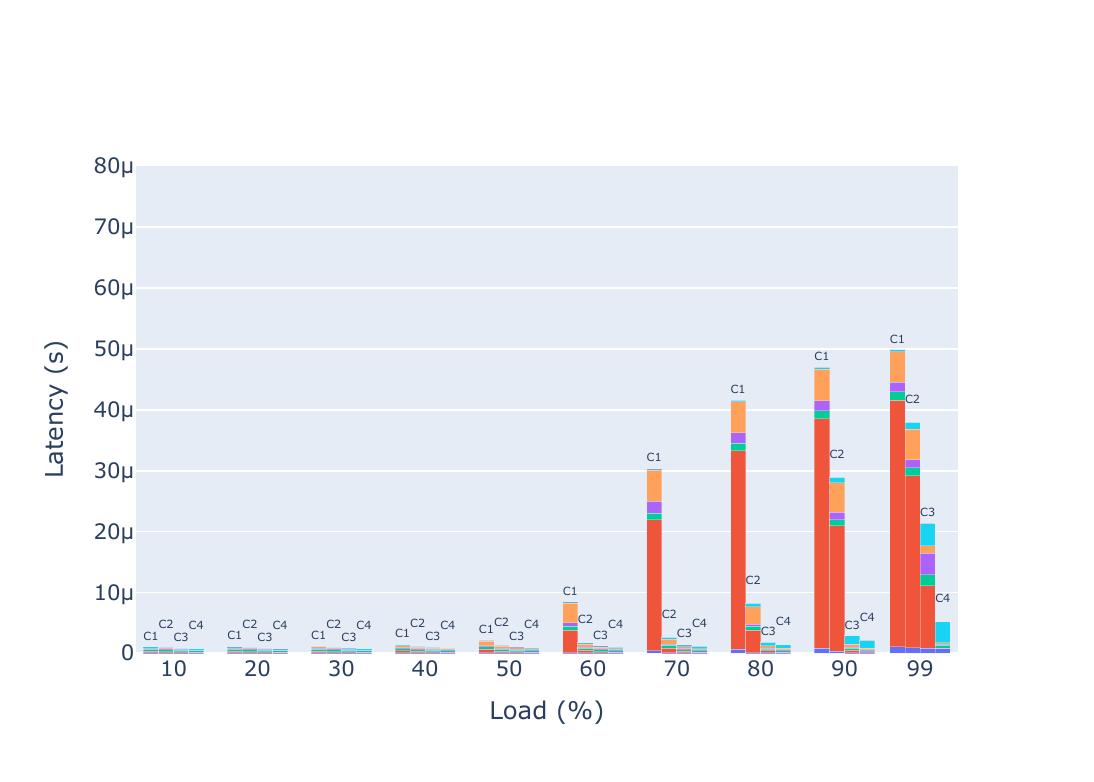}
        \caption{Round Robin - conf \#1.}
        \label{fig:exp:arb:latencias:pcie3:rr}
    \end{subfigure}
    \begin{subfigure}{0.49\textwidth}
        \centering
        \adjincludegraphics[width=1\columnwidth,trim={{.03\width} {.05\width} {.13\width} {.15\width}},clip]{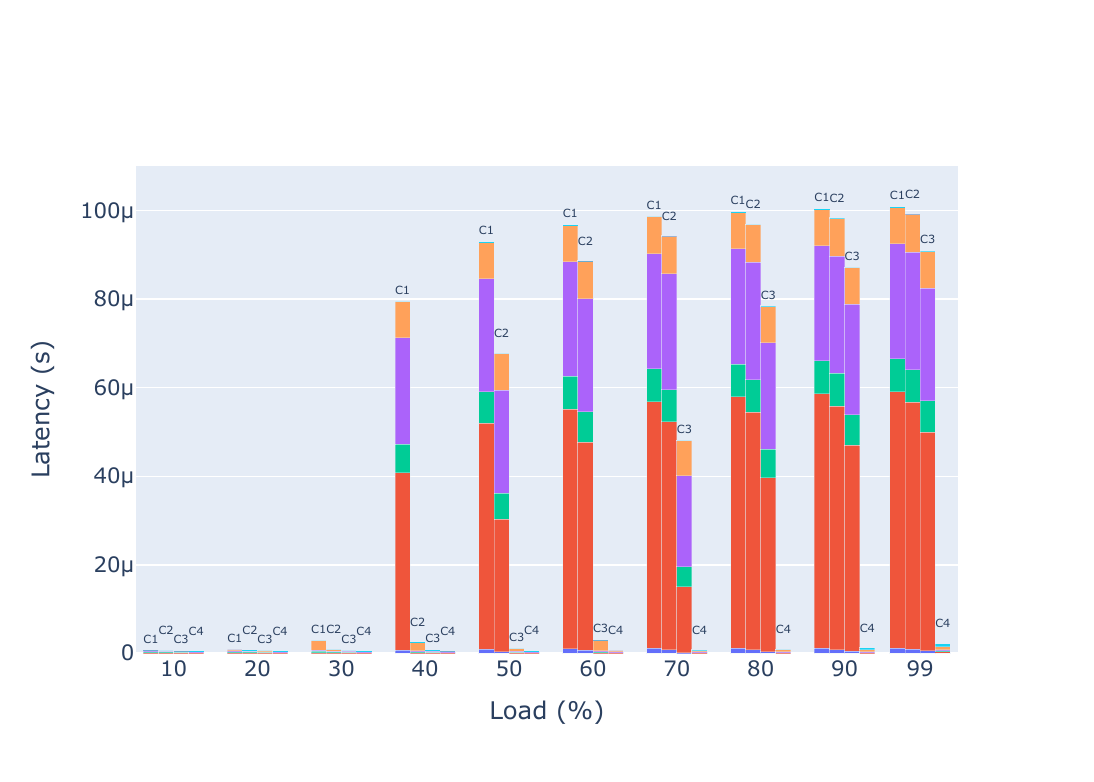}
        \caption{Age Based - conf \#2.}
        \label{fig:exp:arb:latencias:pcie4:agebased}
    \end{subfigure}
    \begin{subfigure}{0.49\textwidth}
        \centering
        \adjincludegraphics[width=1\columnwidth,trim={{.03\width} {.05\width} {.13\width} {.15\width}},clip]{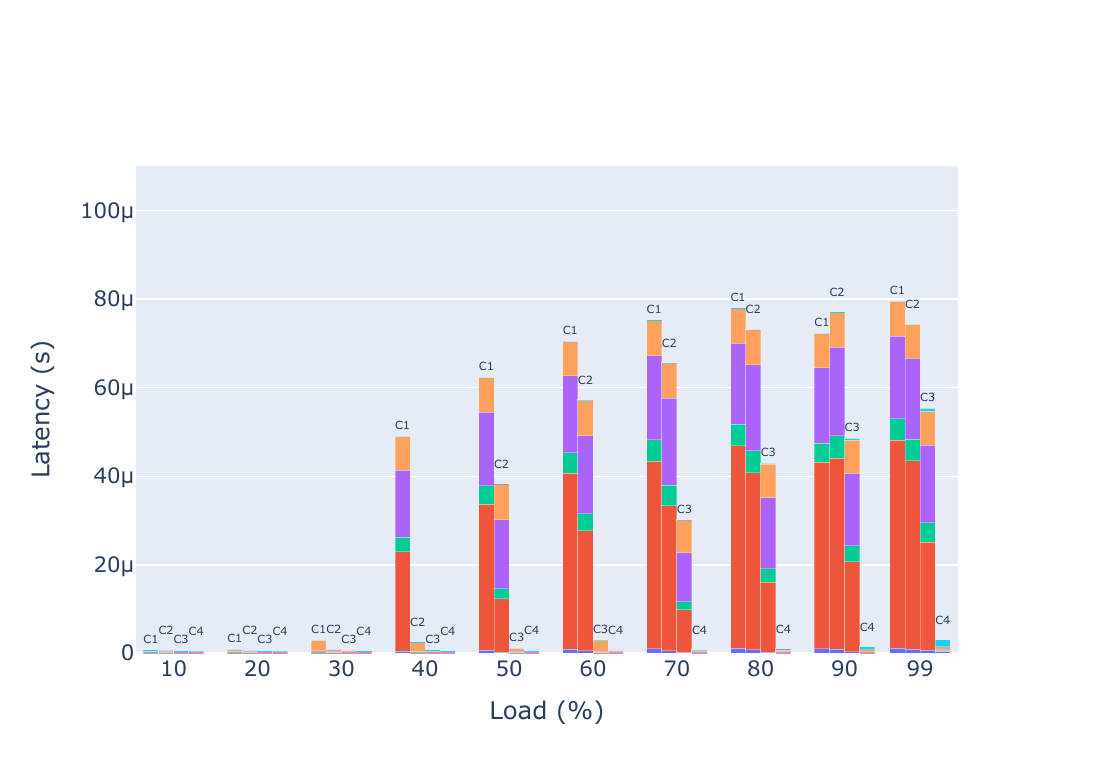}
        \caption{Round Robin - conf \#2.}
        \label{fig:exp:arb:latencias:pcie4:rr}
    \end{subfigure}
    \begin{subfigure}{0.49\textwidth}
        \centering
        \adjincludegraphics[width=1\columnwidth,trim={{.03\width} {.05\width} {.13\width} {.15\width}},clip]{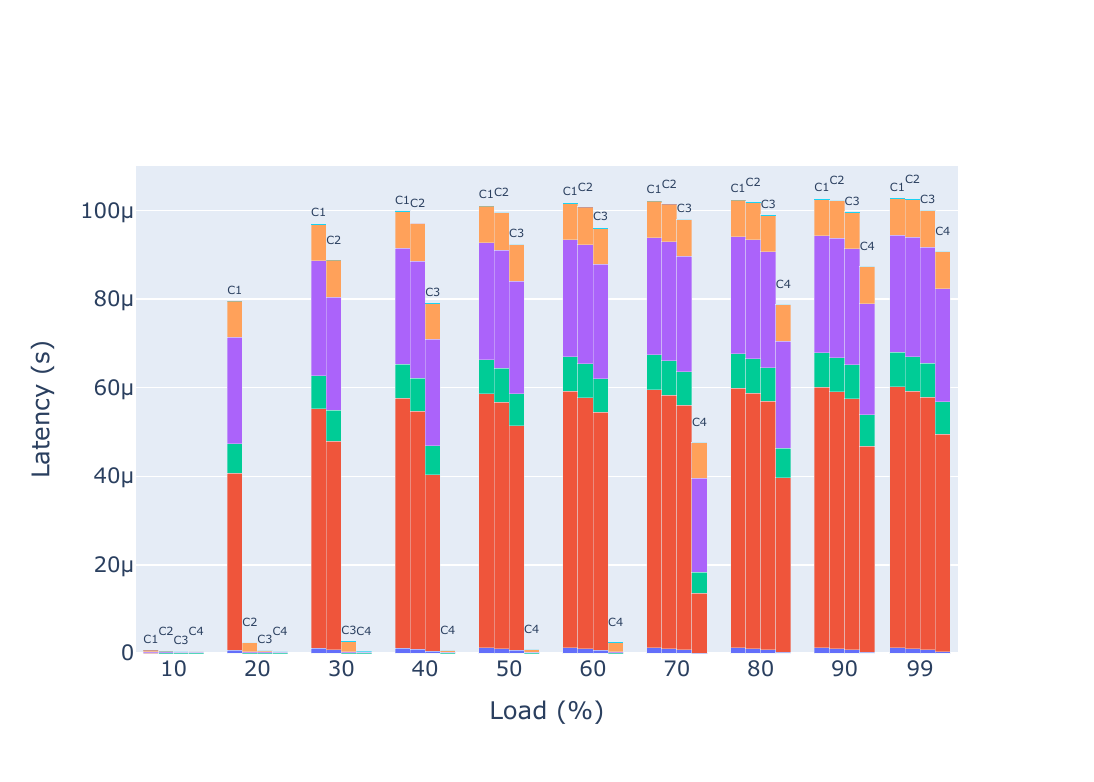}
        \caption{Age Based - conf \#3.}
        \label{fig:exp:arb:latencias:pcie5:agebased}
    \end{subfigure}
    \begin{subfigure}{0.49\textwidth}
        \centering
        \adjincludegraphics[width=1\columnwidth,trim={{.03\width} {.05\width} {.13\width} {.15\width}},clip]{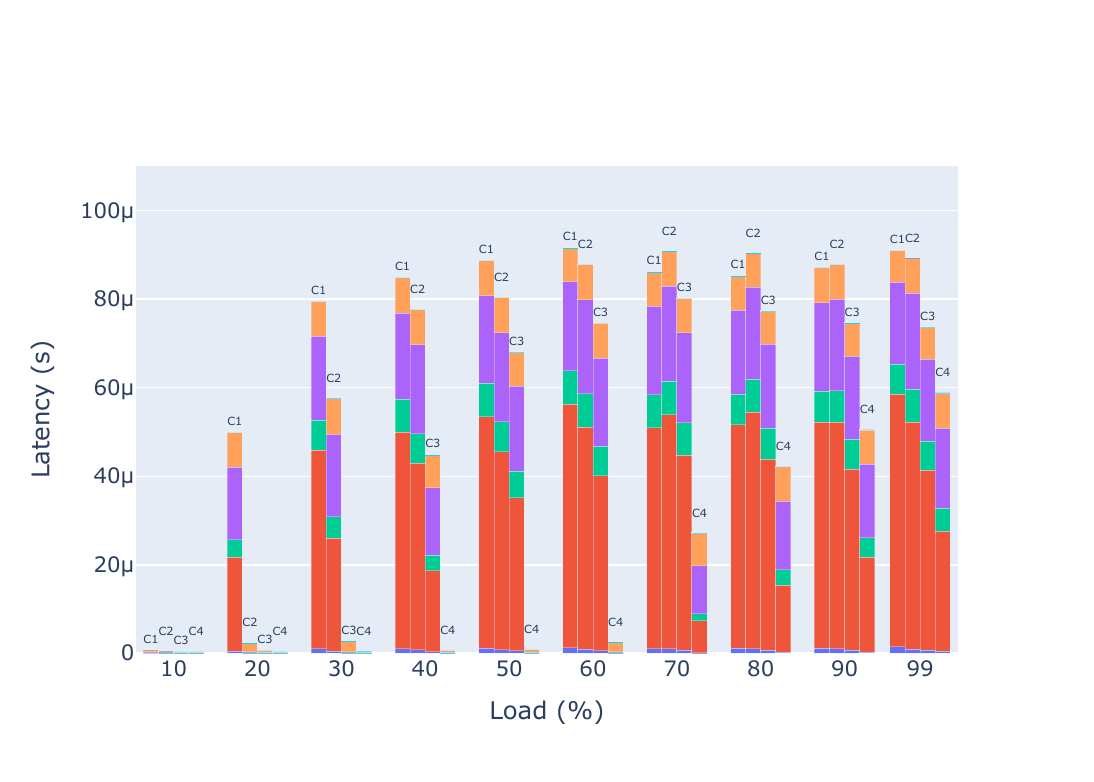}
        \caption{Round Robin - conf \#3.}
        \label{fig:exp:arb:latencias:pcie5:rr}
    \end{subfigure}
    \caption{Packet Latency divided into components versus traffic load for the traffic patterns C1-C4 and network configurations \#1, \#2, and \#3 in Table~\ref{tab:exp:scaleup:config}.}
    \label{fig:exp:arb:latencias}
\end{figure}

Figure~\ref{fig:exp:arb:latencias} shows the aggregated latency results for the scenarios mentioned above. Each vertical bar is divided into seven sections for the seven latency components described in Figure~\ref{fig:evaluation:latencymetrics}. The ``X'' axis shows ten load instants (from 0\% to 100\%) and one bar per traffic pattern C1-C4. Note that the C5 traffic pattern has not been shown since there is no inter-node communication. These barplots also show that most latency is accumulated in the source intra-node network when saturation happens (red color). When the intra-node network speed augments (network configurations \#2 and \#3), the source NIC, inter-node network, and destination NIC latency increase because the amount of traffic the accelerators send through the inter-node network also augments.

Regarding the arbitration policies, there are negligible differences between Age-based and Round Robin policies in terms of network throughput, while there are some variations in latency behavior, which do not significantly impact system performance. We can see that Age-based arbitration scenarios experience 20\% more aggregated latency under saturation compared to Round Robin policies, since the former prioritizes older crossing requests within network switches. Another important aspect, later analyzed, is the biggest portion of the aggregated latency, which is consumed at the Source intra-node network under saturation. But this portion is shared with Inter-node network latency and Source and Destination NIC when the accelerators NIC speed increases (see configurations \#2 and \#3). Since the Round Robin arbitration is commonly used and easy to implement in hardware, we have selected this policy for rest of the experiments of the paper hereafter.

\subsubsection{Analysis of the intra-node network speed}
\label{sec:evaluation:scaleUp:intranodeConf}

\begin{figure}[!htb]
    \centering
    \begin{subfigure}{0.6\textwidth}
        \centering
        \frame{\includegraphics[width=1\textwidth]{Figures/JSC/Legend-modified.pdf}}
    \end{subfigure}
    \\
    \begin{subfigure}{0.24\textwidth}
        \centering
        \includegraphics[width=1\columnwidth]{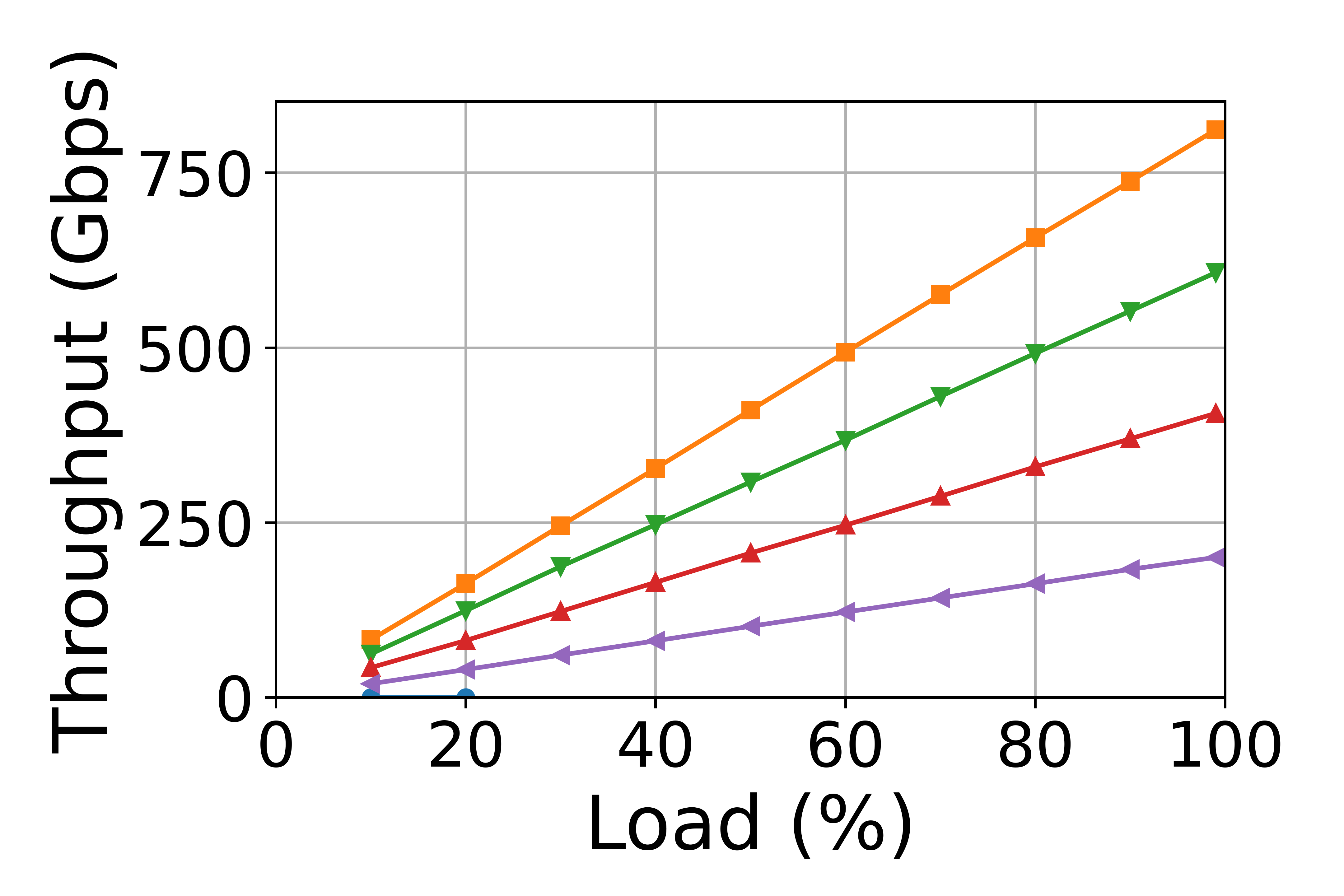}
    \end{subfigure}
    \begin{subfigure}{0.24\textwidth}
        \centering
        \includegraphics[width=1\columnwidth]{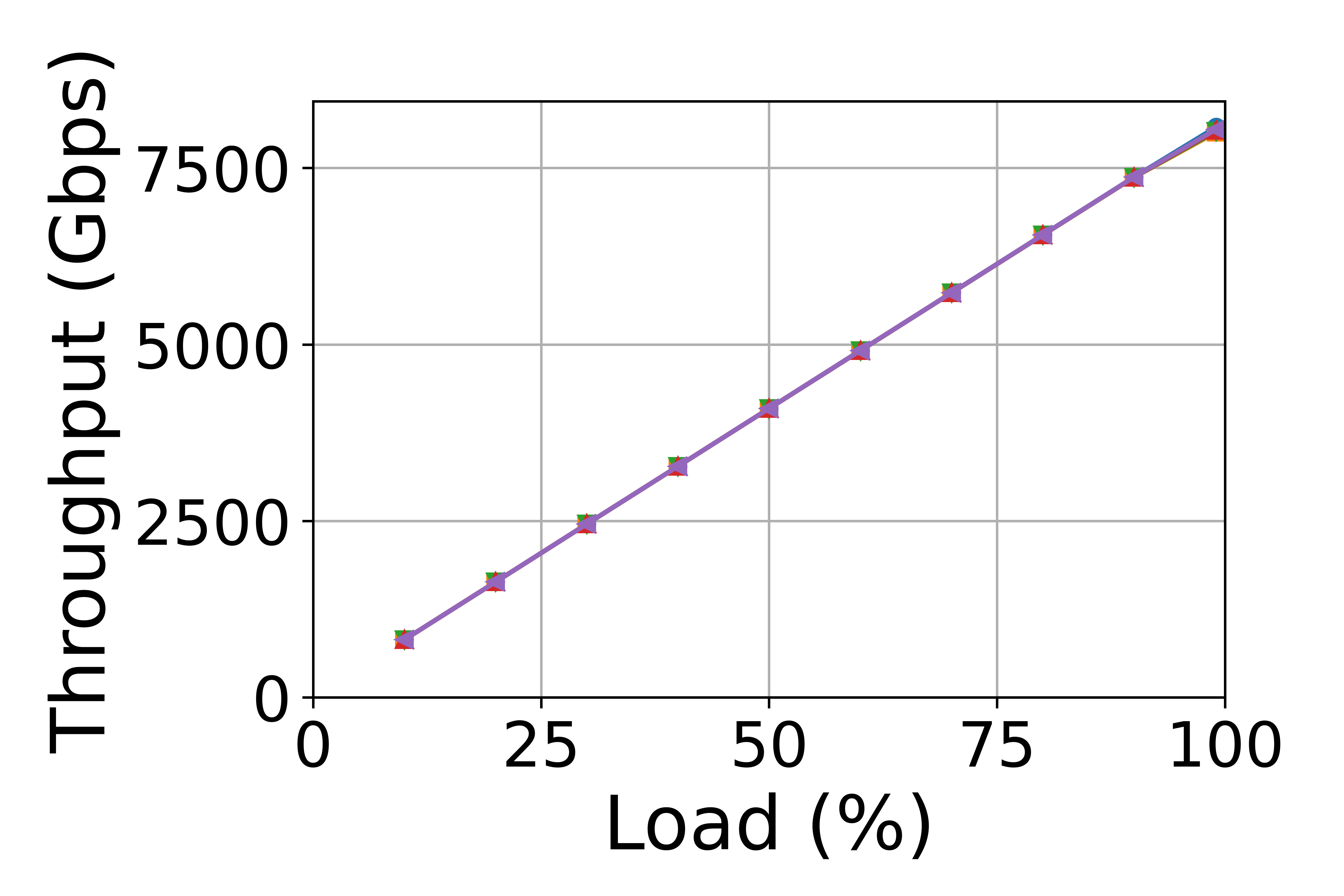}
    \end{subfigure}
    \begin{subfigure}{0.24\textwidth}
        \centering
        \includegraphics[width=1\columnwidth]{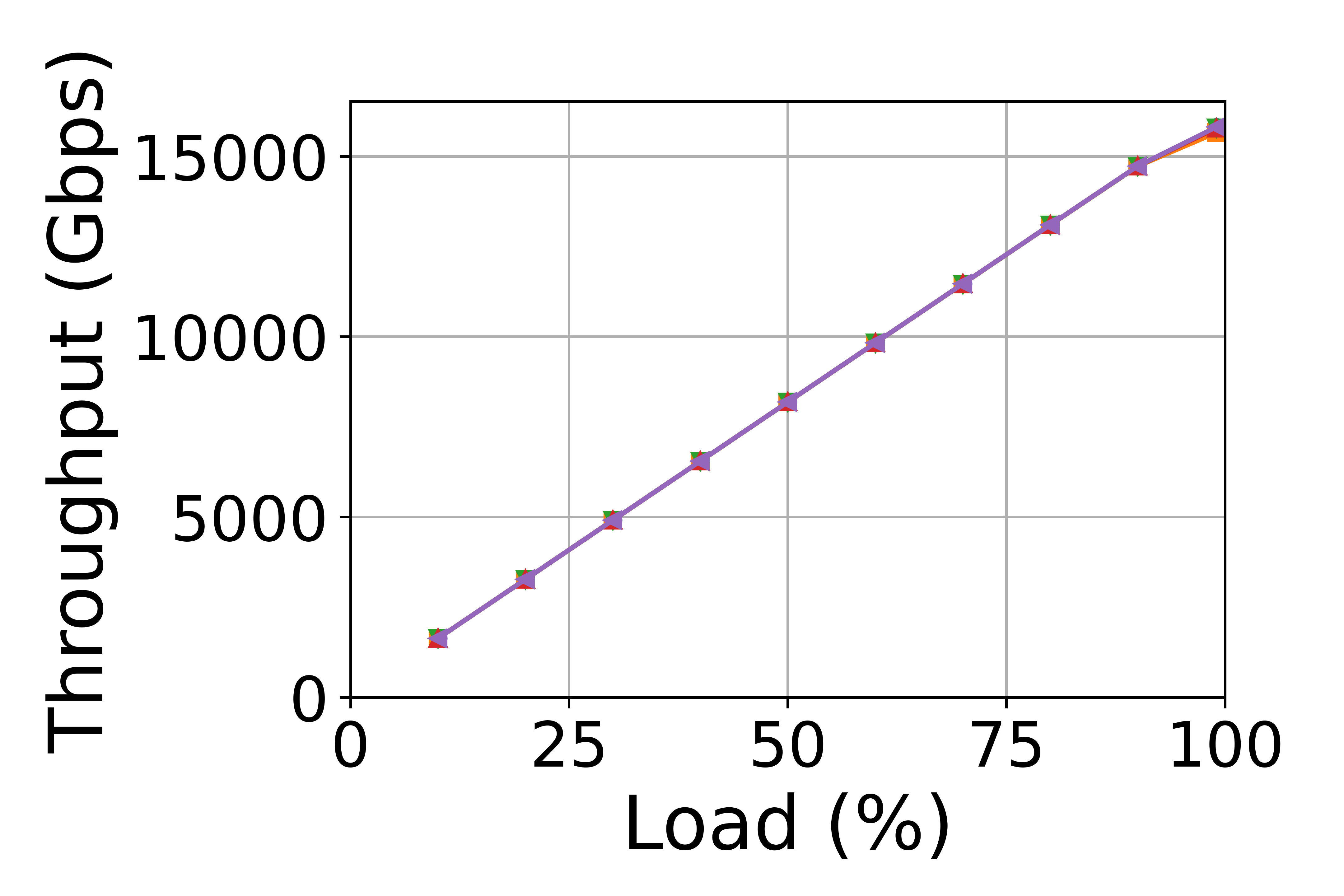}
    \end{subfigure}
    \begin{subfigure}{0.24\textwidth}
        \centering
        \includegraphics[width=1\columnwidth]{Figures/JSC/RR/Scale-Up/PCIe3/8_GPU/Intranode.png}
    \end{subfigure}
    \begin{subfigure}{0.24\textwidth}
        \centering
        \includegraphics[width=1\columnwidth]{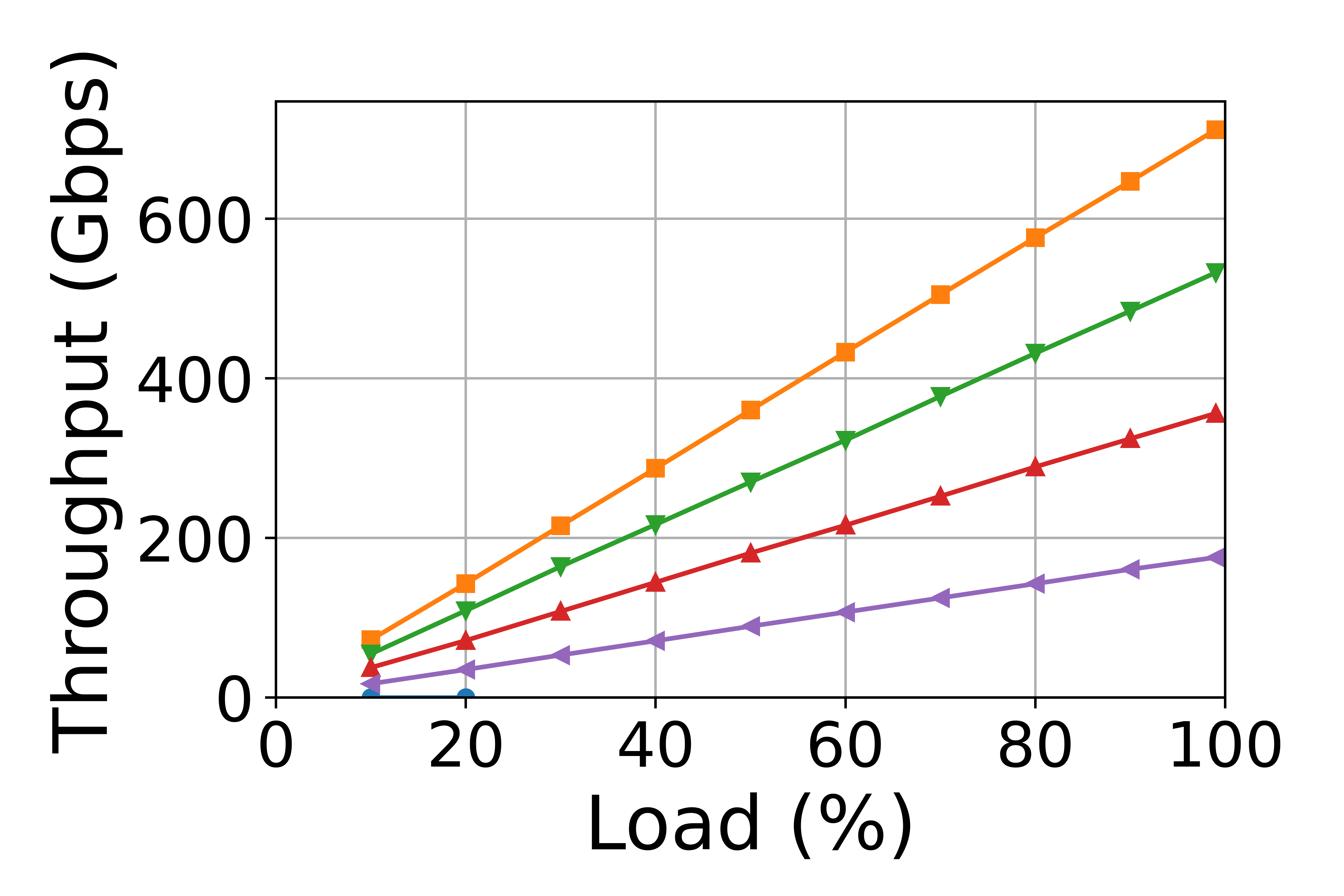}
        \caption{$1$ Acc./Node.}
        \label{fig:exp:scaleup:pcie3:1gpu}
    \end{subfigure}
    \begin{subfigure}{0.24\textwidth}
        \centering
        \includegraphics[width=1\columnwidth]{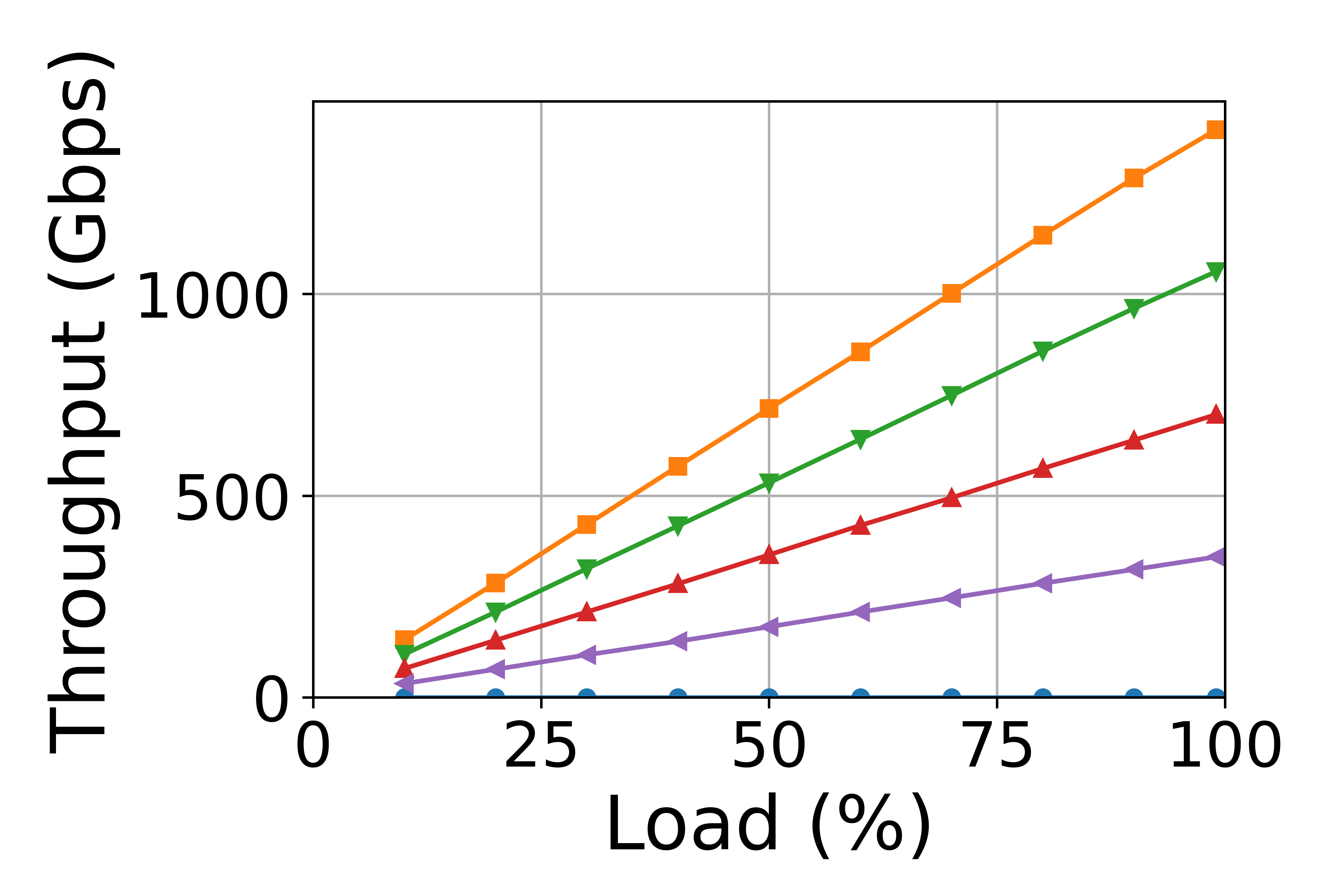}
        \caption{$2$ Acc./Node.}
        \label{fig:exp:scaleup:pcie3:2gpu}
    \end{subfigure}
    \begin{subfigure}{0.24\textwidth}
        \centering
        \includegraphics[width=1\columnwidth]{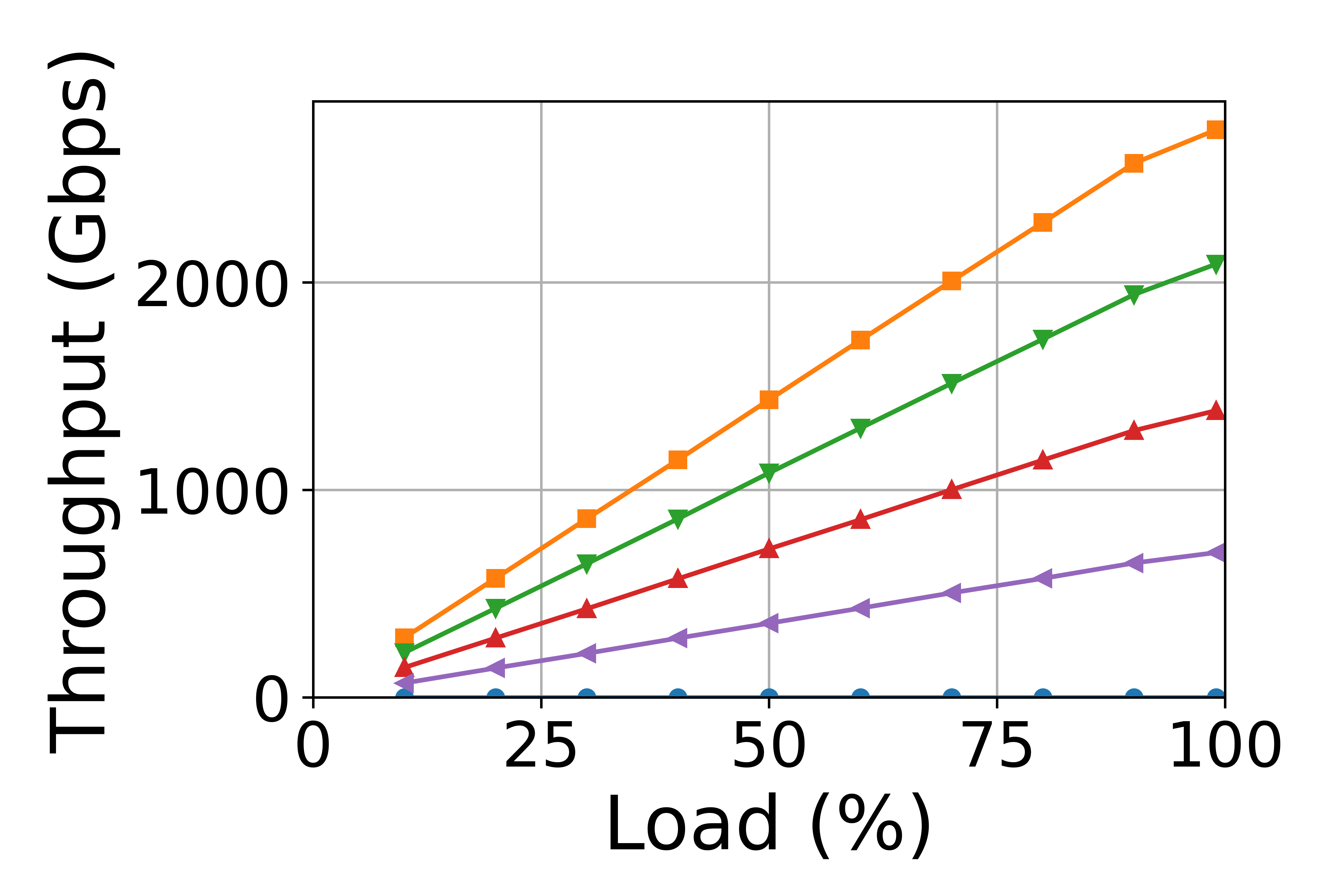}
        \caption{$4$ Acc./Node.}
        \label{fig:exp:scaleup:pcie3:4gpu}
    \end{subfigure}
    \begin{subfigure}{0.24\textwidth}
        \centering
        \includegraphics[width=1\columnwidth]{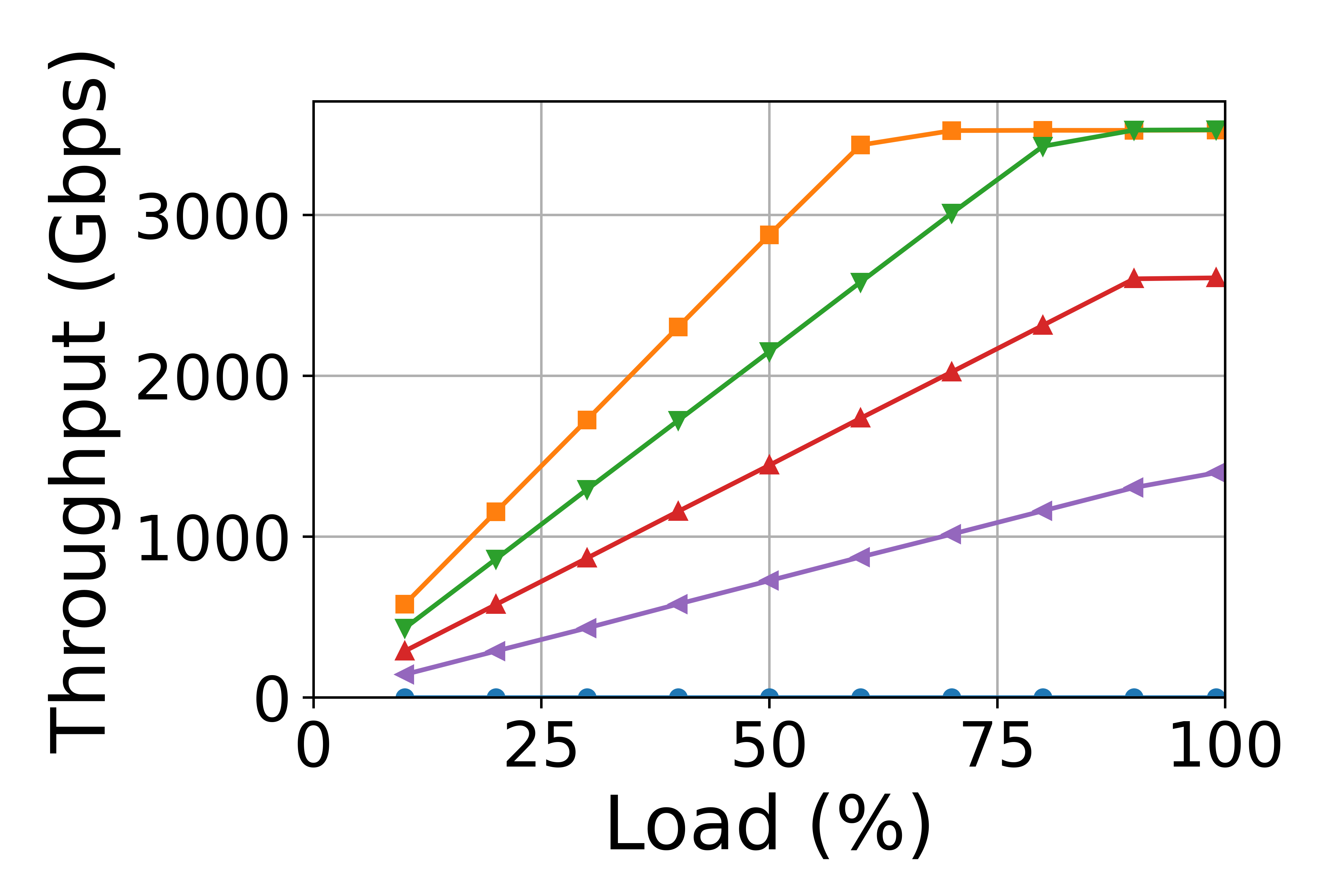}
        \caption{$8$ Acc./Node.}
        \label{fig:exp:scaleup:pcie3:8gpu}
    \end{subfigure}
    \caption{Network throughput as a function of traffic load (\%) in a 32-node RLFT topology (network configuration \#1 in Table~\ref{tab:exp:scaleup:config}). Total network throughput is shown in the top figures, while inter-node throughput is depicted in the bottom ones.}
    \label{fig:exp:scaleup:pcie3}
\end{figure}

This section analyzes the intra- and inter-node network performance when the intra-node network scales up in terms of the number of accelerators per end node and the intra-node link speed. Figure~\ref{fig:exp:scaleup:pcie3} illustrates the intra- and inter-node network throughput of a $32$-node RLFT topology, when C1 to C5 traffic patterns are generated. We assume that the number of accelerators per end node is $1$, $2$, $4$, or $8$, so that the total number of accelerators is $32$, $64$, $128$, and $256$ for the entire system. We also assume that each accelerator can generate up to $128$ Gbps of traffic to the intra-node network.

Figure~\ref{fig:exp:scaleup:pcie3:1gpu} shows throughput results (i.e., the amount of information received per accelerator) with a single accelerator per node (i.e., $32$ accelerators in total and $128$ Gbps per accelerator). In this case there is no possible intra-node communication among accelerators within a single end node. For instance, for C1 there is a 20\% of inter-node traffic, i.e., each accelerator generates $128\times0.2=25.6$~Gbps, meaning that the total generation rate in the network can be up to $25.6\times32$accelerators$=819.2$~Gbps. Therefore, the throughput assuming no saturation at maximum load (in this case $25.6$ Gbps per accelerator) is $819.2$~Gbps. Note that single-accelerator end nodes are no longer practical, and multi-accelerator end nodes have become essential. For instance, one of the reasons is the increasingly popular use of LLMs~\cite{LLMParamsIncrease} and their growing amount of memory requirements, where Tensor Parallelism (TP) is crucial for partitioning models across multiple accelerators. TP is communication-intensive, requiring frequent message exchanges for \emph{All-Reduce} operations.

Increasing the number of accelerators per end node to $2$ or $4$ increments the network throughput as well. Figures~\ref{fig:exp:scaleup:pcie3:2gpu} and~\ref{fig:exp:scaleup:pcie3:4gpu} show that there is no saturation for server nodes with $2$ and $4$ accelerators per node, regardless of the traffic pattern, reaching the maximum bandwidth of \num{8192}~Gbps and \num{16384}~Gbps, respectively. This throughput is achieved with $64$ and $128$ accelerators generating traffic at full speed (i.e., $128$~Gbps) without any bottleneck in the network. The link between the intra-node network and the NIC (see Figure~\ref{fig:evaluation:latencymetrics}) remains unsaturated. Therefore, the intra and inter-node networks achieve the maximum performance for all the traffic pattern configurations (C1-C4), even for C5, which does not involve inter-node traffic.

Figure~\ref{fig:exp:scaleup:pcie3:8gpu} shows the results with $8$ accelerators per server node. For traffic pattern C5 (i.e., 100\% intra-node traffic), the inter-node network is unused, so the system achieves its maximum possible throughput, which is $128~\text{Gbps}/\text{accelerator} \times 8~\text{accelerators}/\text{node} \times 32~\text{nodes} = $\num{32768}$~\text{Gbps}$. In traffic configurations C4 and C3 (i.e., 95\% and 90\% of intra-node traffic, respectively) do not saturate the NIC link. Considering that the maximum traffic an end node can inject into the inter-node network is $128~\text{Gbps}/\text{accelerator} \times 8~\text{accelerators} = \num{1024}~\text{Gbps}$, each end-node NIC only receives packets at a rate of 51.2~Gbps for traffic pattern C4 (i.e., 5\% of inter-node traffic) and 102.4~Gbps for C3 (i.e., 10\% of inter-node traffic), respectively. Thus, the link connecting the intra-node switch to the end-node NIC, whose maximum bandwidth is $128$ Gbps, can accommodate the intra-node load that the $8$ accelerators send to the inter-node network without becoming a bottleneck.

By contrast, traffic patterns C2 and C1 (85\% and 80\% of intra-node traffic, respectively) exceed the intra-node network link maximum capacity (i.e., $128$~Gbps), since $8$ accelerators per end node generate a traffic load of $153.6$~Gbps for C2 (i.e., 15\% of inter-node traffic) and $204.8$~Gbps for C1 (i.e., 20\% of inter-node traffic), respectively. In these scenarios, the intra-node link becomes a bottleneck. When scaling up the number of accelerators per end node, it is essential to consider the intra-node link bandwidth and the amount of traffic the accelerators can generate to the inter-node network, as excessive traffic routed through the end-node NIC (either incoming or outgoing) can degrade overall system performance. Note that, due to the mentioned bottleneck, the inter-node network cannot reach its theoretical maximum throughput of $400~\text{Gbps} \times 32~\text{nodes} =$ \num{12800}$~\text{Gbps}$, as the traffic injection rate from the intra-node network is limited by its maximum link bandwidth, i.e., $128$~Gbps. Therefore, for $128$~Gbps intra-node links, the appropriate number of accelerators per end node is $1$, $2$, or $4$, while $8$ accelerators saturate the intra-node network when there is a small percentage of traffic generated to the inter-node network.

\begin{figure}[!htb]
    \centering
    \begin{subfigure}{0.6\textwidth}
        \centering
        \frame{\includegraphics[width=1\textwidth]{Figures/JSC/Legend-modified.pdf}}
    \end{subfigure}
    \\
        \begin{subfigure}{0.24\textwidth}
        \centering
        \includegraphics[width=1\columnwidth]{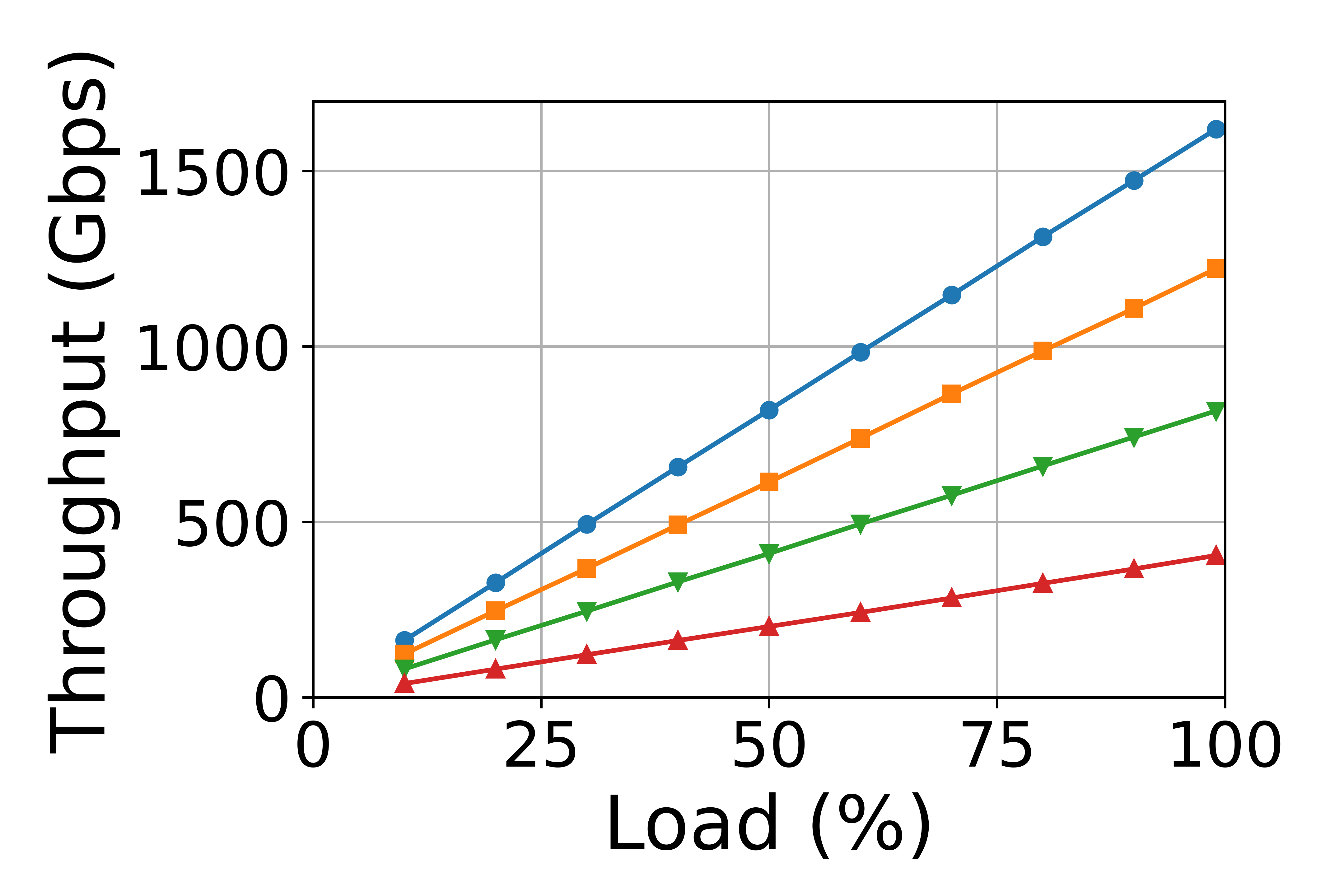}
    \end{subfigure}
    \begin{subfigure}{0.24\textwidth}
        \centering
        \includegraphics[width=1\columnwidth]{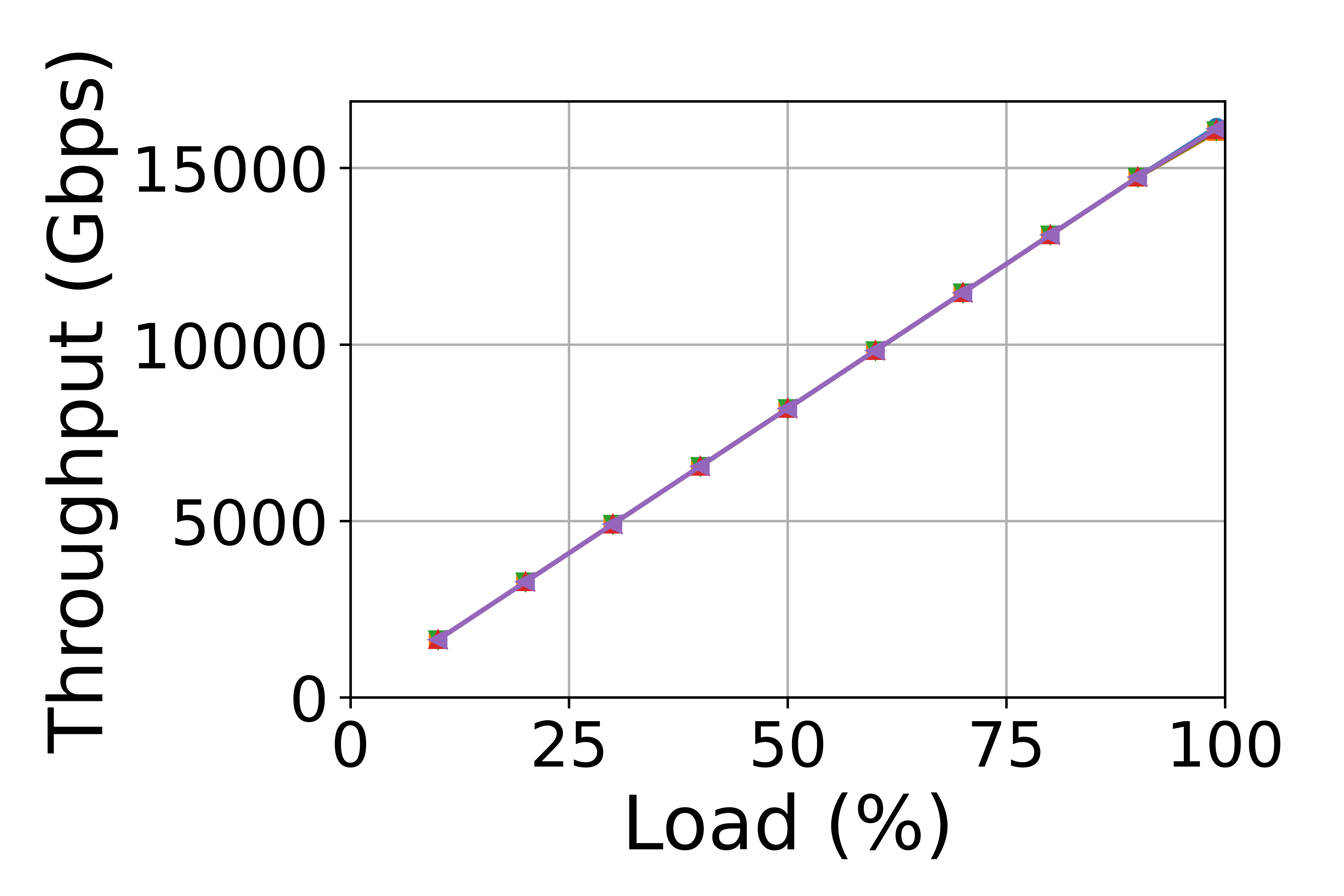}
    \end{subfigure}
    \begin{subfigure}{0.24\textwidth}
        \centering
        \includegraphics[width=1\columnwidth]{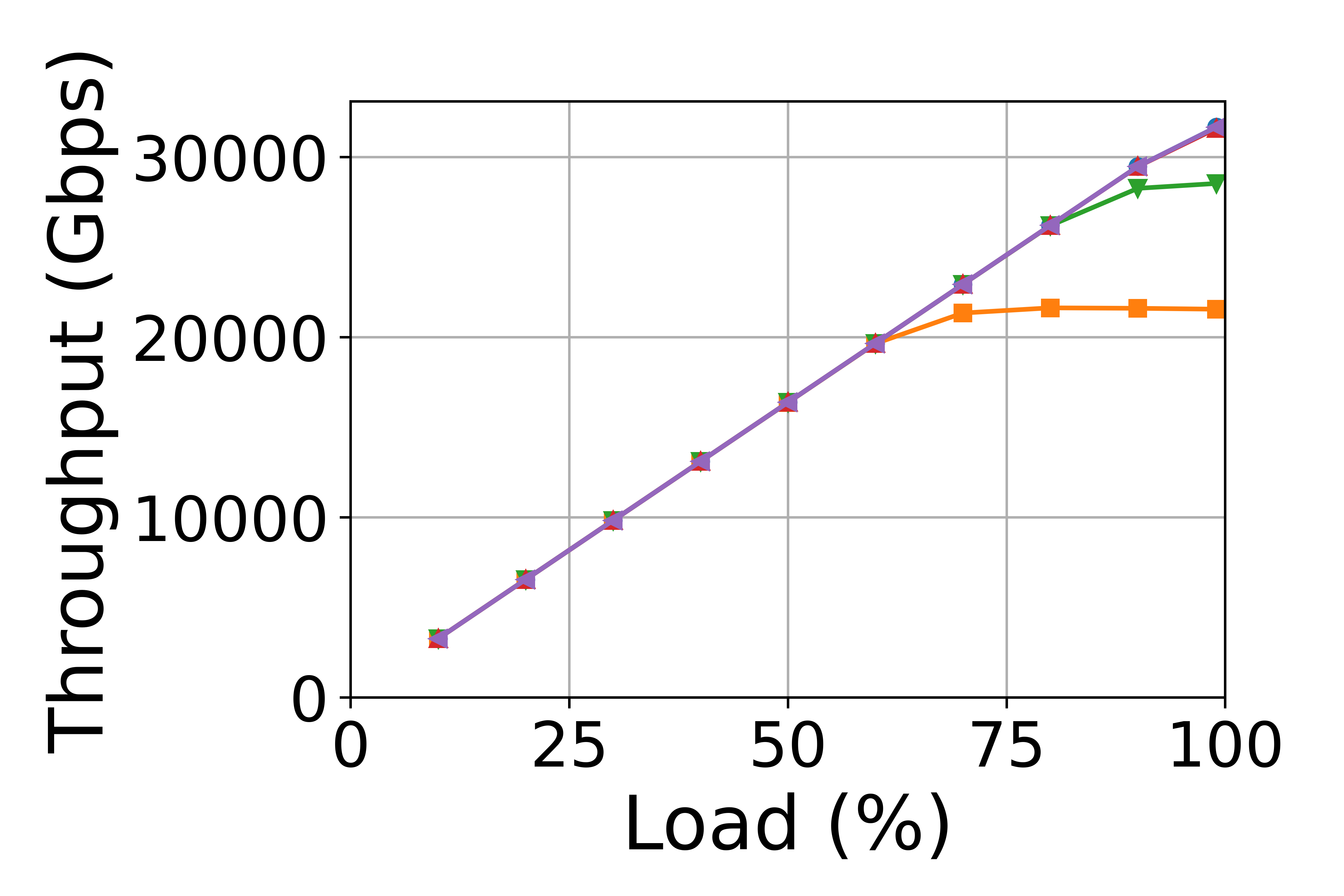}
    \end{subfigure}
    \begin{subfigure}{0.24\textwidth}
        \centering
        \includegraphics[width=1\columnwidth]{Figures/JSC/RR/Scale-Up/PCIe4/8_GPU/Intranode.png}
    \end{subfigure}
    \begin{subfigure}{0.24\textwidth}
        \centering
        \includegraphics[width=1\columnwidth]{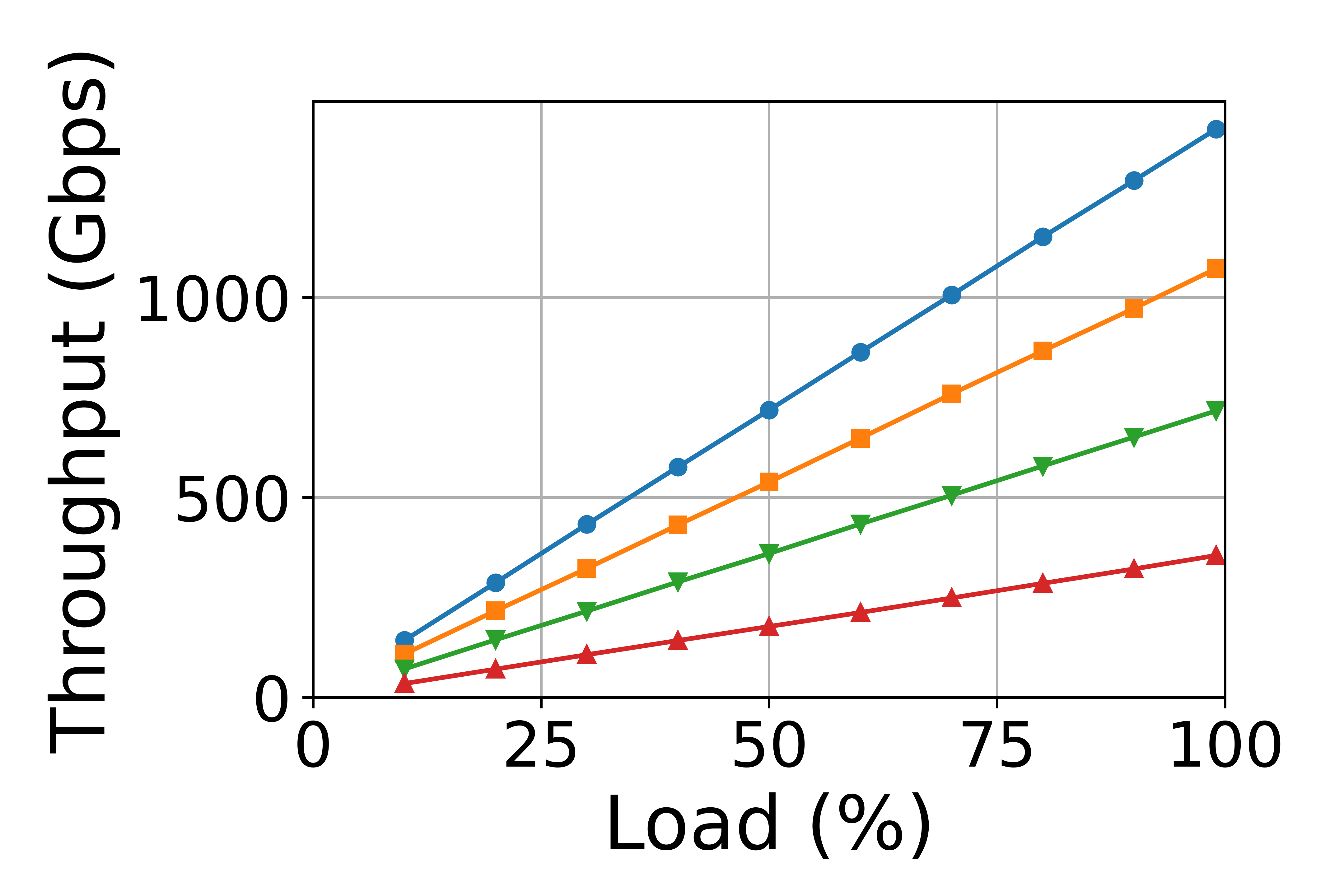}
        \caption{$1$ Acc./Node.}
        \label{fig:exp:scaleup:pcie4:1gpu}
    \end{subfigure}
    \begin{subfigure}{0.24\textwidth}
        \centering
        \includegraphics[width=1\columnwidth]{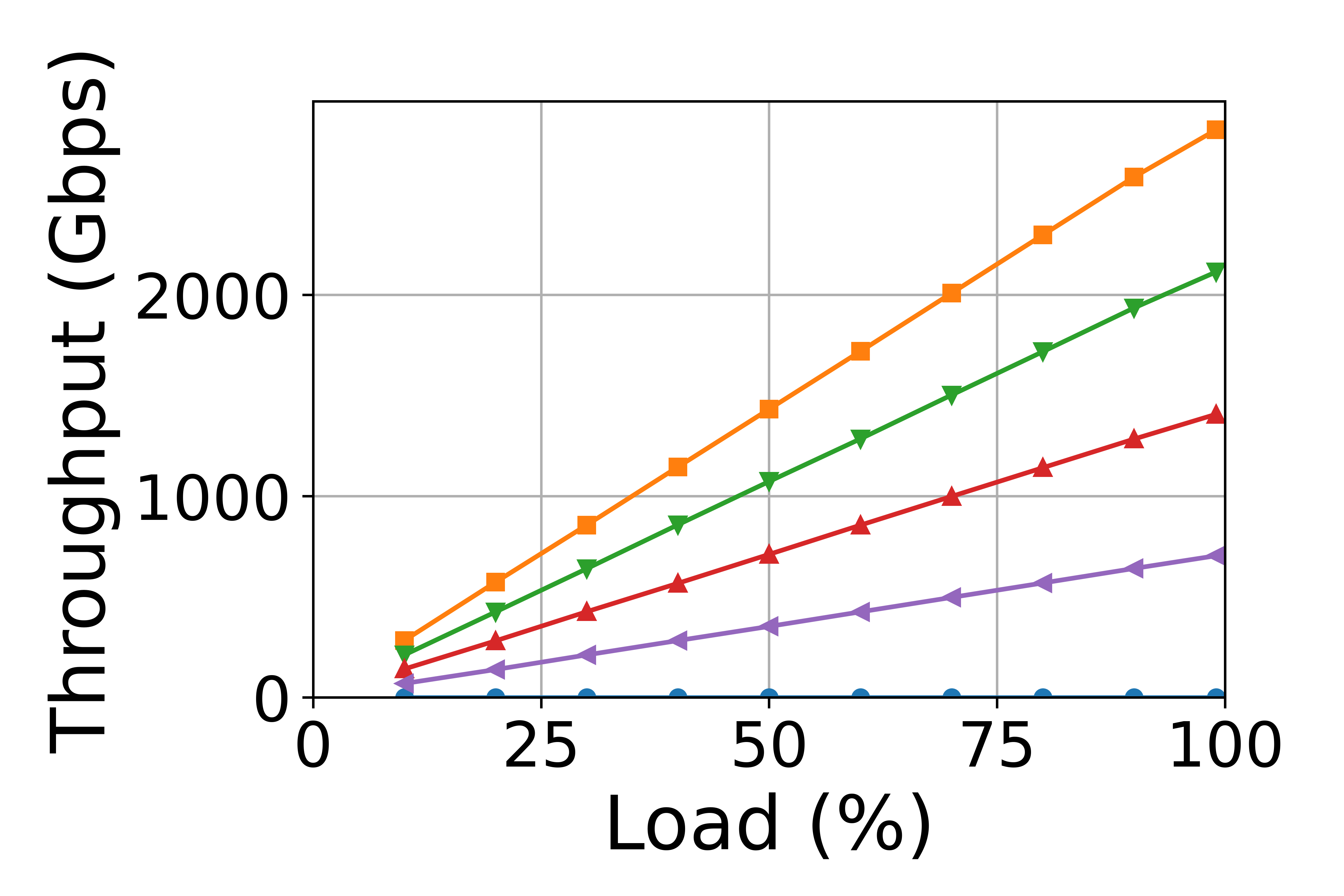}
        \caption{$2$ Acc./Node}
        \label{fig:exp:scaleup:pcie4:2gpu}
    \end{subfigure}
    \begin{subfigure}{0.24\textwidth}
        \centering
        \includegraphics[width=1\columnwidth]{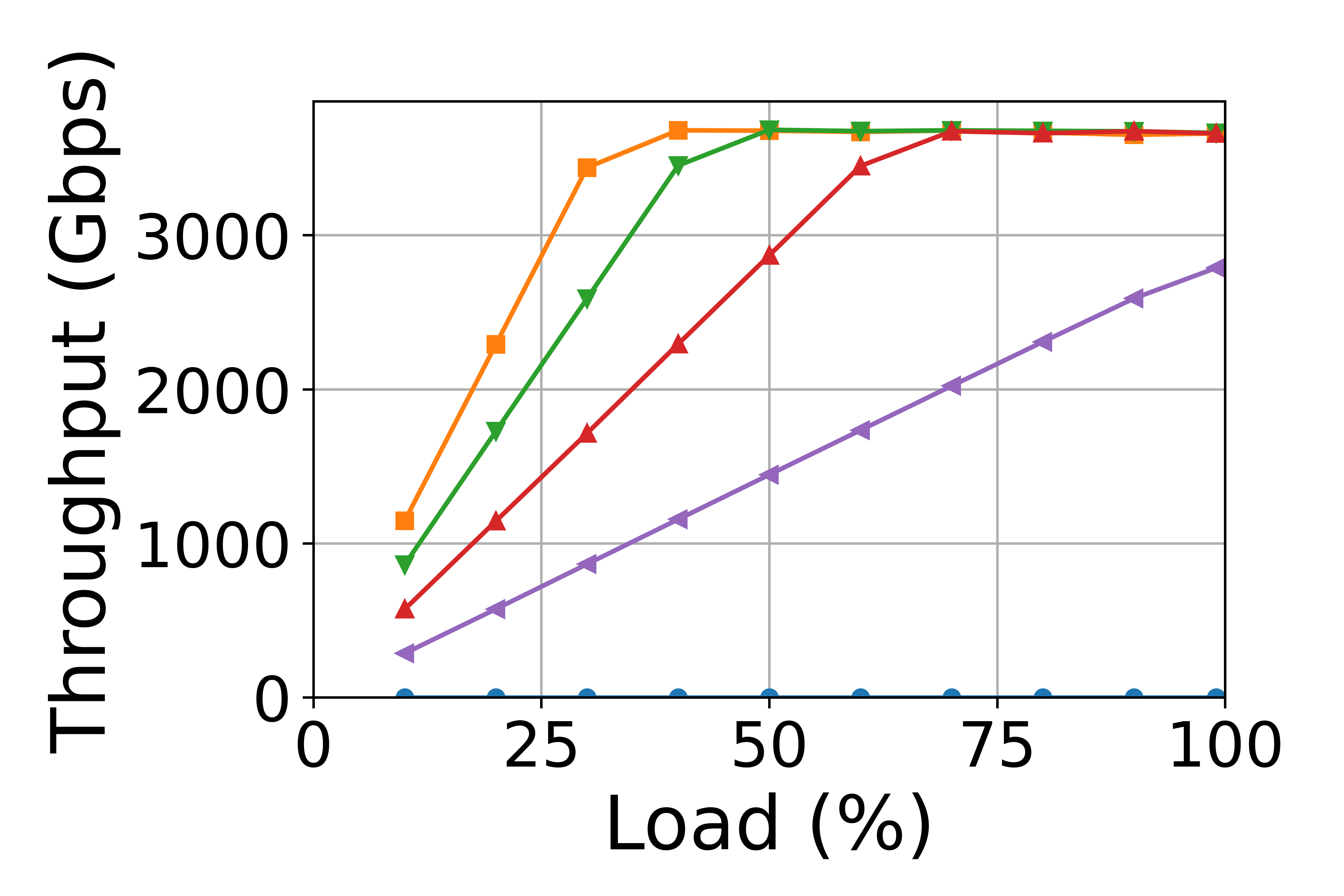}
        \caption{$4$ Acc./Node}
        \label{fig:exp:scaleup:pcie4:4gpu}
    \end{subfigure}
    \begin{subfigure}{0.24\textwidth}
        \centering
        \includegraphics[width=1\columnwidth]{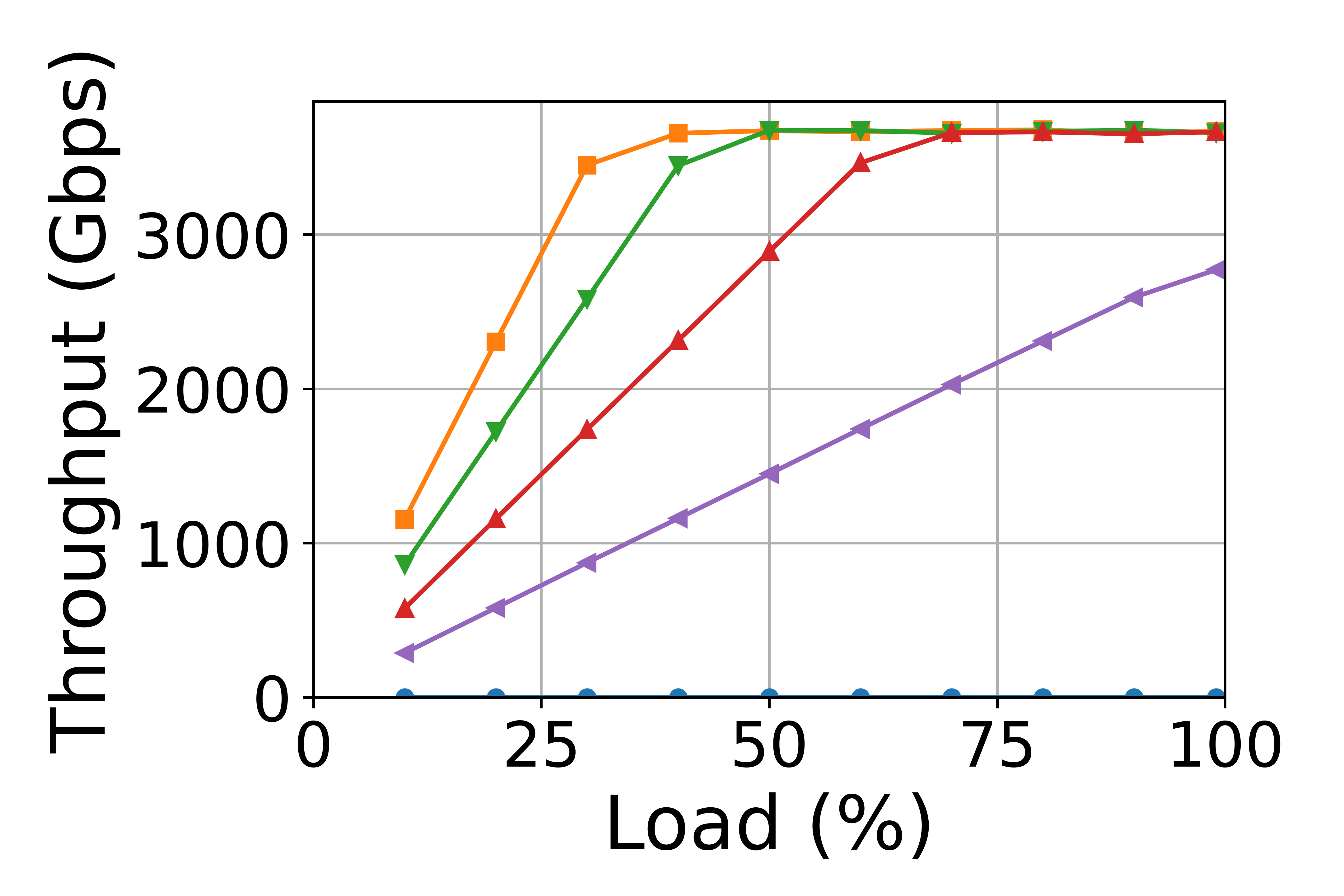}
        \caption{$8$ Acc./Node}
        \label{fig:exp:scaleup:pcie4:8gpu}
    \end{subfigure}
    \caption{Network performance as a function of traffic load (\%) in a 32-node RLFT topology (network config. \#2 in Table~\ref{tab:exp:scaleup:config}). Total network throughput is shown in the top figures, while inter-node throughput is depicted in the bottom ones.}
    \label{fig:exp:scaleup:pcie4}
\end{figure}

Figure~\ref{fig:exp:scaleup:pcie4} shows throughput results for network configuration \#2 (see Table~\ref{tab:exp:scaleup:config}), when intra-node link speed is $256$~Gbps. In this scenario, the saturation point is reached with only $4$ accelerators and traffic pattern C2 (i.e., 15\% of inter-node traffic). Considering the maximum bandwidth all the accelerators of an end node can generate, i.e., $256~\text{Gbps}/\text{accelerator} \times 8~\text{accelerators} = $\num{2048}$~\text{Gbps}$. Note that the bottleneck between the intra-node network and the NIC is still present.  Indeed, system performance degrades more rapidly compared to network configuration \#1.

\begin{figure}[!htb]
    \centering
    \begin{subfigure}{0.6\textwidth}
        \centering
        \frame{\includegraphics[width=1\textwidth]{Figures/JSC/Legend-modified.pdf}}
    \end{subfigure}
    \\
    \begin{subfigure}{0.24\textwidth}
        \centering
        \includegraphics[width=1\columnwidth]{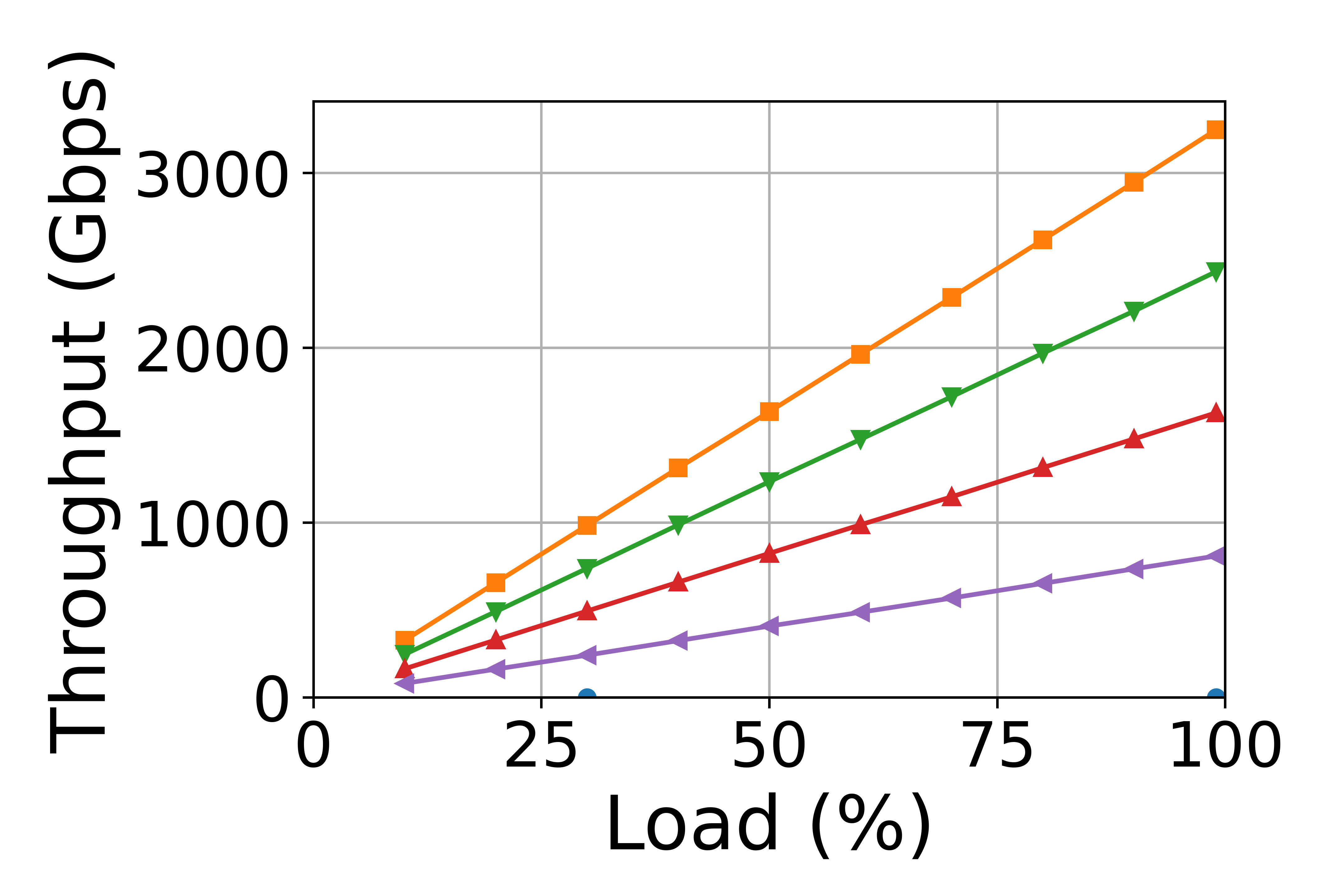}
    \end{subfigure}
    \begin{subfigure}{0.24\textwidth}
        \centering
        \includegraphics[width=1\columnwidth]{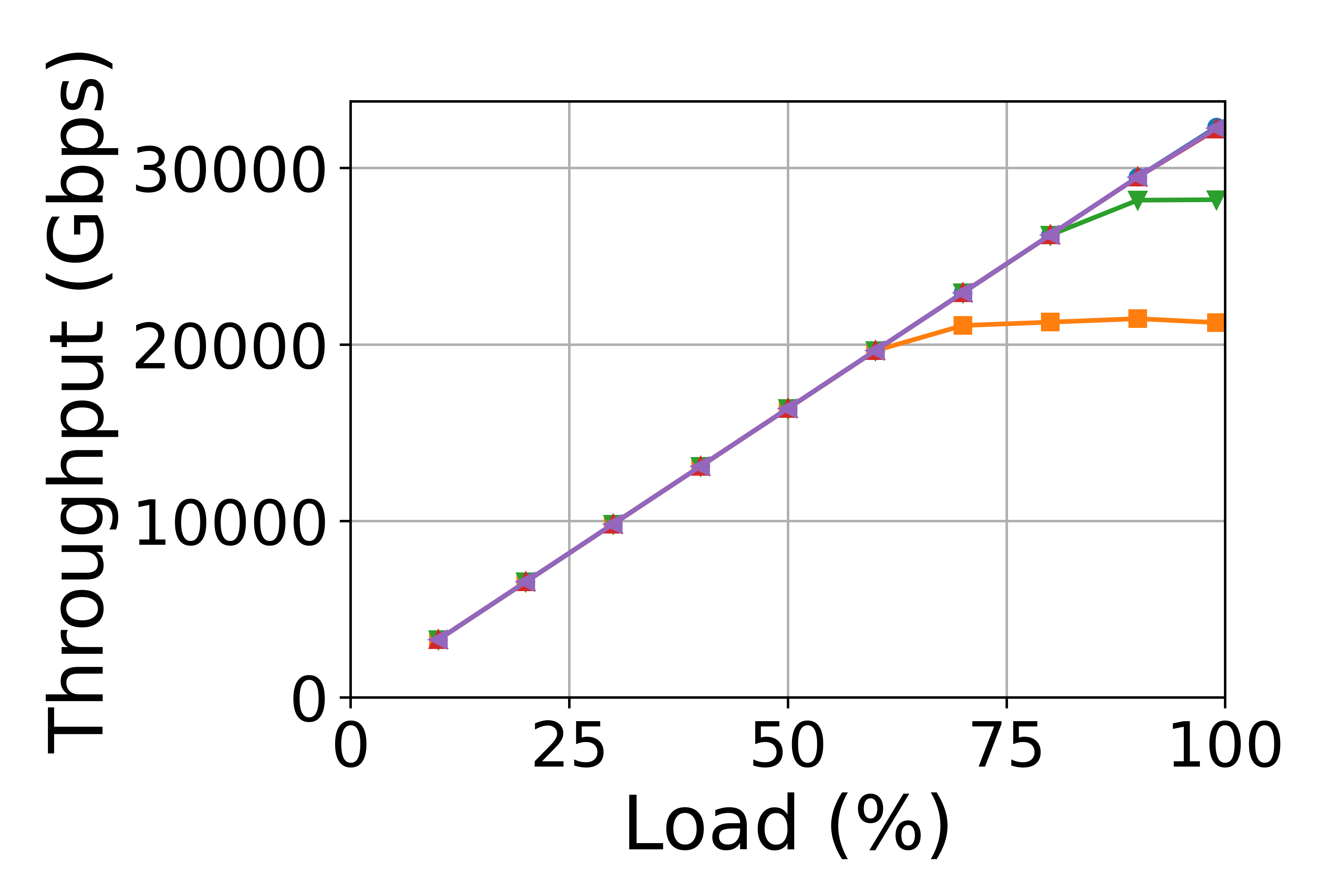}
    \end{subfigure}
    \begin{subfigure}{0.24\textwidth}
        \centering
        \includegraphics[width=1\columnwidth]{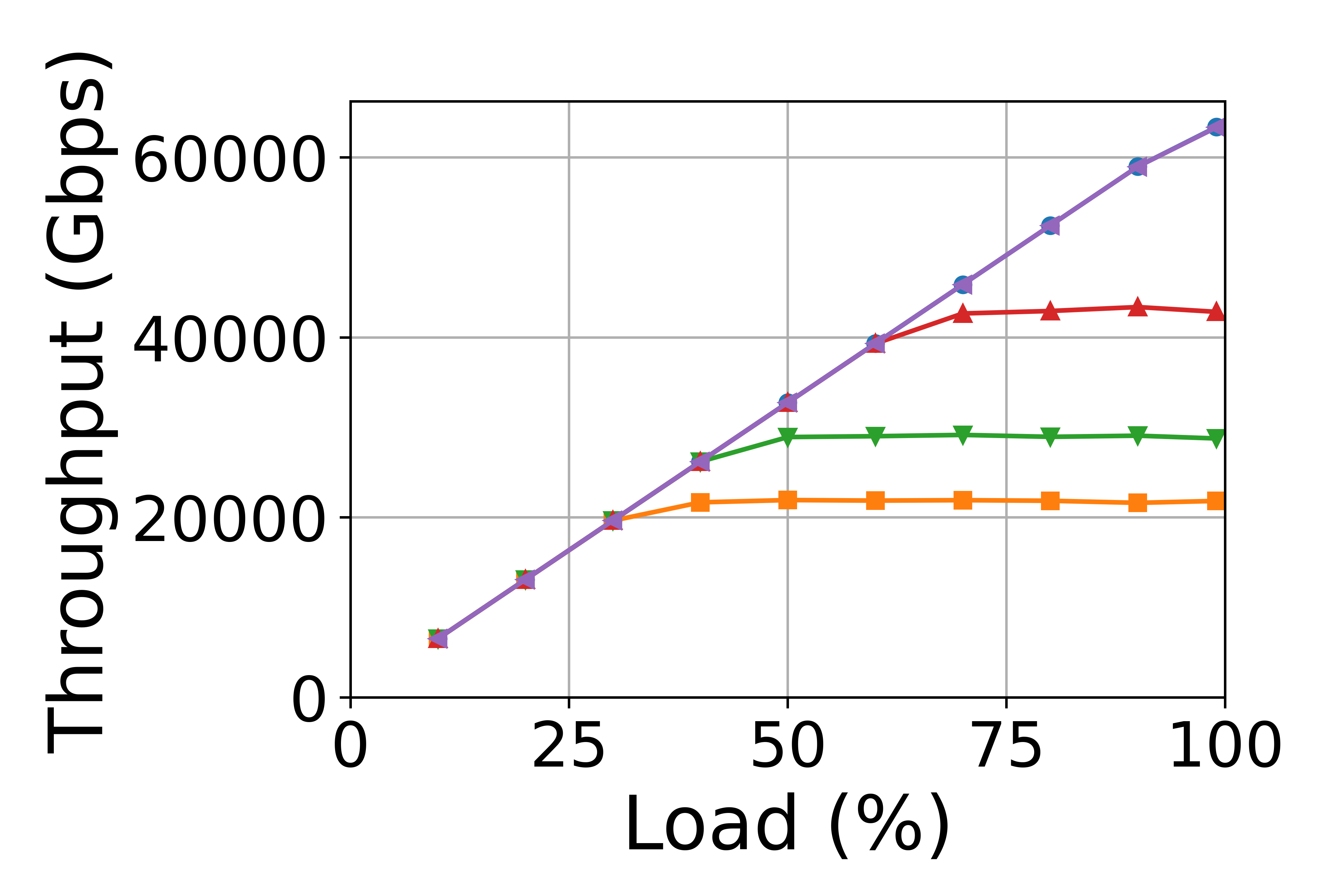}
    \end{subfigure}
    \begin{subfigure}{0.24\textwidth}
        \centering
        \includegraphics[width=1\columnwidth]{Figures/JSC/RR/Scale-Up/PCIe5/8_GPU/Intranode.png}
    \end{subfigure}
    \begin{subfigure}{0.24\textwidth}
        \centering
        \includegraphics[width=1\columnwidth]{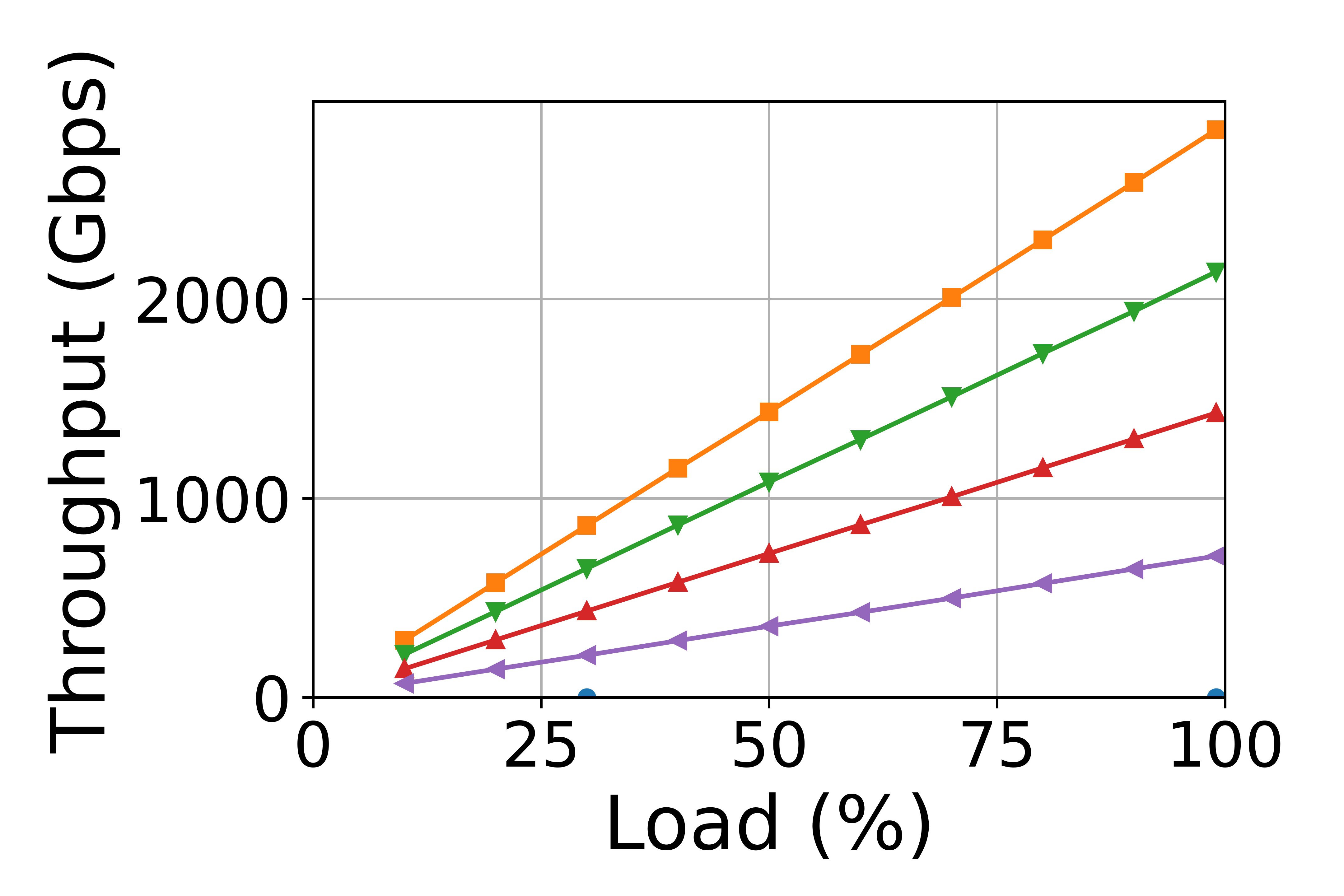}
        \caption{1 Acc./Node}
        \label{fig:exp:scaleup:pcie5:1gpu}
    \end{subfigure}
    \begin{subfigure}{0.24\textwidth}
        \centering
        \includegraphics[width=1\columnwidth]{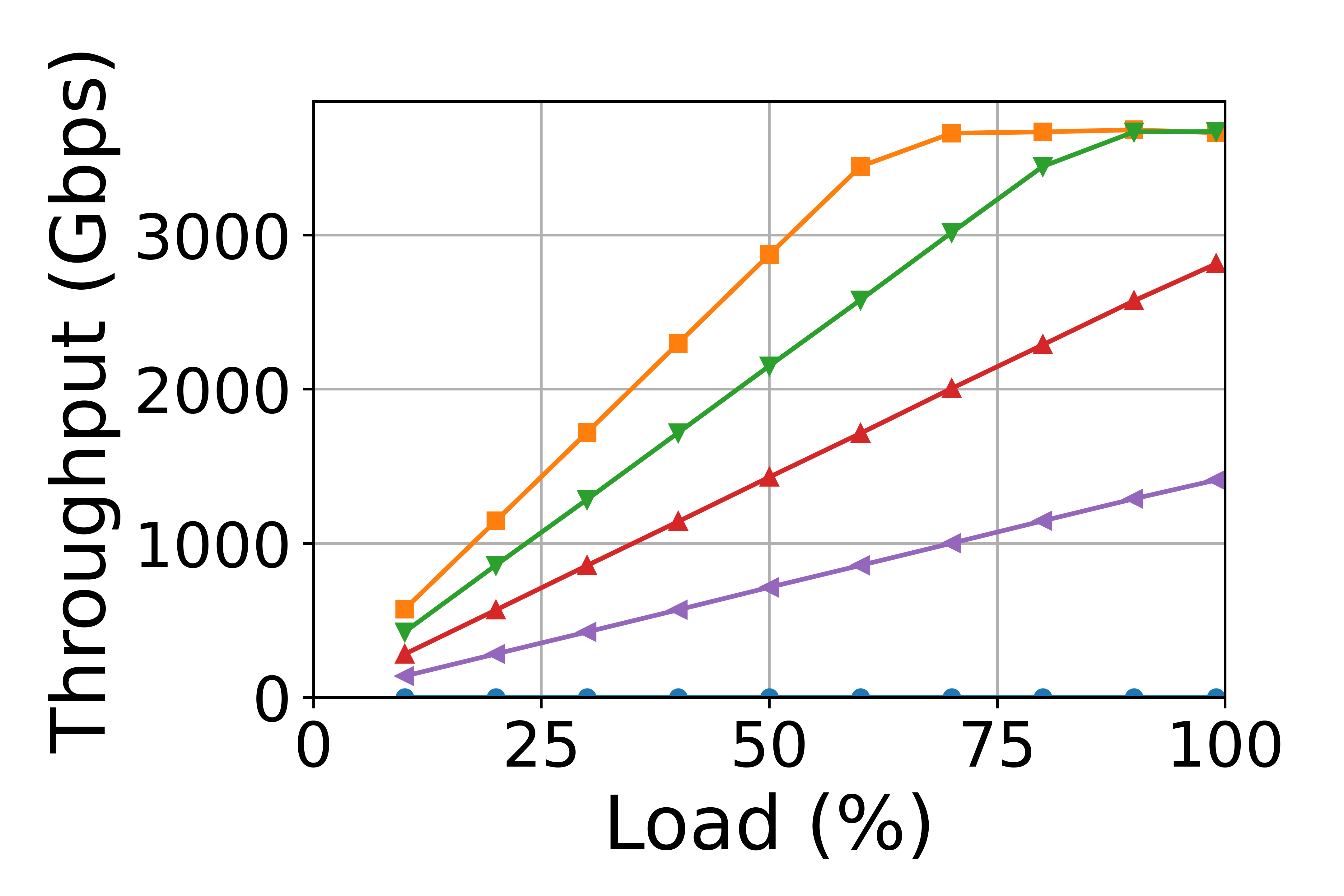}
        \caption{2 Acc./Node.}
        \label{fig:exp:scaleup:pcie5:2gpu}
    \end{subfigure}
    \begin{subfigure}{0.24\textwidth}
        \centering
        \includegraphics[width=1\columnwidth]{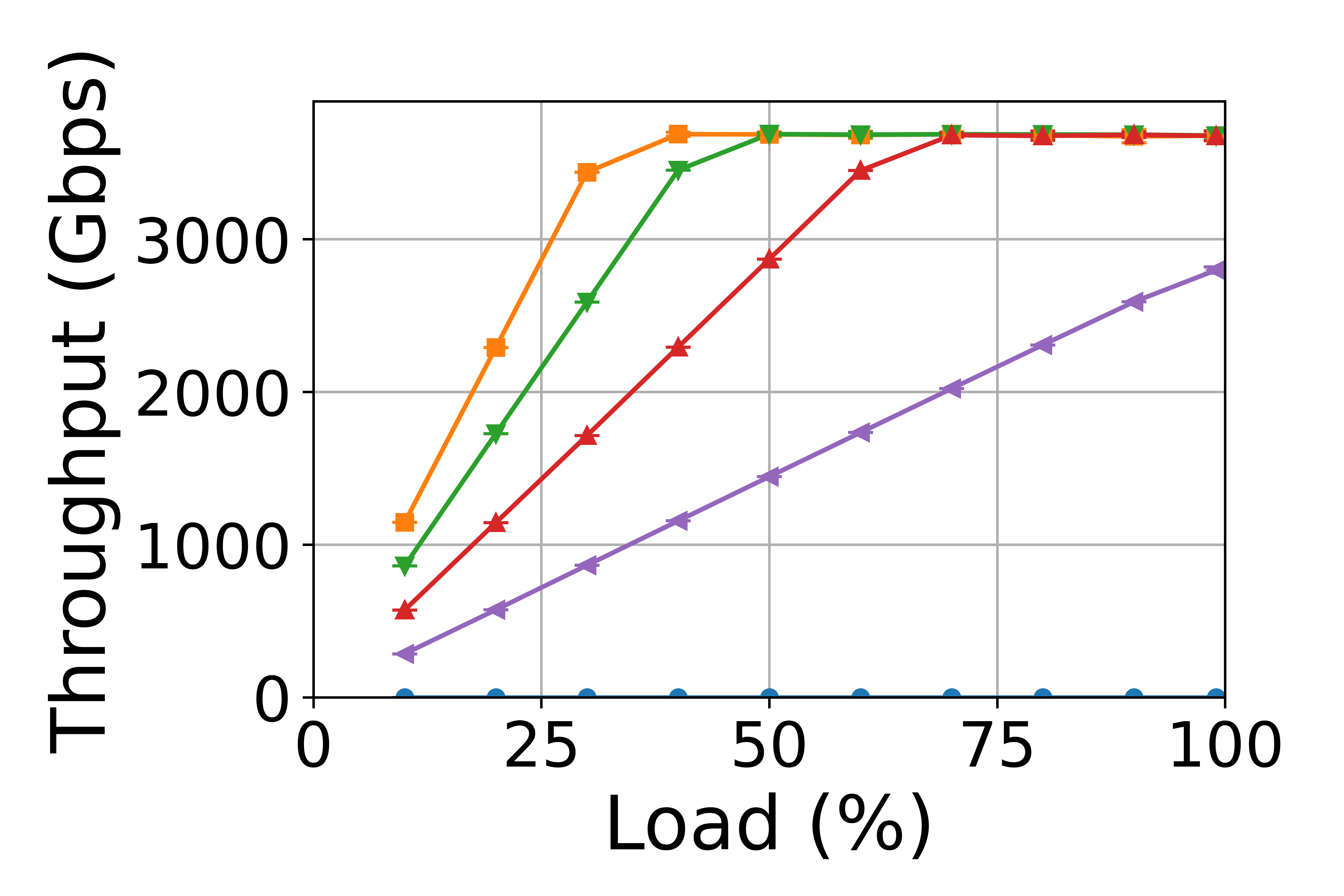}
        \caption{4 Acc./Node.}
        \label{fig:exp:scaleup:pcie5:4gpu}
    \end{subfigure}
    \begin{subfigure}{0.24\textwidth}
        \centering
        \includegraphics[width=1\columnwidth]{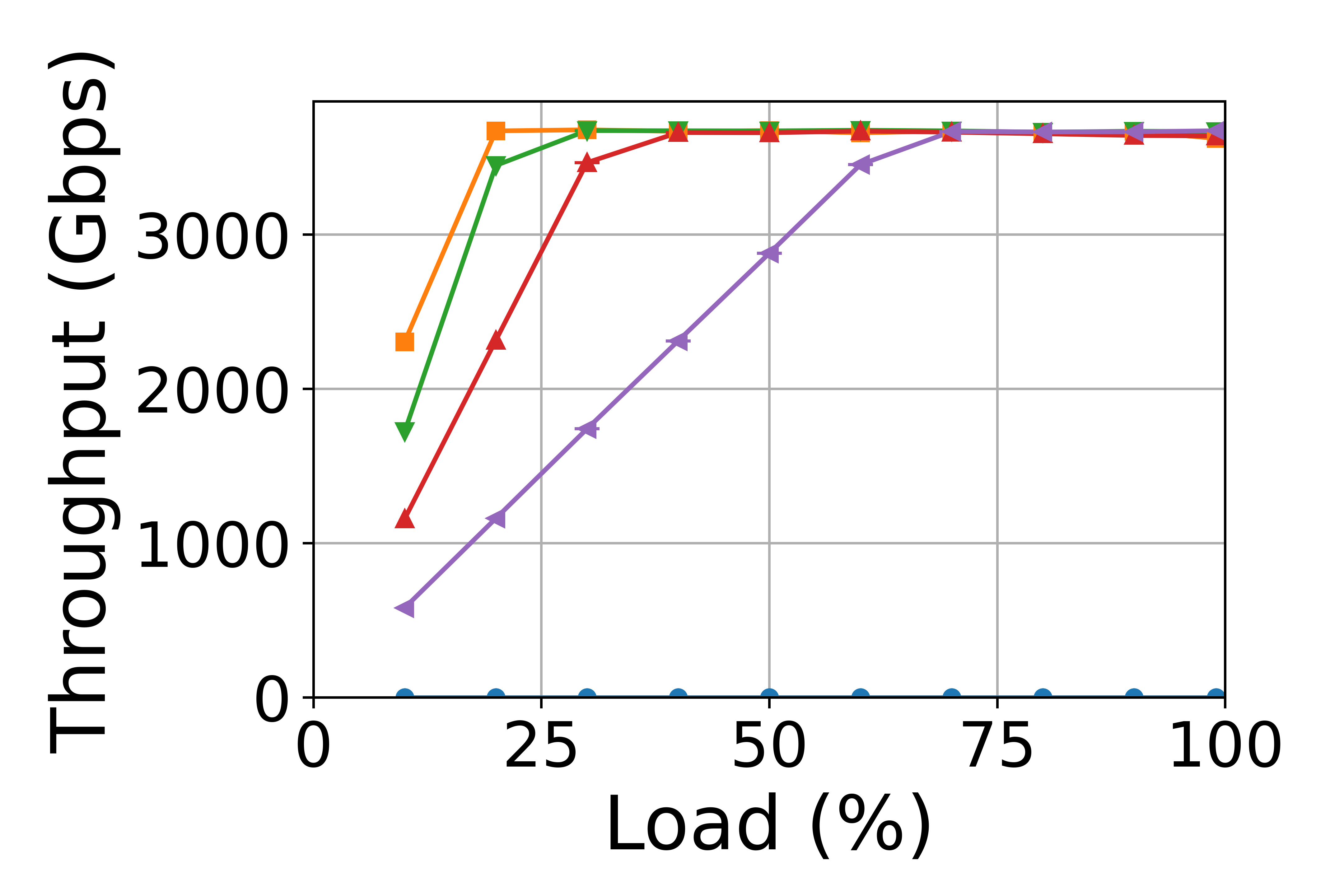}
        \caption{8 Acc./Node.}
        \label{fig:exp:scaleup:pcie5:8gpu}
    \end{subfigure}
    \caption{Network performance as a function of traffic load (\%) in a 32-node RLFT topology (network configuration \#3 from Table~\ref{tab:exp:scaleup:config}). Total network throughput is shown in the top figures, while inter-node throughput is depicted in the bottom ones..}
    \label{fig:exp:scaleup:pcie5}
\end{figure}

Figure~\ref{fig:exp:scaleup:pcie5} shows throughput results for network configuration \#3 (see Table~\ref{tab:exp:scaleup:config}). In this scenario, the link between the intra-node network and the NIC can operate at $512$~Gbps, and the inter-node network links operate at $400$~Gbps which is not enough to alleviate the bottleneck between the inter-node switch and the end-node NIC when accelerators exchange a small amount of traffic. Note that the network saturates with just $2$ accelerators per node generating $60$\% of traffic load and traffic pattern C1.

These experimental results suggest that simply increasing intra-node network bandwidth does not guarantee improved performance. While multiple accelerators per end node are beneficial for minimizing latency during Tensor Parallelism (TP), they also increase the volume of inter-node traffic. If this traffic exceeds the capacity of the inter-node network or the NIC, performance degradation may occur even earlier when intra-node bandwidth is high. 

\subsubsection{Scale-Up conclusions}

As the LLMs continue to grow, it becomes increasingly necessary to distribute model parameters across multiple accelerators, increasing the number of accelerators per end node. We analyzed the implications of scaling up the number of accelerators within an end node. We have observed that increasing the number of accelerators or the intra-node network speed can lead to bottlenecks at the destination NIC, because the intra-node network typically uses smaller payload sizes than the inter-node network, which results in a higher overhead due to conversion.

Therefore, accelerators' scale-up within server nodes is effective where most communication occurs within the same node, such as in setups with intensive Tensor Parallelism (TP) within nodes and only Pipeline Parallelism (PP) or Data Parallelism (DP) across nodes. However, in scenarios where the memory available on a single node is insufficient to store a whole model, and TP must be applied across multiple nodes, inter-node communication becomes more frequent and critical. In such cases, system performance degrades significantly, and the benefits of scale-up diminish.

\subsection{Scale-Out simulation experiments}

In this section, we evaluate system performance as the number of end nodes increases, consequently increasing the total number of accelerators in the system. Table~\ref{tab:exp:scaleout:config} summarizes the network configurations used in this scale-out evaluation. Each end node is equipped with $8$ accelerators. We evaluate two high-speed intra-node network configurations to assess their impact on overall performance. The rest of the network parameters are the same as in the previous experiments.

\begin{table}[!htb]
    \centering
    \caption{Scale-Out network configurations.}
    \begin{tabular}{@{} c c c c c @{}}
        \toprule
        \textbf{\#}         & \textbf{Nodes}        & \textbf{Inter-node switches}  & \textbf{Accelerators}     & \textbf{Intra-node network speed} \\
        \midrule
        1                   & \multirow{2}{*}{128}  & \multirow{2}{*}{24}           & \multirow{2}{*}{\num{1204}}    & 256 Gbps/accelerator      \\
        2                   &                       &                               &                           & 512 Gbps/accelerator      \\
        \midrule
        3                   & \multirow{2}{*}{512}  & \multirow{2}{*}{48}           & \multirow{2}{*}{\num{4096}}    & 256 Gbps/accelerator      \\
        4                   &                       &                               &                           & 512 Gbps/accelerator      \\
        \bottomrule
    \end{tabular}
    \label{tab:exp:scaleout:config}
\end{table}

\begin{figure}[!htb]
    \centering
    \begin{subfigure}{0.6\textwidth}
        \centering
        \frame{\includegraphics[width=1\textwidth]{Figures/JSC/Legend-modified.pdf}}
    \end{subfigure}
    \\
    \begin{subfigure}{0.24\textwidth}
        \centering
        \includegraphics[width=1\columnwidth]{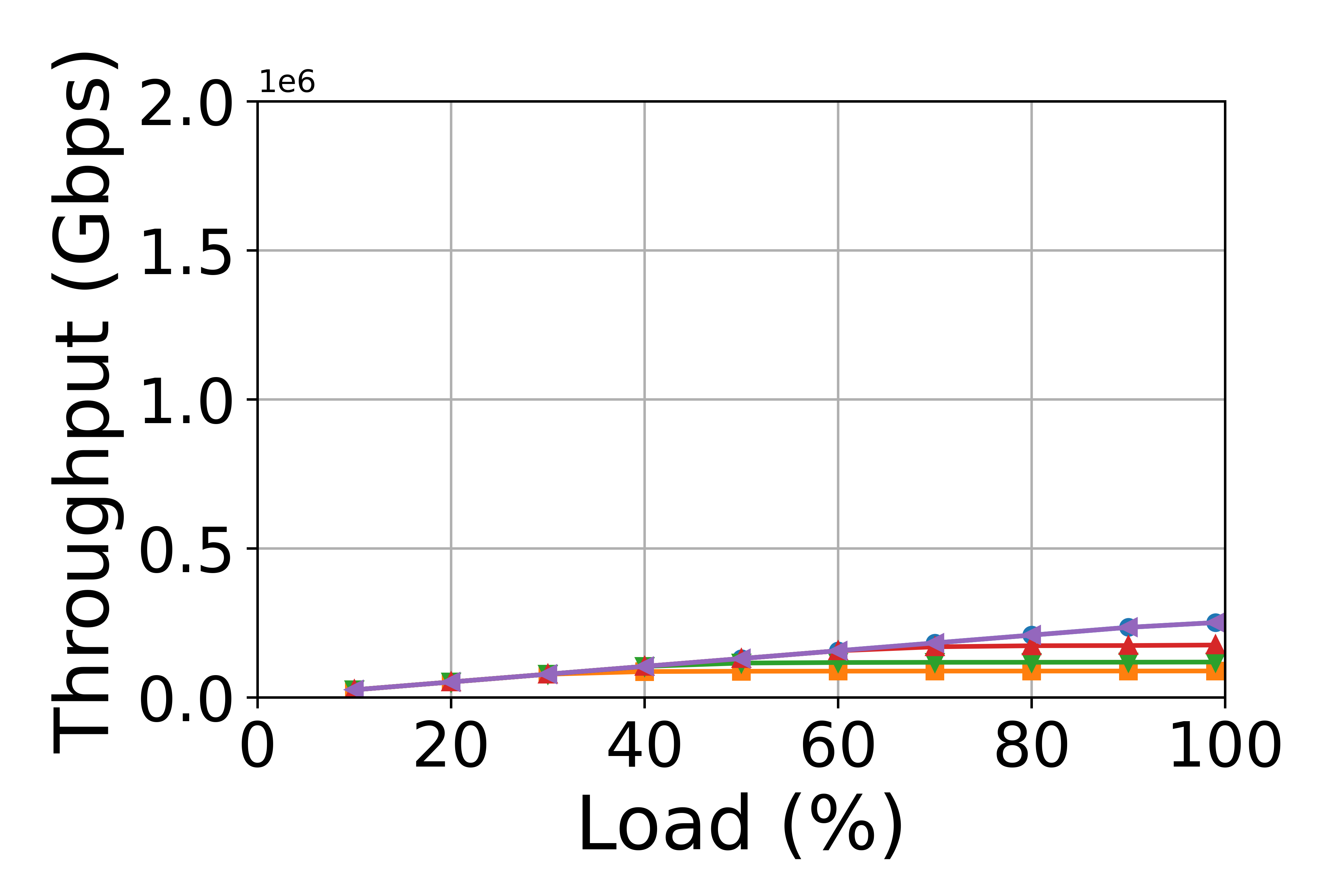}
    \end{subfigure}
    \begin{subfigure}{0.24\textwidth}
        \centering
        \includegraphics[width=1\columnwidth]{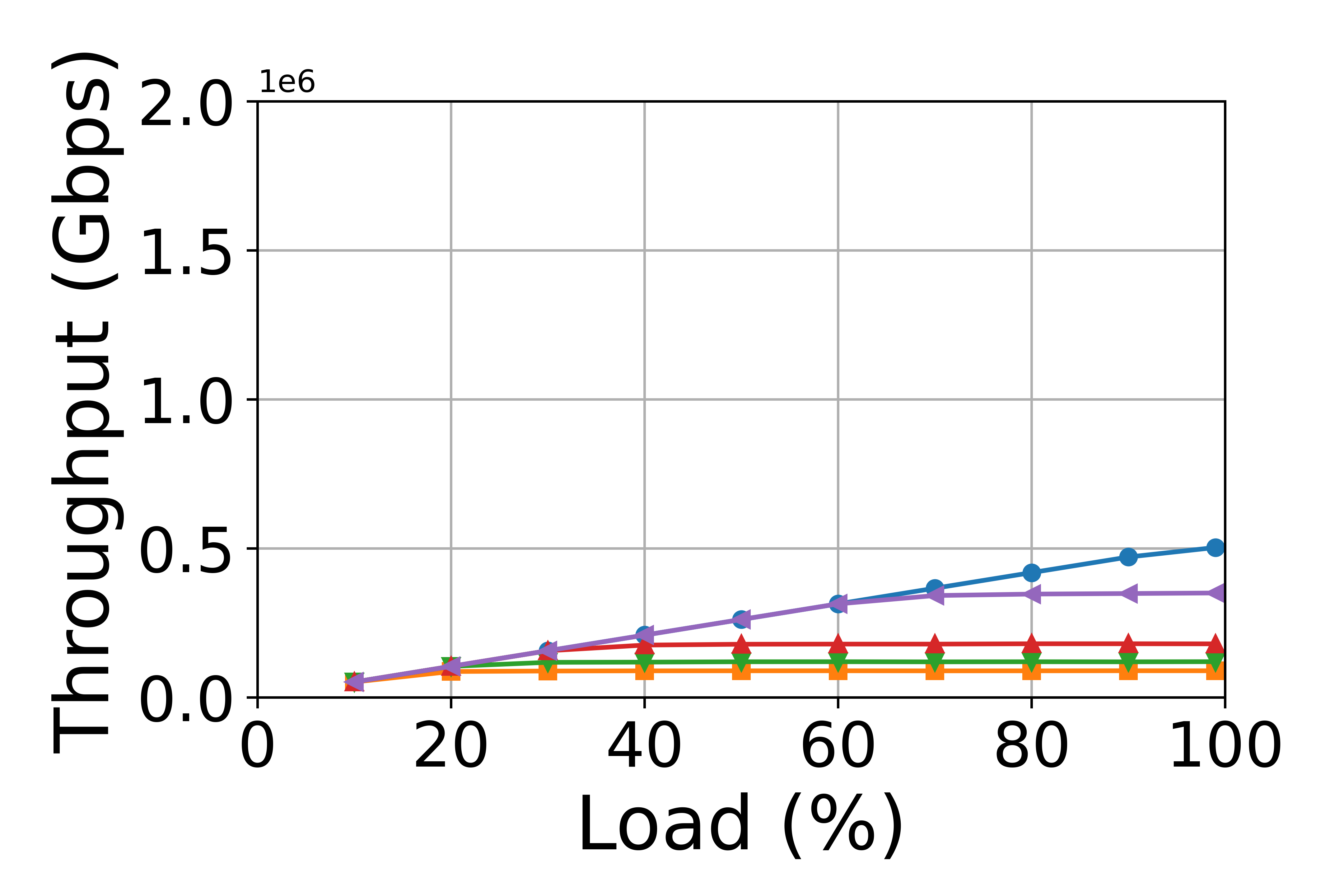}
    \end{subfigure}
    \begin{subfigure}{0.24\textwidth}
        \centering
        \includegraphics[width=1\columnwidth]{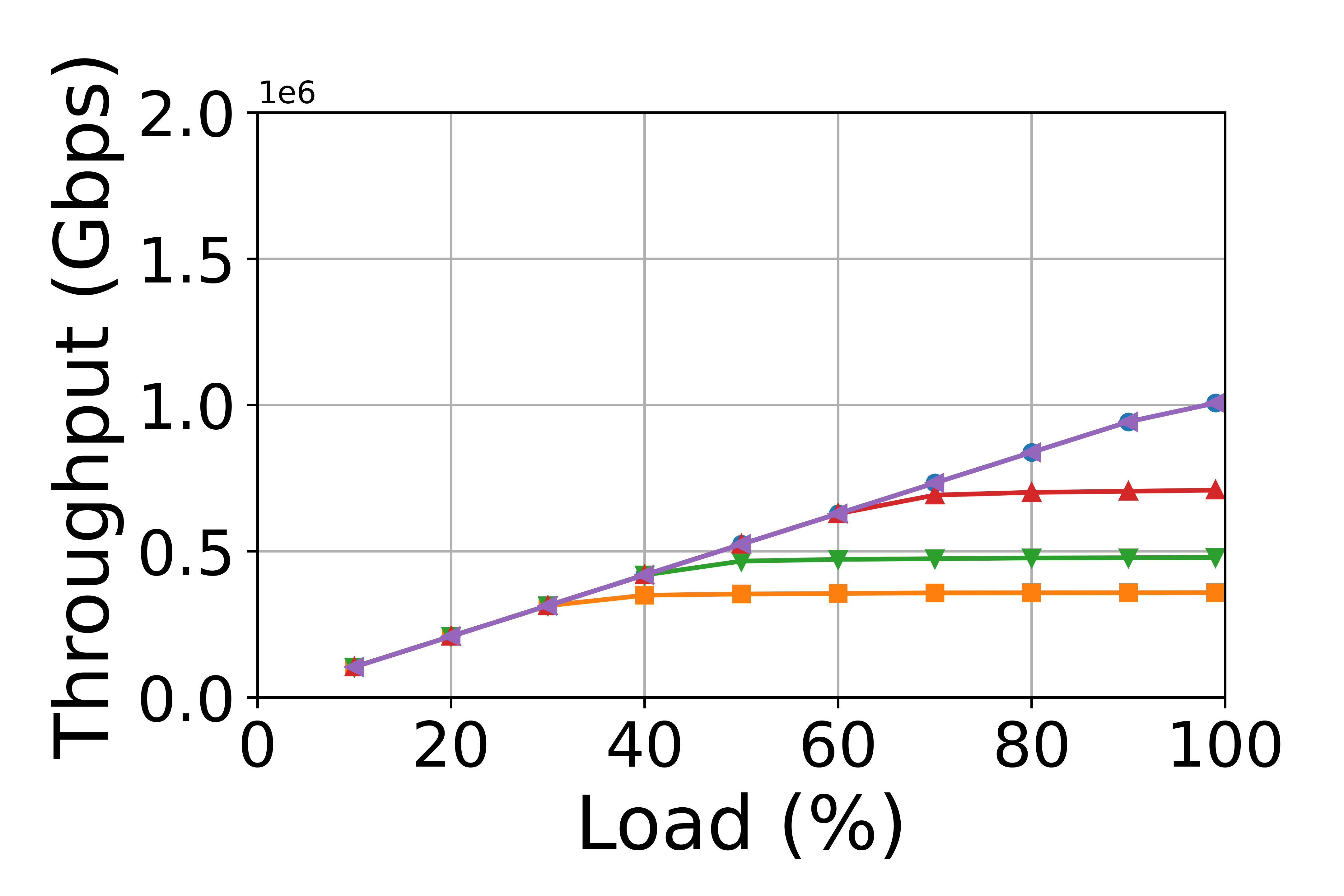}
    \end{subfigure}
    \begin{subfigure}{0.24\textwidth}
        \centering
        \includegraphics[width=1\columnwidth]{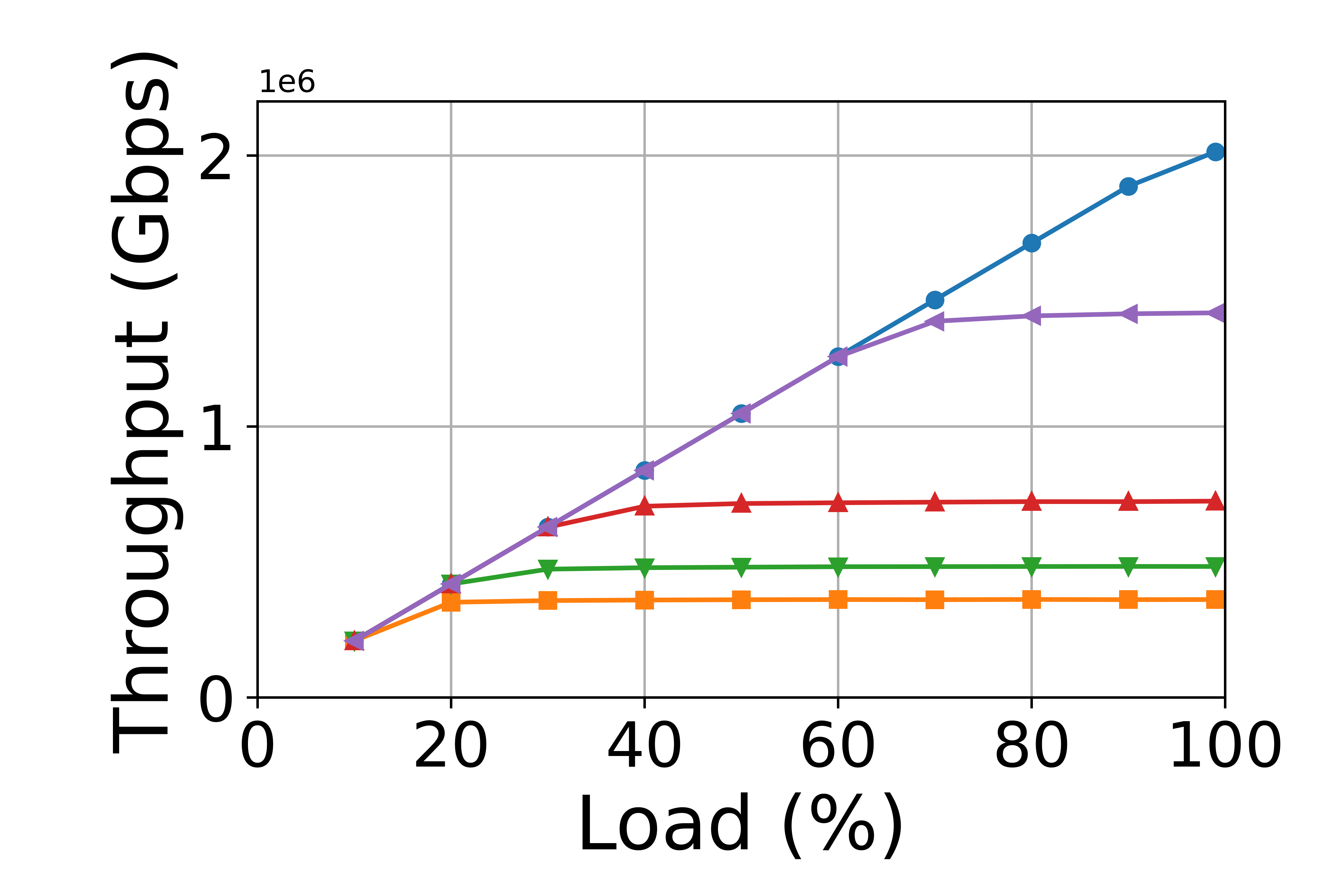}
    \end{subfigure}
    \begin{subfigure}{0.24\textwidth}
        \centering
        \includegraphics[width=1\columnwidth]{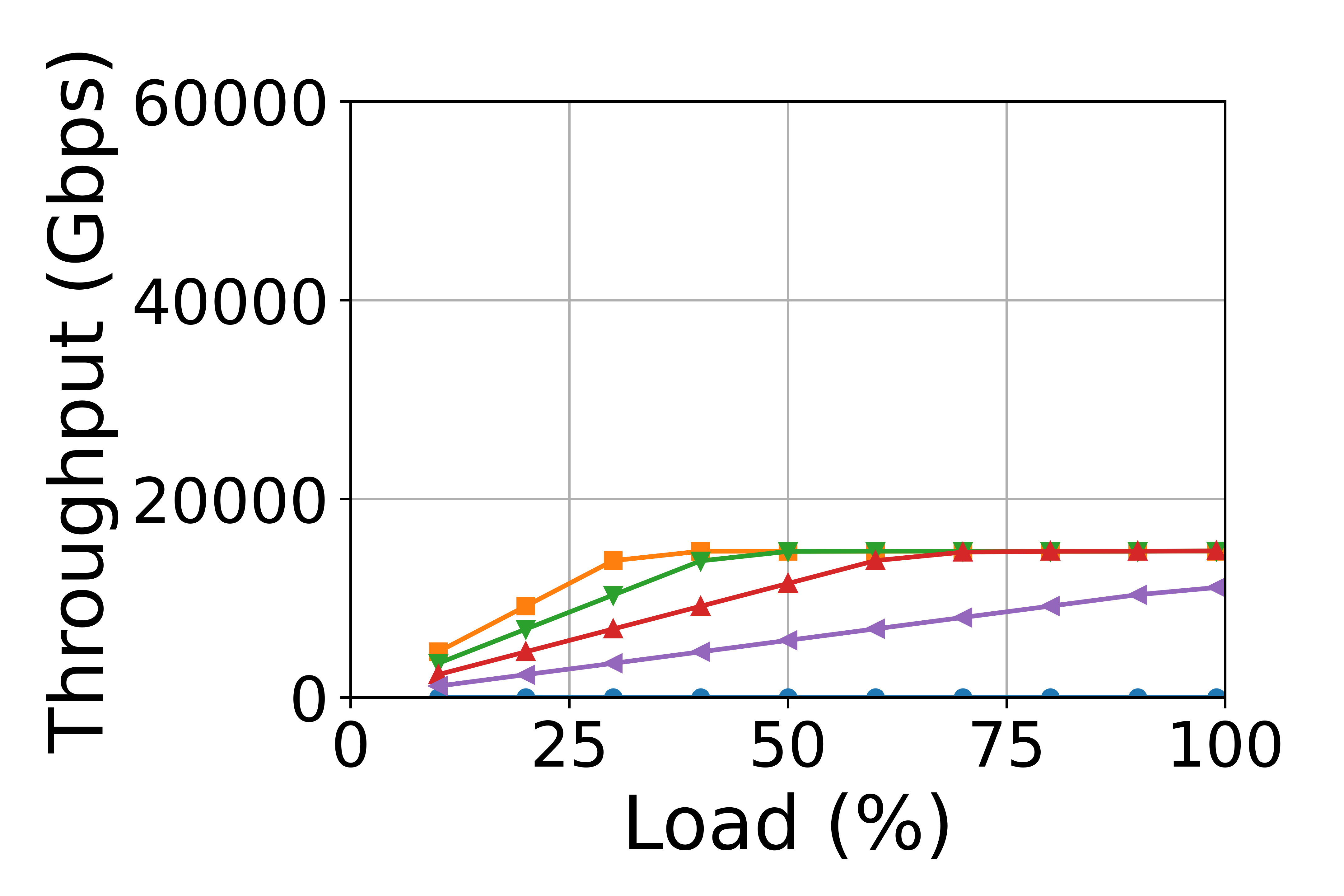}
        \caption{Configuration \#1.}
        \label{fig:exp:scaleout:128n:pcie4}
    \end{subfigure}
    \begin{subfigure}{0.24\textwidth}
        \centering
        \includegraphics[width=1\columnwidth]{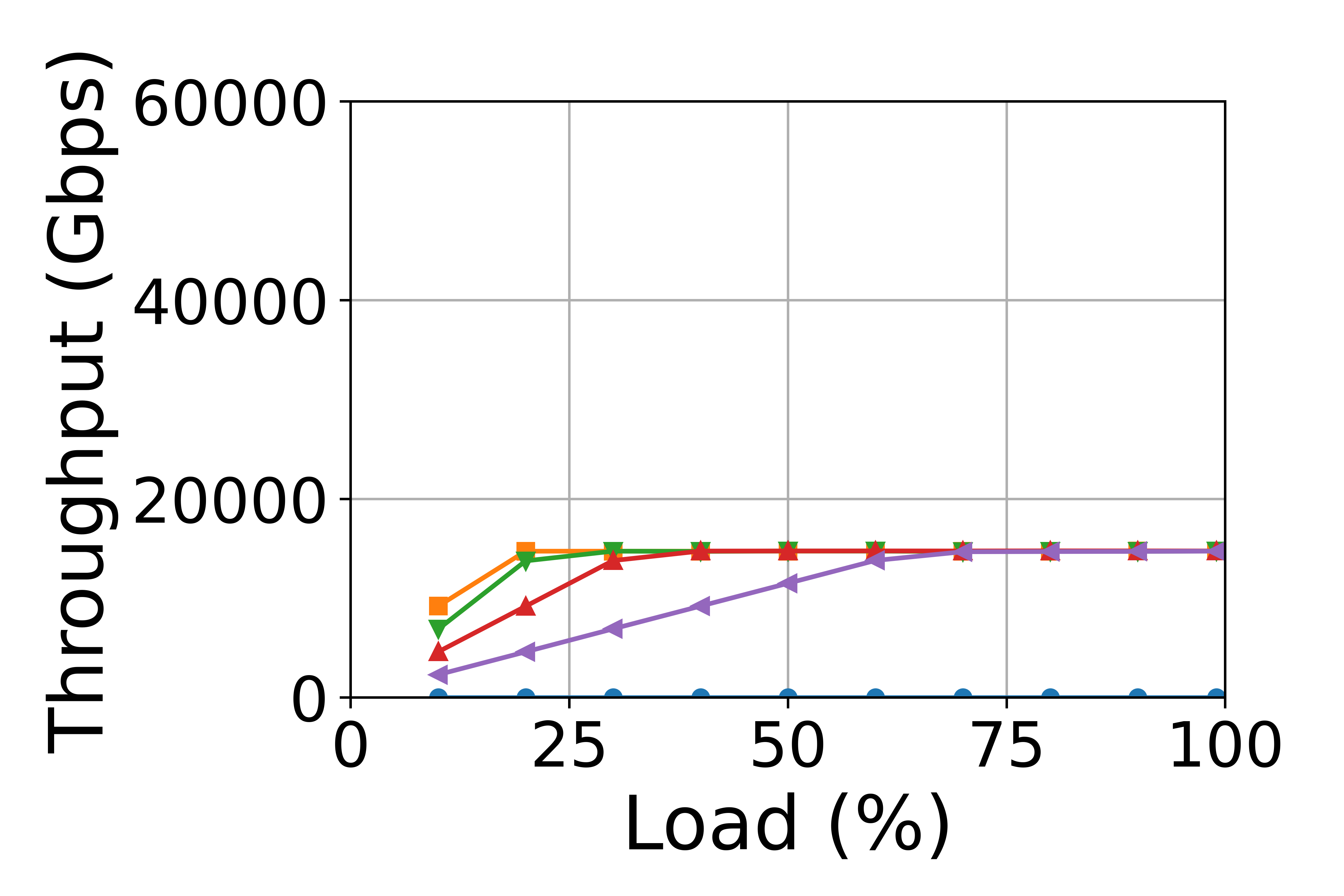}
        \caption{Configuration \#2.}
        \label{fig:exp:scaleout:128n:pcie5}
    \end{subfigure}
    \begin{subfigure}{0.24\textwidth}
        \centering
        \includegraphics[width=1\columnwidth]{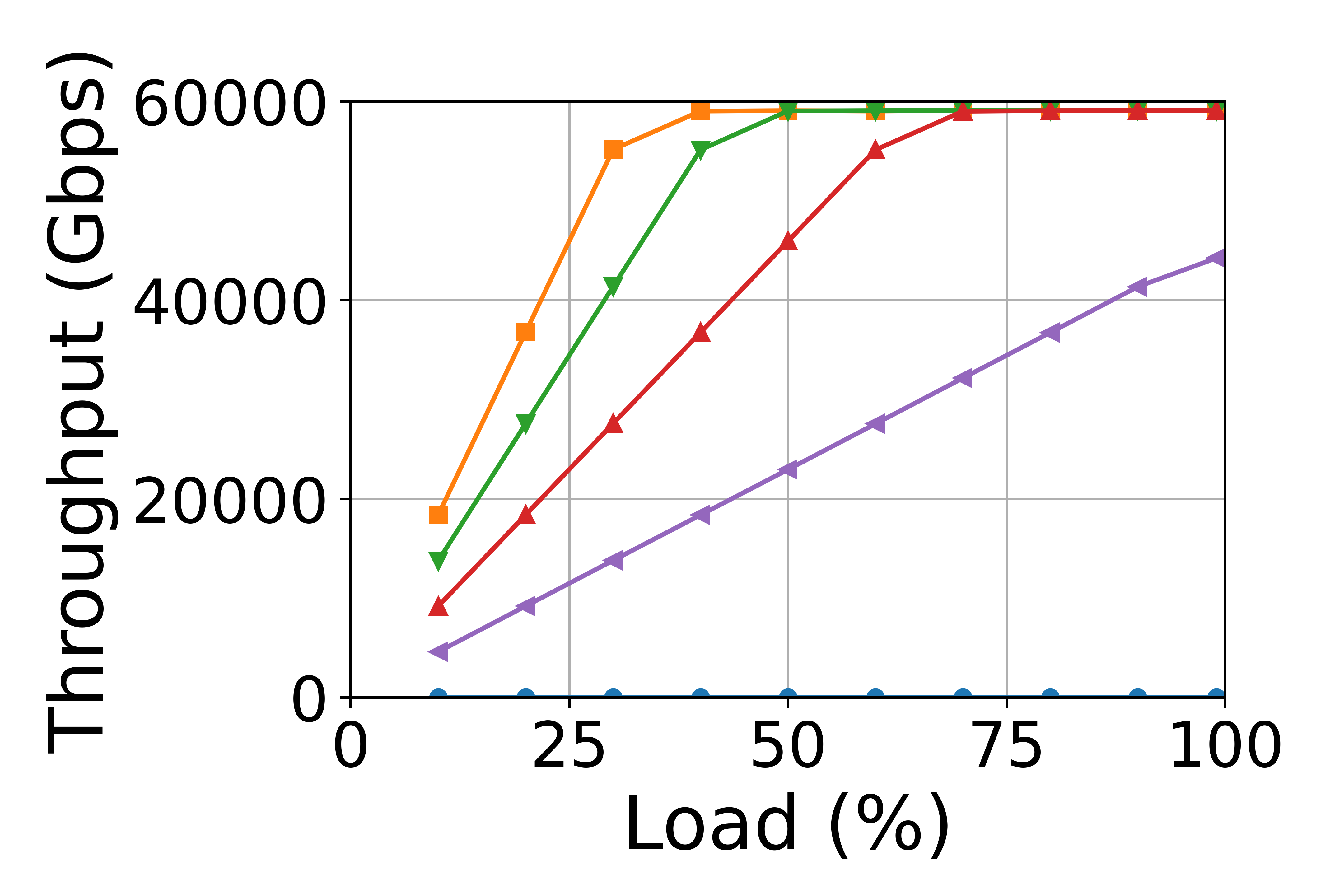}
        \caption{Configuration \#3.}
        \label{fig:exp:scaleout:512n:pcie4}
    \end{subfigure}
    \begin{subfigure}{0.24\textwidth}
        \centering
        \includegraphics[width=1\columnwidth]{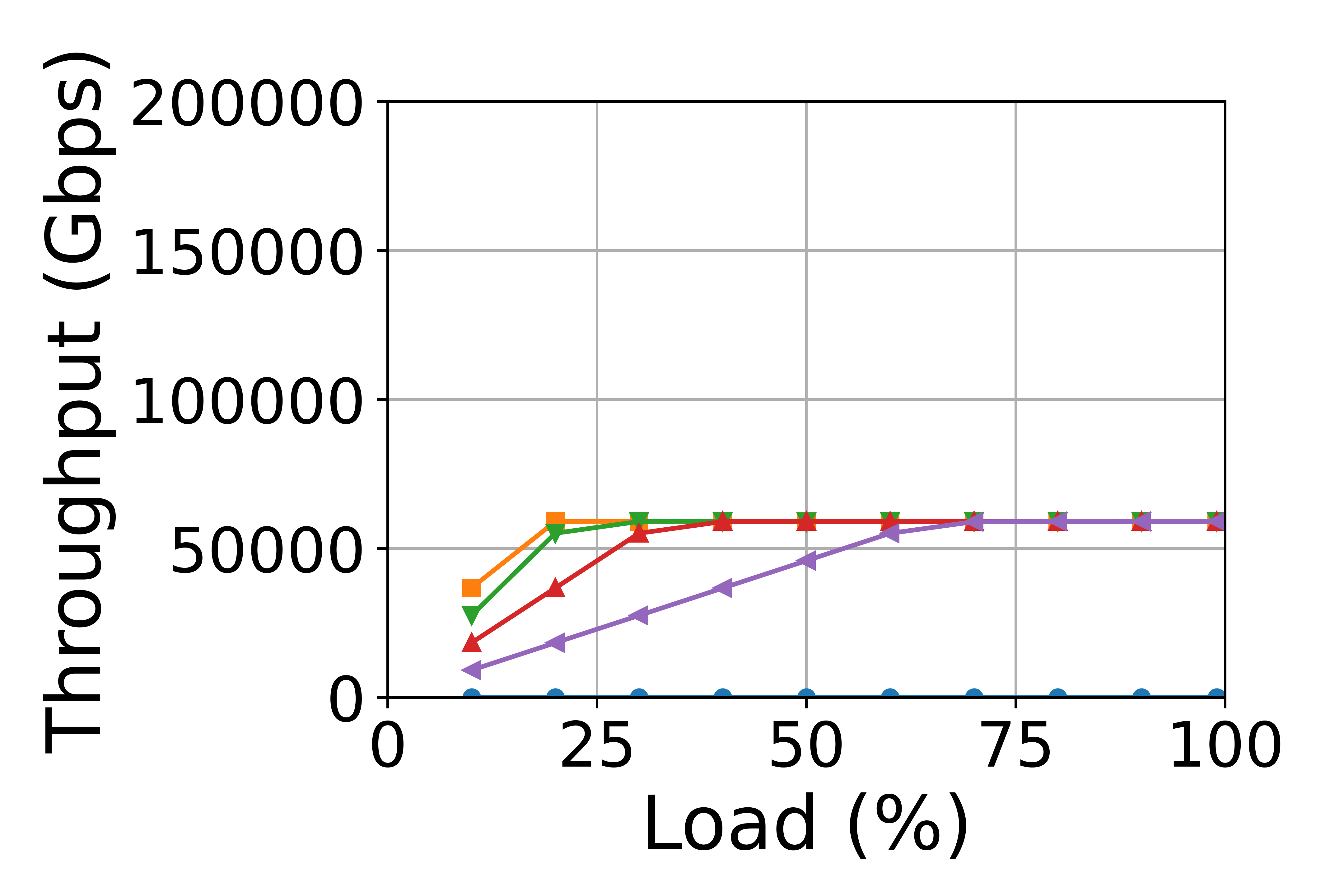}
        \caption{Configuration \#4.}
        \label{fig:exp:scaleout:512n:pcie5}
    \end{subfigure}
    \caption{Network performance as a function of traffic load (\%) (network configurations from Table~\ref{tab:exp:scaleup:config}). Intra-node performance is shown in the upper section, while inter-node performance is depicted in the lower section.}
    \label{fig:exp:scaleout}
\end{figure}

Figure~\ref{fig:exp:scaleout} shows the network performance across the different configurations described in Table~\ref{tab:exp:scaleout:config}. When compared to the results in Section~\ref{sec:evaluation:scaleUp}, we observe that the network behavior follows similar trends, with performance degradation occurring at comparable levels of traffic intensity.

For instance, in network configuration \#2, which supports a theoretical peak throughput of \num{524288}~Gbps, the system achieves only: $\text{Throughput (Gbps)} = \frac{\num{16384}}{1.14 \times 20} \times 128 = \num{91980}~\text{Gbps}$ under the C1 traffic pattern. This corresponds to just 17.54\% of the maximum network capacity. Similarly, in configuration \#4, which has a theoretical peak throughput of \num{2097152}~Gbps, the observed throughput under C1 traffic is \num{367921}~Gbps—again, approximately 17.54\% of peak capacity.

These results reinforce the observation that system scale-out can become critically limited by inter-node traffic patterns. When a significant portion of the traffic is directed between different end nodes, the link between the intra-node network and the NIC can become a severe bottleneck. Additionally, packetization overhead caused by differing payload sizes between intra- and inter-node communications can further degrade performance. As the number of end nodes increases, these limitations persist, making it essential to consider traffic patterns and hardware boundaries when scaling distributed systems for large-scale workloads such as LLM training and inference.

\begin{figure}[!htb]
    \centering
    \begin{subfigure}{0.6\textwidth}
        \centering
        \frame{\includegraphics[width=1\textwidth]{Figures/JSC/leyendaBarras.pdf}}
    \end{subfigure}
    \\
    \begin{subfigure}{0.4\textwidth}
        \centering
        \adjincludegraphics[width=1\columnwidth,trim={{.03\width} {.05\width} {.13\width} {.13\width}},clip]{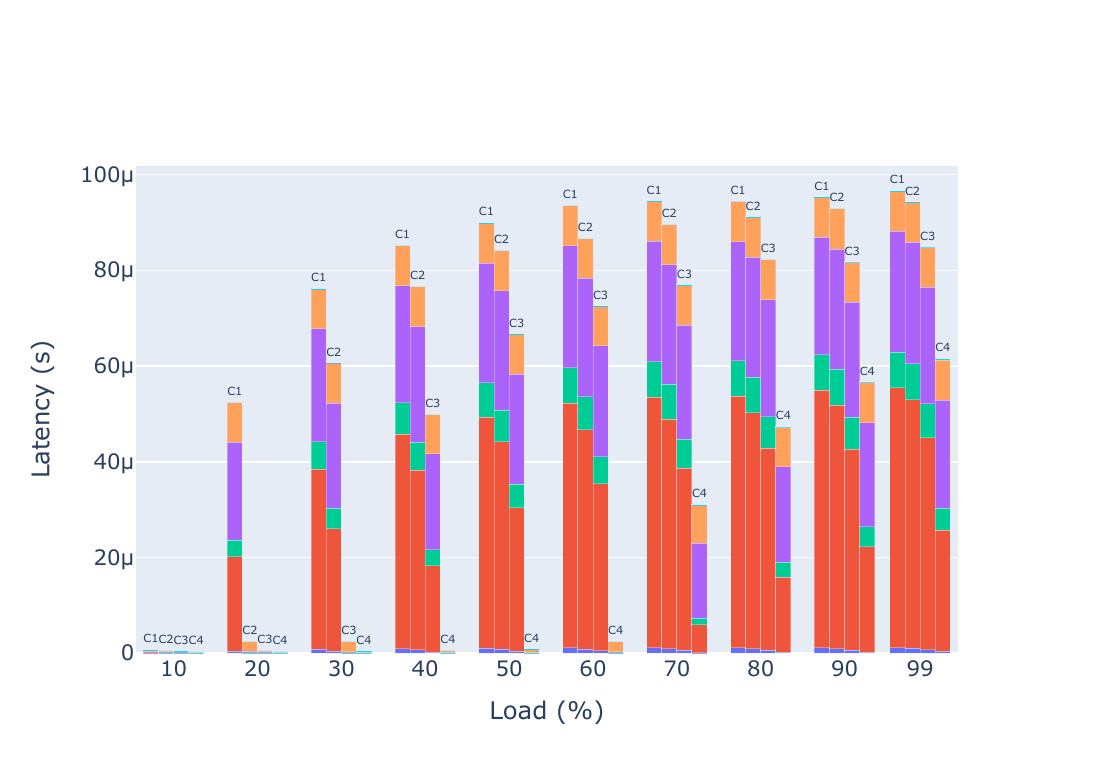}
        \caption{Intra-node packet size of 148\si{\byte}}
        \label{fig:exp:scaleout:latencias:148B}
    \end{subfigure}
    \begin{subfigure}{0.4\textwidth}
        \centering
        \adjincludegraphics[width=1\columnwidth,trim={{.03\width} {.05\width} {.13\width} {.13\width}},clip]{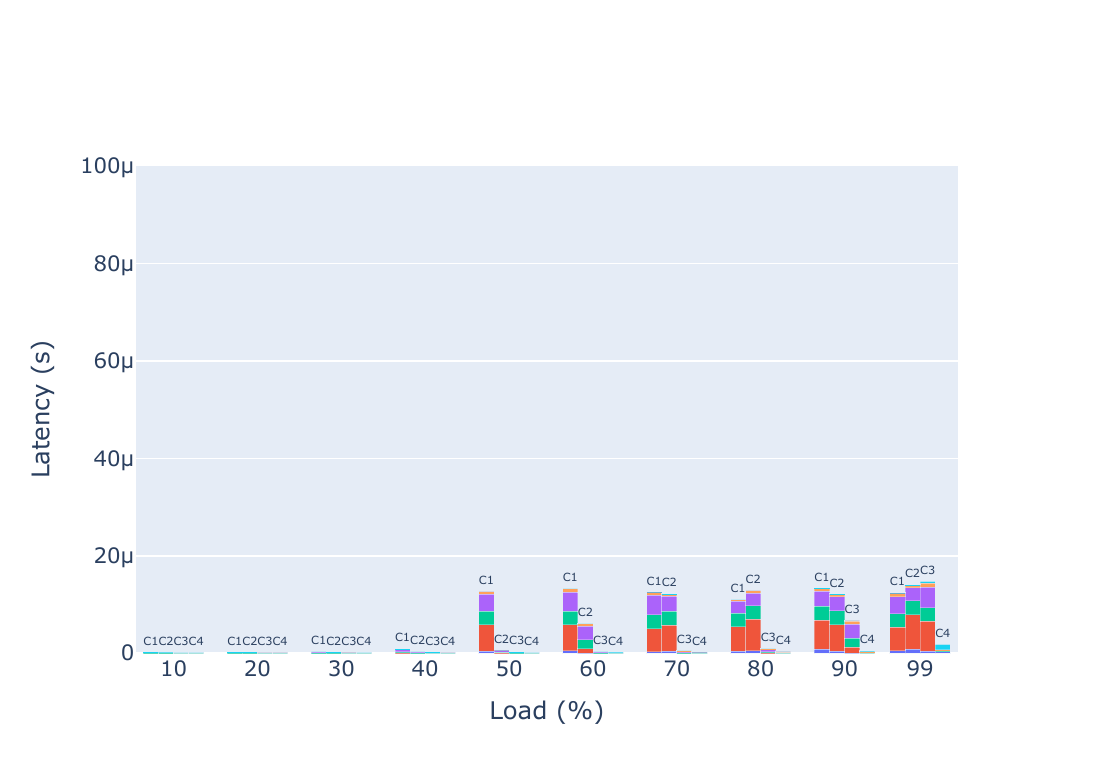}
        \caption{Intra-node packet size of 4\si{\kibi\byte}}
        \label{fig:exp:scaleout:latencias:4kb}
    \end{subfigure}
    \caption{Packet Latency divided into components versus traffic load for the traffic patterns C1-C4 and network configuration \#4 from Table~\ref{tab:exp:scaleup:config}.}
    \label{fig:exp:scaleout:latencias}
\end{figure}

Figure~\ref{fig:exp:scaleout:latencias} shows the packet latency divided into components of network configuration \#4 from Table~\ref{tab:exp:scaleup:config}. We assumed two different configurations, one with intra-node packets of 148\si{\byte} (128\si{\byte} of payload and 20\si{\byte} of header), and another configuration of intra-node packets of 4\si{\kibi\byte} (4032\si{\byte} of payload and 64\si{\byte} of header, as the inter-node packets). In Figure~\ref{fig:exp:scaleout:latencias:148B}, we can see that although we increase the number of end nodes, the latencies do not differ from those in the previous configurations. In this case, as the configuration performance decreases, we can observe how the latencies increase. In these cases, the backpressure from the destination NIC is affecting all previous components on the network. We can conclude that increasing the number of nodes will not resolve the performance issue; it may even exacerbate it, as the throughput difference between C1 and C4 will likely become greater, and the latencies will also increase. Figure~\ref{fig:exp:scaleout:latencias:4kb} shows that if we have not an intra-node overhead, the latency will decrease drastically and there will not affect the performance.

\begin{figure}[!htb]
    \centering
    \begin{subfigure}{0.6\textwidth}
        \centering
        \frame{\includegraphics[width=1\textwidth]{Figures/JSC/Legend-modified.pdf}}
    \end{subfigure}
    \\
    \begin{subfigure}{0.24\textwidth}
        \centering
        \includegraphics[width=1\columnwidth]{Figures/JSC/RR/Scale-Out/512n/PCIe5/Intranode.png}
        \caption{Intra-node thp (packet of 128\si{\byte}).}
        \label{fig:exp:scaleout:overhead:intra:148b}
    \end{subfigure}
    \begin{subfigure}{0.24\textwidth}
        \centering
        \includegraphics[width=1\columnwidth]{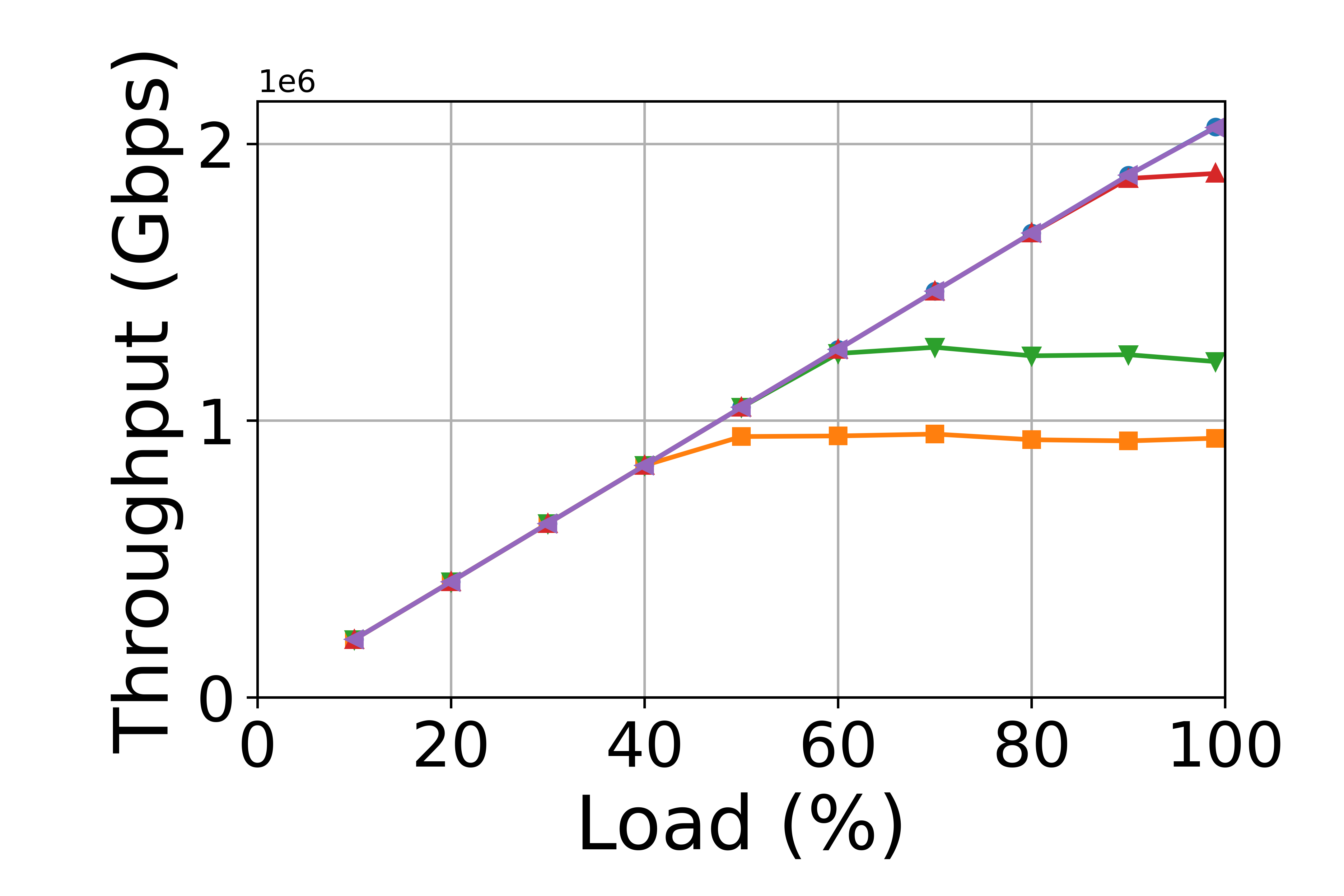}
        \caption{Intra-node thp (packet of 4\si{\kibi\byte}).}
        \label{fig:exp:scaleout:overhead:intra:4kb}
    \end{subfigure}
    \begin{subfigure}{0.24\textwidth}
        \centering
        \includegraphics[width=1\columnwidth]{Figures/JSC/RR/Scale-Out/512n/PCIe5/Internode.png}
        \caption{Inter-node thp (packet of 128\si{\byte}).}
        \label{fig:exp:scaleout:overhead:inter:148b}
    \end{subfigure}
    \begin{subfigure}{0.24\textwidth}
        \centering
        \includegraphics[width=1\columnwidth]{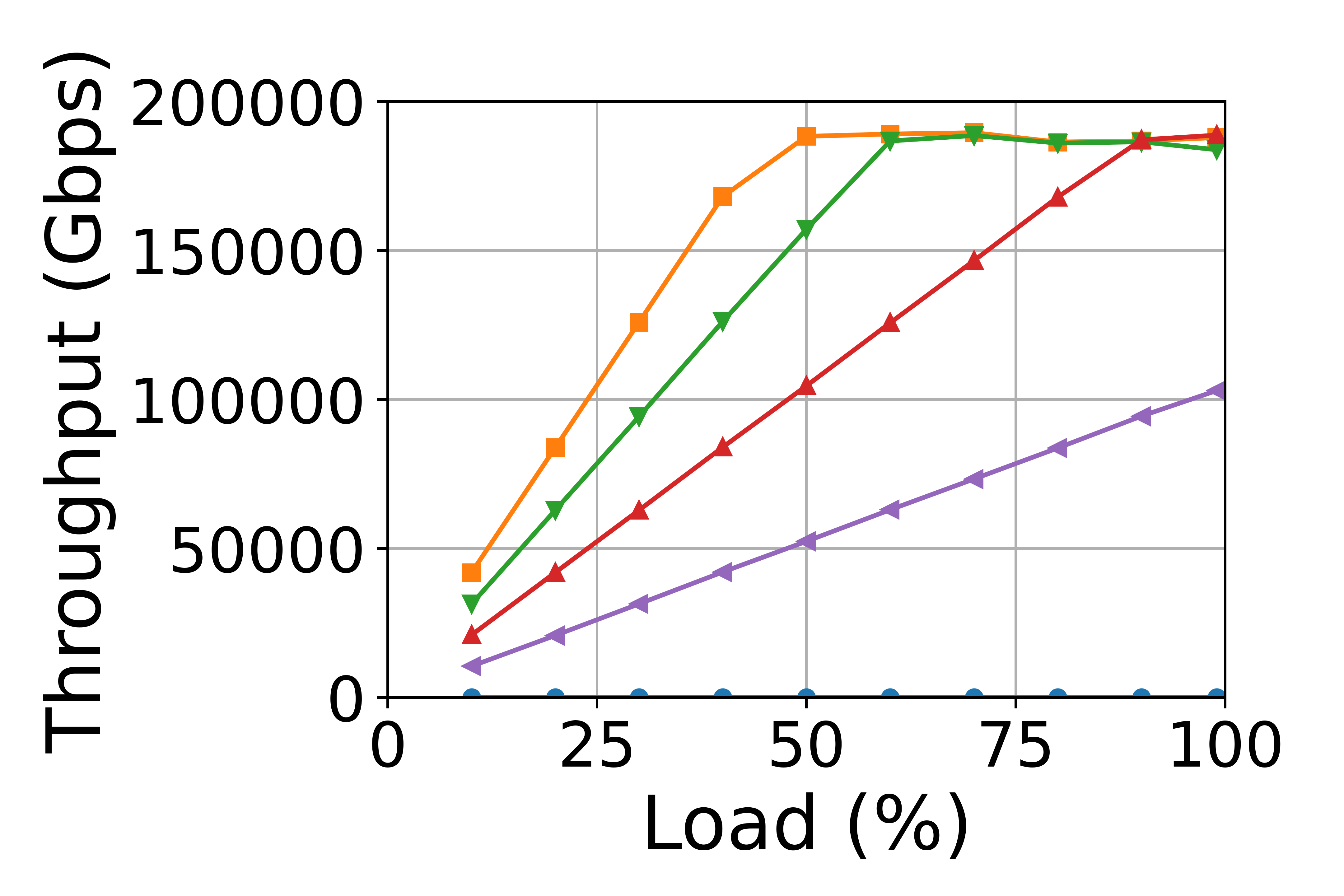}
        \caption{Inter-node thp (packet of 4\si{\kibi\byte}).}
        \label{fig:exp:scaleout:overhead:inter:4kb}
    \end{subfigure}
    \caption{Network throughput vs traffic load (\%) (network config. \#4, Table~\ref{tab:exp:scaleup:config}).}
    \label{fig:exp:scaleout:overhead}
\end{figure}

Figure~\ref{fig:exp:scaleout:overhead} shows the network performance as a function of traffic load (\%) of network configuration \#4 from Table~\ref{tab:exp:scaleup:config}. We assumed the same two different configurations (148\si{\byte} and 4\si{\kibi\byte} intra-node packets). We can see in Figure~\ref{fig:exp:scaleout:overhead:intra:4kb} that when we avoid the overhead in the intra-node network, the performance will increase drastically. In this case, the overhead will be determined by the saturation of the inter-node network (as links work at $400$~Gbps and the intra-node network at $512$~Gbps).

\section{Conclusions}
\label{sec:conclusions}

This paper has analyzed the impact of the interference between intra- and inter-node communication operations generated by emerging applications (e.g., generative AI or LLMs) in Supercomputers and Data Centers. To help with this research, we have developed a detailed packet-level intra- and inter-node network simulation model validated against a real cluster infrastructure using standard micro-benchmarks. We have thoroughly analyzed the latency and overhead effects of inter-to-intra-node packet conversion when different payload sizes are used. Additionally, we conducted scale-out simulation experiments modeling intra- and inter-node networks up to \num{4096} accelerators. We have also modeled realistic communication patterns based on the LLMs programming model, using different types of parallelism (data, pipeline, and tensor parallelism), which have been used to feed our simulation tool. Our analysis has identified a critical bottleneck at the interface between the intra-node network and the NIC link, which significantly impacts the performance of both intra- and inter-node networks and reduces overall system performance. We have also observed that increasing or decreasing the number of server nodes or increasing the speed of intra-node links does not alter this bottleneck's impact, as the intra-node network's limitations remain the dominant factor. Our simulation and the findings related to intra- and inter-node communication dynamics provide valuable insights to improve the design of large-scale intra- and inter-node interconnection networks.

\bmhead{Acknowledgements}

This work is supported by the Spanish Ministry of Science and Universities MCIN/AEI/10.13039/501100011033, the European Union (NextGenertionEU/PRTR) under project TED2021-130233B-C31, the Junta de Comunidades de Castilla-La Mancha and FEDER funds under the project SBPLY/21/180501/000248, the PERTE-Chip grants (UCLM Chair, TSI-069100-2023-0014) funded by the Spanish Ministry of Digital Transformation and Public Service, and the Universidad de Castilla-La Mancha under project 2023-GRIN-34056.

\bibliography{sn-bibliography}% common bib file


\begin{thebibliography}{24}
\ifx \bisbn   \undefined \def \bisbn  #1{ISBN #1}\fi
\ifx \binits  \undefined \def \binits#1{#1}\fi
\ifx \bauthor  \undefined \def \bauthor#1{#1}\fi
\ifx \batitle  \undefined \def \batitle#1{#1}\fi
\ifx \bjtitle  \undefined \def \bjtitle#1{#1}\fi
\ifx \bvolume  \undefined \def \bvolume#1{\textbf{#1}}\fi
\ifx \byear  \undefined \def \byear#1{#1}\fi
\ifx \bissue  \undefined \def \bissue#1{#1}\fi
\ifx \bfpage  \undefined \def \bfpage#1{#1}\fi
\ifx \blpage  \undefined \def \blpage #1{#1}\fi
\ifx \burl  \undefined \def \burl#1{\textsf{#1}}\fi
\ifx \doiurl  \undefined \def \doiurl#1{\url{https://doi.org/#1}}\fi
\ifx \betal  \undefined \def \betal{\textit{et al.}}\fi
\ifx \binstitute  \undefined \def \binstitute#1{#1}\fi
\ifx \binstitutionaled  \undefined \def \binstitutionaled#1{#1}\fi
\ifx \bctitle  \undefined \def \bctitle#1{#1}\fi
\ifx \beditor  \undefined \def \beditor#1{#1}\fi
\ifx \bpublisher  \undefined \def \bpublisher#1{#1}\fi
\ifx \bbtitle  \undefined \def \bbtitle#1{#1}\fi
\ifx \bedition  \undefined \def \bedition#1{#1}\fi
\ifx \bseriesno  \undefined \def \bseriesno#1{#1}\fi
\ifx \blocation  \undefined \def \blocation#1{#1}\fi
\ifx \bsertitle  \undefined \def \bsertitle#1{#1}\fi
\ifx \bsnm \undefined \def \bsnm#1{#1}\fi
\ifx \bsuffix \undefined \def \bsuffix#1{#1}\fi
\ifx \bparticle \undefined \def \bparticle#1{#1}\fi
\ifx \barticle \undefined \def \barticle#1{#1}\fi
\bibcommenthead
\ifx \bconfdate \undefined \def \bconfdate #1{#1}\fi
\ifx \botherref \undefined \def \botherref #1{#1}\fi
\ifx \url \undefined \def \url#1{\textsf{#1}}\fi
\ifx \bchapter \undefined \def \bchapter#1{#1}\fi
\ifx \bbook \undefined \def \bbook#1{#1}\fi
\ifx \bcomment \undefined \def \bcomment#1{#1}\fi
\ifx \oauthor \undefined \def \oauthor#1{#1}\fi
\ifx \citeauthoryear \undefined \def \citeauthoryear#1{#1}\fi
\ifx \endbibitem  \undefined \def \endbibitem {}\fi
\ifx \bconflocation  \undefined \def \bconflocation#1{#1}\fi
\ifx \arxivurl  \undefined \def \arxivurl#1{\textsf{#1}}\fi
\csname PreBibitemsHook\endcsname

%%% 1
\bibitem[\protect\citeauthoryear{Narayanan et~al.}{2021}]{EfficientLargeScale}
\begin{botherref}
\oauthor{\bsnm{Narayanan}, \binits{D.}},
\oauthor{\bsnm{Shoeybi}, \binits{M.}},
\oauthor{\bsnm{Casper}, \binits{J.}},
\oauthor{\bsnm{LeGresley}, \binits{P.}},
\oauthor{\bsnm{Patwary}, \binits{M.}},
\oauthor{\bsnm{Korthikanti}, \binits{V.A.}},
\oauthor{\bsnm{Vainbrand}, \binits{D.}},
\oauthor{\bsnm{Kashinkunti}, \binits{P.}},
\oauthor{\bsnm{Bernauer}, \binits{J.}},
\oauthor{\bsnm{Catanzaro}, \binits{B.}},
\oauthor{\bsnm{Phanishayee}, \binits{A.}},
\oauthor{\bsnm{Zaharia}, \binits{M.}}:
Efficient {Large}-{Scale} {Language} {Model} {Training} on {GPU} {Clusters} {Using} {Megatron}-{LM}.
arXiv.
arXiv:2104.04473 [cs]
(2021).
\url{http://arxiv.org/abs/2104.04473}
Accessed 2024-10-22
\end{botherref}
\endbibitem

%%% 2
\bibitem[\protect\citeauthoryear{De~Sensi et~al.}{2024}]{DeSensi24}
\begin{bchapter}
\bauthor{\bsnm{De~Sensi}, \binits{D.}},
\bauthor{\bsnm{Pichetti}, \binits{L.}},
\bauthor{\bsnm{Vella}, \binits{F.}},
\bauthor{\bsnm{De~Matteis}, \binits{T.}},
\bauthor{\bsnm{Ren}, \binits{Z.}},
\bauthor{\bsnm{Fusco}, \binits{L.}},
\bauthor{\bsnm{Turisini}, \binits{M.}},
\bauthor{\bsnm{Cesarini}, \binits{D.}},
\bauthor{\bsnm{Lust}, \binits{K.}},
\bauthor{\bsnm{Trivedi}, \binits{A.}},
\bauthor{\bsnm{Roweth}, \binits{D.}},
\bauthor{\bsnm{Spiga}, \binits{F.}},
\bauthor{\bsnm{Di~Girolamo}, \binits{S.}},
\bauthor{\bsnm{Hoefler}, \binits{T.}}:
\bctitle{Exploring gpu-to-gpu communication: Insights into supercomputer interconnects}.
In: \bbtitle{SC24: International Conference for High Performance Computing, Networking, Storage and Analysis},
pp. \bfpage{1}--\blpage{15}
(\byear{2024}).
\doiurl{10.1109/SC41406.2024.00039}
\end{bchapter}
\endbibitem

%%% 3
\bibitem[\protect\citeauthoryear{Binkert et~al.}{2011}]{gem5}
\begin{barticle}
\bauthor{\bsnm{Binkert}, \binits{N.}},
\bauthor{\bsnm{Beckmann}, \binits{B.}},
\bauthor{\bsnm{Black}, \binits{G.}},
\bauthor{\bsnm{Reinhardt}, \binits{S.K.}},
\bauthor{\bsnm{Saidi}, \binits{A.}},
\bauthor{\bsnm{Basu}, \binits{A.}},
\bauthor{\bsnm{Hestness}, \binits{J.}},
\bauthor{\bsnm{Hower}, \binits{D.R.}},
\bauthor{\bsnm{Krishna}, \binits{T.}},
\bauthor{\bsnm{Sardashti}, \binits{S.}},
\bauthor{\bsnm{Sen}, \binits{R.}},
\bauthor{\bsnm{Sewell}, \binits{K.}},
\bauthor{\bsnm{Shoaib}, \binits{M.}},
\bauthor{\bsnm{Vaish}, \binits{N.}},
\bauthor{\bsnm{Hill}, \binits{M.D.}},
\bauthor{\bsnm{Wood}, \binits{D.A.}}:
\batitle{The gem5 simulator}.
\bjtitle{SIGARCH Comput. Archit. News}
\bvolume{39}(\bissue{2}),
\bfpage{1}--\blpage{7}
(\byear{2011})
\doiurl{10.1145/2024716.2024718}
\end{barticle}
\endbibitem

%%% 4
\bibitem[\protect\citeauthoryear{Bhowmik et~al.}{2021}]{Bhowmik21}
\begin{bchapter}
\bauthor{\bsnm{Bhowmik}, \binits{S.}},
\bauthor{\bsnm{Jain}, \binits{N.}},
\bauthor{\bsnm{Yuan}, \binits{X.}},
\bauthor{\bsnm{Bhatele}, \binits{A.}}:
\bctitle{A simulation study of hardware parameters for future gpu-based hpc platforms}.
In: \bbtitle{2021 IEEE International Performance, Computing, and Communications Conference (IPCCC)},
pp. \bfpage{1}--\blpage{10}
(\byear{2021}).
\doiurl{10.1109/IPCCC51483.2021.9679359}
\end{bchapter}
\endbibitem

%%% 5
\bibitem[\protect\citeauthoryear{Jain et~al.}{2016}]{Jain16}
\begin{bchapter}
\bauthor{\bsnm{Jain}, \binits{N.}},
\bauthor{\bsnm{Bhatele}, \binits{A.}},
\bauthor{\bsnm{White}, \binits{S.}},
\bauthor{\bsnm{Gamblin}, \binits{T.}},
\bauthor{\bsnm{Kale}, \binits{L.V.}}:
\bctitle{Evaluating hpc networks via simulation of parallel workloads}.
In: \bbtitle{Proceedings of the International Conference for High Performance Computing, Networking, Storage and Analysis}.
\bsertitle{SC '16}.
\bpublisher{IEEE Press},
\blocation{Salt Lake City, Utah}
(\byear{2016})
\end{bchapter}
\endbibitem

%%% 6
\bibitem[\protect\citeauthoryear{}{2024}]{H100}
\begin{botherref}
{NVIDIA H100 Tensor Core GPU Architecture Overview}.
[Online; accessed 18. Dec. 2024]
(2024).
\url{https://resources.nvidia.com/en-us-tensor-core/gtc22-whitepaper-hopper}
\end{botherref}
\endbibitem

%%% 7
\bibitem[\protect\citeauthoryear{}{2024}]{GraceHopper}
\begin{botherref}
{NVIDIA Grace Hopper Superchip Architecture Whitepaper}.
[Online; accessed 18. Dec. 2024]
(2024).
\url{https://resources.nvidia.com/en-us-grace-cpu/nvidia-grace-hopper}
\end{botherref}
\endbibitem

%%% 8
\bibitem[\protect\citeauthoryear{}{2024}]{Gaudi3}
\begin{botherref}
{Intel{\ifmmode\circledR\else\textregistered\fi} Gaudi{\ifmmode\circledR\else\textregistered\fi} 3 AI Accelerator White Paper}.
[Online; accessed 18. Dec. 2024]
(2024).
\url{https://www.intel.com/content/www/us/en/content-details/817486/intel-gaudi-3-ai-accelerator-white-paper.html}
\end{botherref}
\endbibitem

%%% 9
\bibitem[\protect\citeauthoryear{Hoefler et~al.}{2022}]{Hoefler22Convergence}
\begin{barticle}
\bauthor{\bsnm{Hoefler}, \binits{T.}},
\bauthor{\bsnm{Hendel}, \binits{A.}},
\bauthor{\bsnm{Roweth}, \binits{D.}}:
\batitle{The convergence of hyperscale data center and high-performance computing networks}.
\bjtitle{Computer}
\bvolume{55}(\bissue{7}),
\bfpage{29}--\blpage{37}
(\byear{2022})
\doiurl{10.1109/MC.2022.3158437}
\end{barticle}
\endbibitem

%%% 10
\bibitem[\protect\citeauthoryear{Biagioni and et~al.}{2022}]{Exascale}
\begin{bchapter}
\bauthor{\bsnm{Biagioni}, \binits{A.}},
\bauthor{\bsnm{al.}}:
\bctitle{Red-sea: Network solution for exascale architectures}.
In: \bbtitle{2022 25th Euromicro Conference on Digital System Design (DSD)},
pp. \bfpage{712}--\blpage{719}
(\byear{2022}).
\doiurl{10.1109/DSD57027.2022.00100}
\end{bchapter}
\endbibitem

%%% 11
\bibitem[\protect\citeauthoryear{Borkar and Chien}{2019}]{Borkar19Heterogeneous}
\begin{barticle}
\bauthor{\bsnm{Borkar}, \binits{S.}},
\bauthor{\bsnm{Chien}, \binits{A.A.}}:
\batitle{The future of microprocessors}.
\bjtitle{Communications of the ACM}
\bvolume{62}(\bissue{5}),
\bfpage{44}--\blpage{52}
(\byear{2019})
\doiurl{10.1145/3292341}
\end{barticle}
\endbibitem

%%% 12
\bibitem[\protect\citeauthoryear{Shallue et~al.}{2018}]{DataParallelism}
\begin{botherref}
\oauthor{\bsnm{Shallue}, \binits{C.J.}},
\oauthor{\bsnm{Lee}, \binits{J.}},
\oauthor{\bsnm{Antognini}, \binits{J.}},
\oauthor{\bsnm{Sohl-Dickstein}, \binits{J.}},
\oauthor{\bsnm{Frostig}, \binits{R.}},
\oauthor{\bsnm{Dahl}, \binits{G.E.}}:
Measuring the effects of data parallelism on neural network training.
arXiv preprint arXiv:1811.03600
(2018)
\end{botherref}
\endbibitem

%%% 13
\bibitem[\protect\citeauthoryear{Shoeybi et~al.}{2019}]{TensorParallelism}
\begin{botherref}
\oauthor{\bsnm{Shoeybi}, \binits{M.}},
\oauthor{\bsnm{Patwary}, \binits{M.}},
\oauthor{\bsnm{Puri}, \binits{R.}},
\oauthor{\bsnm{LeGresley}, \binits{P.}},
\oauthor{\bsnm{Casper}, \binits{J.}},
\oauthor{\bsnm{Catanzaro}, \binits{B.}}:
Megatron-lm: Training multi-billion parameter language models using model parallelism.
arXiv preprint:1909.08053
(2019)
\end{botherref}
\endbibitem

%%% 14
\bibitem[\protect\citeauthoryear{Huang et~al.}{2019}]{PipelineParallelism}
\begin{botherref}
\oauthor{\bsnm{Huang}, \binits{Y.}},
\oauthor{\bsnm{Firat}, \binits{O.}},
\oauthor{\bsnm{Lee}, \binits{H.}},
\oauthor{\bsnm{Wu}, \binits{Y.}},
\oauthor{\bsnm{Cheng}, \binits{Y.}},
\oauthor{\bsnm{XuChen}, \binits{M.}},
\oauthor{\bsnm{Ngiam}, \binits{J.}},
\oauthor{\bsnm{Bapna}, \binits{A.}},
\oauthor{\bsnm{Chen}, \binits{D.}},
\oauthor{\bsnm{Le}, \binits{Q.V.}},
\oauthor{\bsnm{Chen}, \binits{Z.}}:
Gpipe: Easy scaling with micro-batch pipeline parallelism.
arXiv preprint arXiv:1811.06965
(2019)
\end{botherref}
\endbibitem

%%% 15
\bibitem[\protect\citeauthoryear{Yébenes et~al.}{2016}]{SAURON}
\begin{bchapter}
\bauthor{\bsnm{Yébenes}, \binits{P.}},
\bauthor{\bsnm{Escudero-Sahuquillo}, \binits{J.}},
\bauthor{\bsnm{García}, \binits{P.J.}},
\bauthor{\bsnm{Alfaro}, \binits{F.J.}},
\bauthor{\bsnm{Quiles}, \binits{F.J.}}:
\bctitle{Providing differentiated services, congestion management, and deadlock freedom in dragonfly networks}.
In: \bbtitle{2016 2nd IEEE International Workshop on High-Performance Interconnection Networks in the Exascale and Big-Data Era (HiPINEB)},
pp. \bfpage{33}--\blpage{40}
(\byear{2016}).
\doiurl{10.1109/HIPINEB.2016.11}
\end{bchapter}
\endbibitem

%%% 16
\bibitem[\protect\citeauthoryear{Suzuki et~al.}{2014}]{HeaderOverhead}
\begin{bchapter}
\bauthor{\bsnm{Suzuki}, \binits{J.}},
\bauthor{\bsnm{Hayashi}, \binits{Y.}},
\bauthor{\bsnm{Kan}, \binits{M.}},
\bauthor{\bsnm{Miyakawa}, \binits{S.}},
\bauthor{\bsnm{Yoshikawa}, \binits{T.}}:
\bctitle{End-to-end adaptive packet aggregation for high-throughput i/o bus network using ethernet}.
In: \bbtitle{2014 IEEE 22nd Annual Symposium on High-Performance Interconnects},
pp. \bfpage{17}--\blpage{24}
(\byear{2014}).
\doiurl{10.1109/HOTI.2014.16}
\end{bchapter}
\endbibitem

%%% 17
\bibitem[\protect\citeauthoryear{NVIDIA}{2023}]{perftest}
\begin{botherref}
\oauthor{\bsnm{NVIDIA}}:
{Perftest Package}.
[Online; accessed 11. Mar. 2024]
(2023).
\url{https://enterprise-support.nvidia.com/s/article/perftest-package}
\end{botherref}
\endbibitem

%%% 18
\bibitem[\protect\citeauthoryear{Schieffer et~al.}{2024}]{schieffer2024}
\begin{botherref}
\oauthor{\bsnm{Schieffer}, \binits{G.}},
\oauthor{\bsnm{Shi}, \binits{R.}},
\oauthor{\bsnm{Markidis}, \binits{S.}},
\oauthor{\bsnm{Herten}, \binits{A.}},
\oauthor{\bsnm{Faj}, \binits{J.}},
\oauthor{\bsnm{Peng}, \binits{I.}}:
Understanding Data Movement in AMD Multi-GPU Systems with Infinity Fabric
(2024).
\url{https://arxiv.org/abs/2410.00801}
\end{botherref}
\endbibitem

%%% 19
\bibitem[\protect\citeauthoryear{McKeown}{1999}]{iSlip}
\begin{barticle}
\bauthor{\bsnm{McKeown}, \binits{N.}}:
\batitle{The islip scheduling algorithm for input-queued switches}.
\bjtitle{IEEE/ACM Transactions on Networking}
\bvolume{7}(\bissue{2}),
\bfpage{188}--\blpage{201}
(\byear{1999})
\doiurl{10.1109/90.769767}
\end{barticle}
\endbibitem

%%% 20
\bibitem[\protect\citeauthoryear{Mahanta et~al.}{2015}]{RLFT}
\begin{bchapter}
\bauthor{\bsnm{Mahanta}, \binits{H.J.}},
\bauthor{\bsnm{Biswas}, \binits{A.}},
\bauthor{\bsnm{Hussain}, \binits{A.}}:
\bctitle{An architecture based routing for heterogeneous fat tree network on chip}.
In: \bbtitle{2015 International Symposium on Advanced Computing and Communication (ISACC)},
pp. \bfpage{341}--\blpage{345}
(\byear{2015}).
\doiurl{10.1109/ISACC.2015.7377366}
\end{bchapter}
\endbibitem

%%% 21
\bibitem[\protect\citeauthoryear{Zahavi}{2012}]{Zahavi12dmodk}
\begin{barticle}
\bauthor{\bsnm{Zahavi}, \binits{E.}}:
\batitle{Fat-tree routing and node ordering providing contention free traffic for mpi global collectives}.
\bjtitle{Journal of Parallel and Distributed Computing}
\bvolume{72}(\bissue{11}),
\bfpage{1423}--\blpage{1432}
(\byear{2012})
\doiurl{10.1016/j.jpdc.2012.01.018} .
\bcomment{Communication Architectures for Scalable Systems}
\end{barticle}
\endbibitem

%%% 22
\bibitem[\protect\citeauthoryear{Gomez et~al.}{2007}]{Routing}
\begin{bchapter}
\bauthor{\bsnm{Gomez}, \binits{C.}},
\bauthor{\bsnm{Gilabert}, \binits{F.}},
\bauthor{\bsnm{Gomez}, \binits{M.E.}},
\bauthor{\bsnm{Lopez}, \binits{P.}},
\bauthor{\bsnm{Duato}, \binits{J.}}:
\bctitle{Deterministic versus adaptive routing in fat-trees}.
In: \bbtitle{2007 IEEE International Parallel and Distributed Processing Symposium},
pp. \bfpage{1}--\blpage{8}
(\byear{2007}).
\doiurl{10.1109/IPDPS.2007.370482}
\end{bchapter}
\endbibitem

%%% 23
\bibitem[\protect\citeauthoryear{Isaev et~al.}{2023}]{calculon}
\begin{bchapter}
\bauthor{\bsnm{Isaev}, \binits{M.}},
\bauthor{\bsnm{Mcdonald}, \binits{N.}},
\bauthor{\bsnm{Dennison}, \binits{L.}},
\bauthor{\bsnm{Vuduc}, \binits{R.}}:
\bctitle{{Calculon: a methodology and tool for high-level co-design of systems and large language models}}.
In: \bbtitle{{ACM Conferences}},
pp. \bfpage{1}--\blpage{14}.
\bpublisher{Association for Computing Machinery},
\blocation{New York, NY, USA}
(\byear{2023}).
\doiurl{10.1145/3581784.3607102}
\end{bchapter}
\endbibitem

%%% 24
\bibitem[\protect\citeauthoryear{Villalobos et~al.}{2022}]{LLMParamsIncrease}
\begin{botherref}
\oauthor{\bsnm{Villalobos}, \binits{P.}},
\oauthor{\bsnm{Sevilla}, \binits{J.}},
\oauthor{\bsnm{Besiroglu}, \binits{T.}},
\oauthor{\bsnm{Heim}, \binits{L.}},
\oauthor{\bsnm{Ho}, \binits{A.}},
\oauthor{\bsnm{Hobbhahn}, \binits{M.}}:
Machine Learning Model Sizes and the Parameter Gap
(2022).
\url{https://arxiv.org/abs/2207.02852}
\end{botherref}
\endbibitem

\end{thebibliography}

\end{document}